\documentclass[twocolumn]{aastex63} 

\usepackage{amsmath, bm}

\newcommand{\about}{\mbox{$\sim$}}               
\newcommand{\frest}{\mbox{$f_{\rm rest}$}}      
\newcommand{\Lbol}{\mbox{$L_{\rm bol}$}}                    
\newcommand{\Lir}{\mbox{$L_{\rm IR}$}}  
\newcommand{\Lsun}{\mbox{$L_\odot$}}            
\newcommand{\NH}{\mbox{$N_{\rm H}$}}             
\newcommand{\Tb}{\mbox{$T_{\rm b}$}}               
\newcommand{\Vmin}{\mbox{$V_{\rm min}$}}        
\newcommand{\Vabs}{\mbox{$V_{\rm abs}$}}        
\newcommand{\Vsys}{\mbox{$V_{\rm sys}$}}        
\newcommand{\perbeam}{\mbox{beam$^{-1}$}}                         
\newcommand{\persquarecm}{\mbox{cm$^{-2}$}}                      
\newcommand{\kms}{\mbox{km s$^{-1}$}}                                    

\newcommand{\persquare}[1]{\mbox{{#1}$^{-2}$}}

\newcommand{\plus}{\mbox{$+$}}    
\newcommand{\minus}{\mbox{$-$}}  

\newcommand{\twelveCO}{\mbox{$^{12}$CO}}                  
\newcommand{\thirteenCO}{\mbox{$^{13}$CO}}                
\newcommand{\CeighteenO}{\mbox{C$^{18}$O}}              
\newcommand{\sixteenO}{\mbox{$^{16}$O}}                      
\newcommand{\eighteenO}{\mbox{$^{18}$O}}                    
\newcommand{\HH}{\mbox{H$_2$}}                                      
\newcommand{\water}{\mbox{H$_2$O}}                              
\newcommand{\HtwoS}{\mbox{H$_2$S}}                             
\newcommand{\CtwoH}{\mbox{C$_2$H}}                             
\newcommand{\HthirteenCN}{\mbox{H$^{13}$CN}}          
\newcommand{\HNthirteenC}{\mbox{HN$^{13}$C}}          
\newcommand{\HCthreeN}{\mbox{HC$_{3}$N}}                 
\newcommand{\HtwoCO}{\mbox{H$_2$CO}}                       
\newcommand{\HCOplus}{\mbox{HCO$^{+}$}}                    
\newcommand{\HthirteenCOplus}{\mbox{H$^{13}$CO$^{+}$}}     
\newcommand{\NHtwoCN}{\mbox{NH$_{2}$CN}}             
                 
\newcommand{\CthirtyfourS}{\mbox{C$^{34}$S}}              
\newcommand{\thirteenCS}{\mbox{$^{13}$CS}}                 
\newcommand{\HthreeOplus}{\mbox{H$_{3}$O$^{+}$}}        
\newcommand{\HtwoO}{\mbox{H$_{2}$O}}                         

\newcommand{\vtwo}{\mbox{$v_2$}}
\newcommand{\vseven}{\mbox{$v_7$}}
\newcommand{\rotJ}{\mbox{$J$}}

\newcommand{\tnm}[1]{\tablenotemark{#1}}
\newcommand{\citest}[1]{\citeauthor*{#1}}
\newcommand{\citesp}[1]{(\citeauthor*{#1})}

\newcommand{\arpE}{Arp~220E}
\newcommand{\arpW}{Arp~220W}

\shorttitle{ALMA 1.4--0.4 mm Spectral Scan on NGC 4418 and Arp 220: II. Lines }
\shortauthors{SAKAMOTO et al.}

\graphicspath{{./}{figures/}}

\begin{document}
\title{Deeply Buried Nuclei in the Infrared-Luminous Galaxies NGC 4418 and Arp 220 \\
II. Line Forests at $\lambda = $1.4--0.4 mm and Circumnuclear Gas Observed with ALMA}

\author{Kazushi Sakamoto}
\affiliation{Academia Sinica, Institute of Astronomy and Astrophysics, Taipei, Taiwan}

\author{Sergio Mart\'{i}n}
\affiliation{European Southern Observatory, Alonso de C\'{o}rdova 3107, Vitacura Casilla 763 0355, Santiago, Chile}
\affiliation{Joint ALMA Observatory, Alonso de C\'{o}rdova 3107, Vitacura 763 0355, Santiago, Chile}

\author{David J. Wilner}
\affiliation{Harvard-Smithsonian Center for Astrophysics, 60 Garden Street, Cambridge, MA 02138, USA}

\author{Susanne Aalto}
\affiliation{Department of Earth and Space Sciences, Chalmers University of Technology, Onsala Observatory, 439 92 Onsala, Sweden}

\author{Aaron S. Evans}
\affiliation{Department of Astronomy, University of Virginia, P.O. Box 400325, Charlottesville, VA 22904, USA}
\affiliation{National Radio Astronomy Observatory, 520 Edgemont Road, Charlottesville, VA 22903, USA}

\author{Nanase Harada}
\affiliation{Academia Sinica, Institute of Astronomy and Astrophysics, Taipei, Taiwan}
\affiliation{National Astronomical Observatory of Japan, Mitaka, Tokyo, 181-8588, Japan}

\begin{abstract}
We present the line observations in our ALMA imaging spectral scan toward  
three deeply buried nuclei in \object{NGC~4418} and \object{Arp~220}. 
We cover 67 GHz in \frest=215--697 GHz at about 0\farcs2 (30, 80 pc) resolution.
All the nuclei show dense line forests; we report our initial line identification using 55 species.
The line velocities generally indicate gas rotation around each nucleus, tracing nuclear disks of \about100 pc sizes.
We confirmed the counter-rotation of the nuclear disks in Arp 220 and
that of the nuclear disk and the galactic disk in NGC 4418.
While the brightest lines exceed 100 K, 
most of the major lines and many $^{13}$C isotopologues show absorption against even brighter 
continuum cores of the nuclei.
The lines with higher upper-level energies, including those from vibrationally-excited molecules,  
tend to arise from smaller areas, indicating radially varying conditions in these nuclei. 
The outflows from the two Arp 220 nuclei cause blueshifted line absorption below the continuum level.
The absorption mostly has small spatial offsets from the continuum peaks to indicate the outflow orientations.
The bipolar outflow from the western nucleus is also imaged in multiple emission lines, 
showing the extent of \about1\arcsec\ (400 pc). 
Redshifted line absorption against the nucleus of NGC 4418 indicates either an inward gas motion 
or a small collimated outflow slanted to the nuclear disk.   
We also resolved some previous confusions due to line blending and misidentification. 
\end{abstract}

       
\begin{flushright}
Accepted for publication in ApJ
\end{flushright}

\section{Introduction}  
\label{s.introduction}
The infrared-luminous galaxies NGC 4418 and Arp 220 have local prototypes of extremely obscured, 
luminous, and compact galactic nuclei.
NGC 4418 ($D = 34$ Mpc; 1\arcsec=165 pc) is an Sa-type galaxy 
having 8--1000 \micron\ luminosity \Lir\ of $10^{11.2}$ \Lsun\ \citep{Armus09}. 
Arp 220 ($D = 85$ Mpc; 1\arcsec=412 pc) has  $\Lir = 10^{12.3}$ \Lsun\ \citep{Armus09}
and is a galaxy merger whose two nuclei are \about1\arcsec\ apart from each other;
we call the eastern nucleus \arpE\ and the western one \arpW.
NGC 4418 is a luminous infrared galaxy (LIRG, $\log \Lir/\Lsun = $11--12), 
and Arp 220 is the nearest ultraluminous infrared galaxy \citep[ULIRG, $\log \Lir/\Lsun = $12--13;][]{SM96}.

These nuclei have obscuring column densities as large as  
\about$10^{25}$--$10^{26}$ H \persquarecm\ 
\citep{GA04, GA12, Downes07, Sakamoto08, Sakamoto13, Paper1, Scoville17, GS19, Dwek20}.
They are bright and compact ($\lesssim 100$ pc) in imaging 
in mid-IR \citep{WWB93, Soifer99, Evans03}, submillimeter \citep{Sakamoto08, Sakamoto13, Wilson14}, 
millimeter \citep{Downes98,Scoville17,Sakamoto17}, and centimeter wavelengths \citep{Norris88, Condon90, Condon91, Barcos-Munoz15}.
The compact nuclei generate the bulk of the vast infrared luminosities of the individual galaxies:
it has been suggested by the high brightness temperature of the submillimeter dust emission, 
modeling of the spectral energy distribution \citep{Soifer99, Dwek20}, 
and that of far-IR spectroscopy \citep{GA12}.

Compact starburst and active galactic nucleus (AGN) have been the most plausible luminosity sources of the nuclei
\citep[see][for early proposals]{Baan82, Soifer84, Rieke85, Roche86, RRC89}.
Vigorous star formation is evident by now in all three nuclei through groups of  supernovae, supernova remnants, 
and super-star clusters seen as compact radio sources \citep{Smith98,Varenius14,Varenius19}. 
In contrast, AGNs in these extinguished nuclei have been difficult to detect and more so to quantify. 
For example, \citet{Veilleux09} used six mid-IR diagnostics for the fractional AGN contribution to the bolometric
luminosity in nearby ULIRGs and quasars. 
The AGN fraction ranges between 0 and $<$37\% with a mean of $<$18\% for Arp 220,
while it is $>$50\% in two of the diagnostics applied to NGC 4418 \citep{Veilleux09,Veilleux13,Rosenberg15}.
Meanwhile, the two galaxies are starburst templates in another infrared diagnostic of AGN \citep{Fadda14}.
We conservatively view AGNs in the three nuclei as likely but unproven possibilities.
In any case, the nuclei of NGC 4418 and Arp 220 are our local prototypes of either
the most extreme starbursts, 
most heavily obscured AGN that are still in dusty cocoons and hence young,
or possible sites of starburst-AGN coevolution.

High-resolution imaging spectroscopy of these nuclei at millimeter and submillimeter wavelengths have
proven useful to study their properties and evolution. 
It required sub-arcsec resolution CO images to find twin compact ($\lesssim$ 100 pc) peaks of molecular gas at the two
radio nuclei of Arp 220 \citep{Scoville97, Downes98}.
A 0\farcs5 resolution spectro-imaging was necessary to find them to be counter-rotating nuclear disks \citep{Sakamoto99}.
Molecular outflows from both nuclei were found from blueshifted absorption lines (below the continuum level) 
in each nucleus at 0\farcs3 resolution \citep{Sakamoto09}. 
The outflow from \arpW\ was estimated to be in the polar direction of the nucleus disk 
on the basis of a small spatial offset of SiO absorption \citep{Tunnard15}. 
This configuration is consistent with the one based on a 150 MHz image of the nucleus \citep{Varenius16}.
The bipolar shape of this outflow has been imaged in the 3 mm continuum \citep{Sakamoto17} 
and molecular lines \citep{Barcos-Munoz18,Wheeler20}.
NGC 4418 also has a massive central reservoir of molecular gas strongly concentrated 
to the central $\lesssim$100 pc \citep{Kawara90, Imanishi04, Sakamoto13}. 
A kpc scale polar outflow has been found in dust extinction while the molecular line profiles toward the nucleus were found skewed to suggest redshifted absorption \citep{Sakamoto13}, which is
also seen in infrared lines \citep{GA12, Veilleux13, Rosenberg15}.
The detection of vibrationally-excited HCN toward the nucleus of NGC 4418 
\citep{Sakamoto10, Sakamoto13} and the two nuclei of Arp 220
\citep{Aalto15, Martin16} suggested hot dust and strong mid-IR radiation inside these nuclei
for excitation to the levels \about1000 K above the ground.

The presence and properties of warm gas in the nuclei of NGC 4418 and Arp 220 have been also 
traced in far-IR and submillimeter lines using space-borne telescopes \citep[e.g.,][]{GA04, GA12, Rangwala11, Rosenberg15}.
These observations with large apertures do not resolve the nuclei but recover some spatial information
by the modeling of high-excitation molecular and atomic lines detected in a wide bandwidth, typically spanning \about0.5 dex.
For example, radial gas motions are seen in Doppler-shifted absorption in NGC 4418 and Arp 220.
The high-resolution imaging spectroscopy with ground-based radio interferometers is complementary to 
the far-infrared spectroscopy from space. 
The former would be more powerful if it has bandwidth as wide as in the latter.

We set out for high-resolution and wide-band imaging spectroscopy of NGC 4418 and Arp 220 
using the Atacama Millimeter-Submillimeter Array (ALMA).
In the observations reported here, we used ten frequency tunings to cover major lines and continuum emission 
in ALMA Bands 6, 7, and 9,
sampling 67 GHz from the rest-frequency range of 215--697 GHz ($\lambda_\mathrm{rest}$=1.4--0.43 mm) 
with a target resolution of 0\farcs2.
With the high angular resolution, about 30 and 80 pc at each galaxy, we can not only separately observe the two nuclei of Arp 220
and isolate the circumnuclear gas from the gas far out in the galaxy disks but also start to resolve the structures 
inside the three nuclei.  
\citet[hereafter \citest{Paper1}]{Paper1} reported in detail our ALMA observations, 
data reduction, and analysis of continuum emission of the three nuclei.
The continuum emission is strongly concentrated in all three nuclei, and each nucleus needs a few concentric
Gaussians to fit the intensity profile.
Their core components are found to be as small as \about0\farcs05 (8--20 pc) and as bright as 200--500 K in 
deconvolved peak brightness temperature. 
Each nucleus also has an oval feature that is several times larger than the core and has its major axis aligned to
the known gas velocity gradient in the nucleus; we identified it with the nuclear disk. 
We also detected extended features around the nuclei in dust emission at 0.4 mm.
The most notable was a faint bipolar feature nearly perpendicular to the nuclear disk of Arp 220W. 
We identified it with the known bipolar outflow from the nucleus. 
Our continuum analysis also provided the dust opacities and through which the column densities in each nucleus.
The nuclei of NGC 4418 and Arp 220W were estimated to be opaque at 1 mm.  
Their 1 mm dust opacities, between the center and the surface, are on the order of 1, 
which translates to  \NH\ \about $10^{26}$ \persquare{cm} for the dust-opacity law we adopted.
Arp 220E was found to be several times less opaque.

This paper presents line observations from our spectral scan toward the three deeply buried nuclei in NGC 4418 and Arp 220.
It follows up our interferometric spectral scans on these galaxies at lower resolution \citep{Costagliola15, Martin11} and
our high-resolution sub/millimeter studies of these nuclei \citep[and those already mentioned]{Sakamoto08,Costagliola13,Aalto15,Martin16, Sakamoto17}. 
There is also a companion ALMA spectral scan of Arp 220 at \about0\farcs6 in Bands 6 and 7;
a part of it is in \citet{Martin16}, and full results will be reported elsewhere.
Our high spatial resolution allows a systematic line census in the immediate vicinities of the individual luminous nuclei,
and our wide bandwidth helps line identification using multiple transitions of individual species. 
Features (e.g., outflow signatures) that are seen in multiple lines are more credible, 
and trends across many lines inform us about the gas conditions in the nuclei.
With the combination of NGC 4418 and Arp 220, we expect that NGC 4418 having narrower lines than Arp 220 guide us 
to de-blend the broad lines in Arp 220. We also anticipate the comparison of the three nuclei to be informative.
Beyond our initial line identification, we analyze such basic properties of the nuclei as their gas kinematics, nuclear disk and outflow structures, and relative configurations between the Arp 220 nuclei and between the nucleus and the main disk of NGC 4418. 
Some earlier observations based on limited frequency coverage are also reevaluated.

The outline of this paper is as follows. 
Section \ref{s.obs} recapitulates our observations and data reduction reported in \citest{Paper1}.
Section \ref{s.Vsys_lineID} describes our choice of fiducial (systemic) velocities for the nuclei and
our initial line identification in the data. 
Section \ref{s.lineIm} presents continuum-subtracted images of the brightest lines,
and describes the most noticeable features and properties of the three nuclei.
They include the size-excitation relation among lines, velocity structure around each nucleus,
outflows from the nuclei, and vibrationally excited lines.
Section \ref{s.followup} revisits some previous observations in which line-blending adversely affected the interpretation.
Section \ref{s.summary} is for a summary of our findings and concluding remarks.

\section{ALMA Observations and Data Reduction}
\label{s.obs}
Table \ref{t.obsSummary} is a summary of our ALMA observations; see \citest{Paper1} for a full description.
We scanned 67 GHz from within \frest=215--697 GHz in ALMA Bands 6, 7, and 9.
Table \ref{t.frestCoverage} has the exact frequency coverages. 
The coverages on NGC 4418 and Arp 220 match in Band 7 and have minor differences in Band 6 and 9
mainly to help continuum measurements. 
We observed a single position in each galaxy.
The full width at half maximum (FWHM) of our primary beam ranges between 24\arcsec\ to 8\arcsec\
in our observing frequencies.
Our target resolution was 0\farcs2 and the achieved resolution ranged between 0\farcs14 and 0\farcs28
while the largest beam major-axis was 0\farcs35.
The data were recorded at the frequency resolution of 0.98 MHz so that they can be flexibly binned later 
for individual purposes. 
The data sensitivity for a 50 \kms\ resolution is typically 0.3--0.6 K at the native resolution (of 0\farcs14--0\farcs28)
and 0.1--0.2 K in the data smoothed to 0\farcs35 resolution.

We calibrated the data of ten tunings for each galaxy with special emphasis on consistency.
We aligned astrometry through self-calibration using common models 
and aligned flux scaling through flux self-calibration (i.e., adjustment of flux scales of adjacent tunings 
using small overlaps between them.) 
We constructed continuum models using measurements at rare gaps in the line forest
and subtracted the models from our data for consistent continuum subtraction.
Details of these calibrations are in \citest{Paper1}.

\section{Systemic Velocities and Line Identification} 
\label{s.Vsys_lineID}
The first step toward line identification is to determine the systemic velocities of the targets.
The observed spectra are then presented in the rest frequency and compared with simulated spectra
for line identification.

\subsection{Systemic Velocities}
\label{s.Vsys_lineID.vsys}
\begin{figure*}[!th]
\plottwo{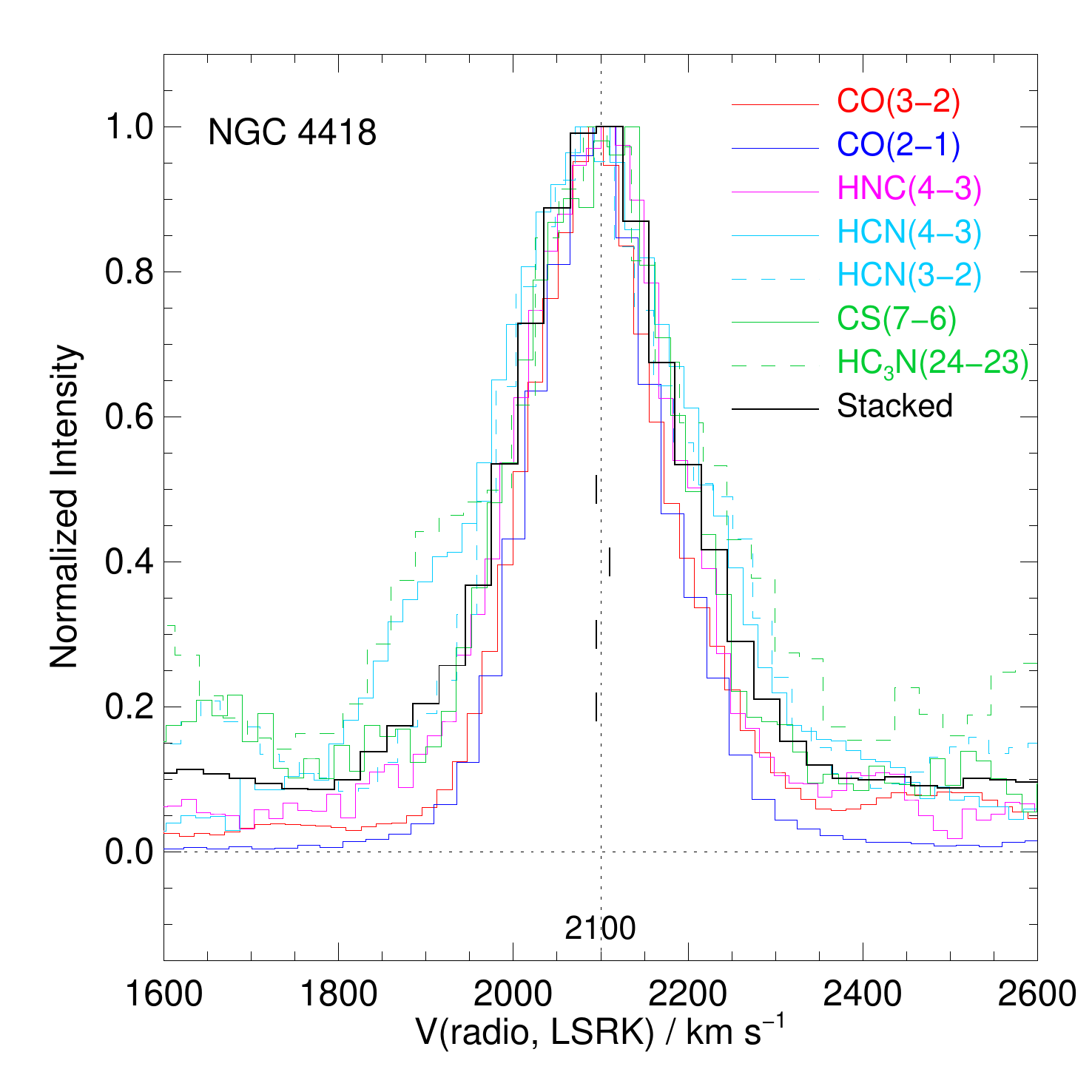}{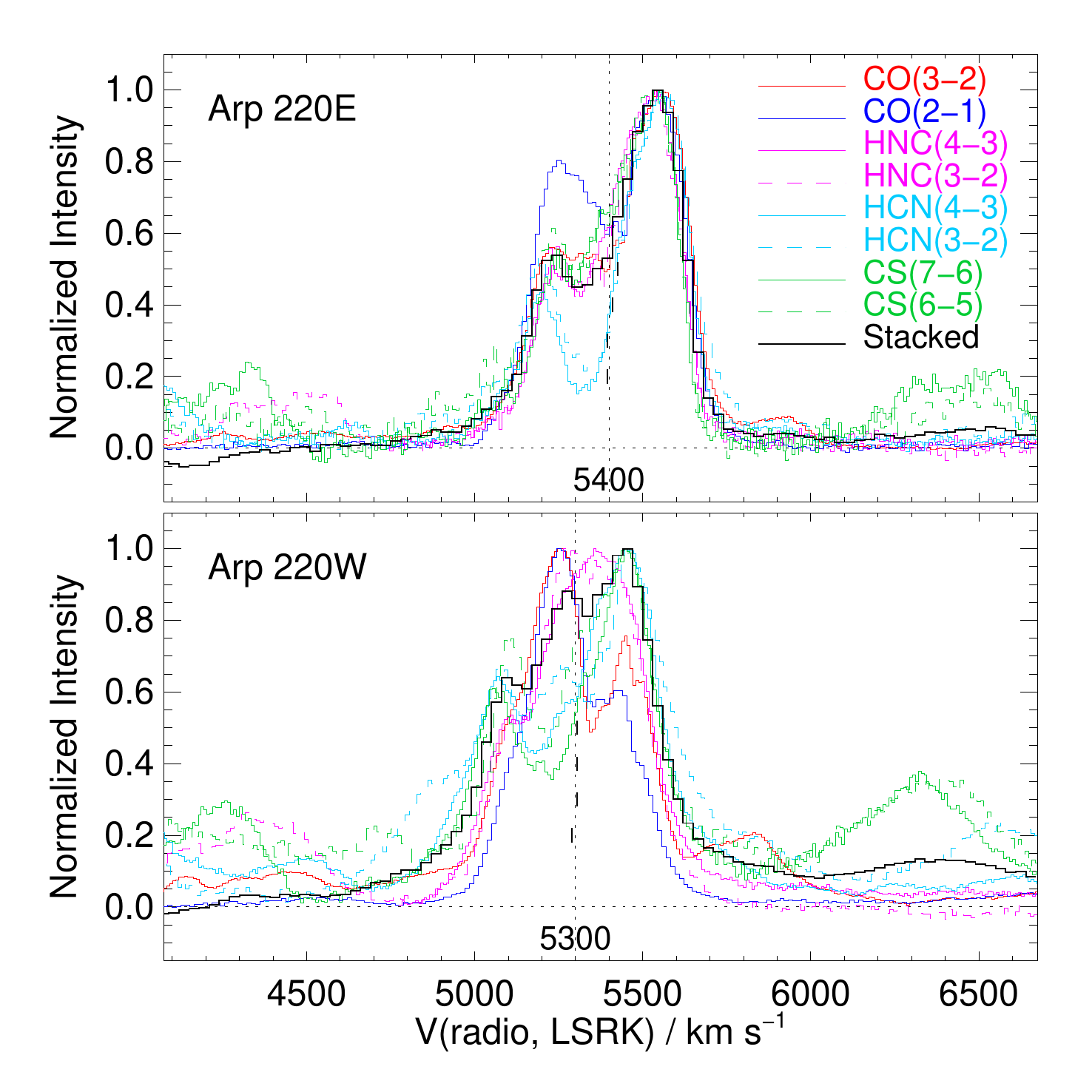} 
\caption{ \label{f.vsys}
Continuum-subtracted line profiles of NGC 4418 (left) and Arp 220 nuclei (right) to estimate their velocities.
The data of NGC 4418 are from this work and are from the 0\farcs35 resolution data sampled in a 1\arcsec-diameter aperture.
Arp 220 data are from our companion spectral scan at resolutions of 0\farcs65 -- 0\farcs82.
Individual line profiles are normalized with their peak values and are averaged to produce the mean line profile
whose normalized shape is shown in black.
The central velocities of the mean profile at the levels of 20, 30, 40, and 50\% of its peak are shown with black vertical bars.
The fiducial velocities of the nuclei that we adopted are indicated with vertical dotted lines.
}
\end{figure*}

We adopt fiducial systemic velocities of 2100 \kms\ for NGC 4418, 5400 \kms\ for Arp 220E, and 5300 \kms\ for Arp 220W.
They are in the LSRK frame and the radio convention, i.e., $v=(\nu_{\rm obs} - \nu_{\rm rest})/\nu_{\rm rest}$.
We decided these velocities by averaging the spectra of several brighter lines to minimize the biases due to
blended lines and skewed line profiles caused by, e.g., self-absorption.
Figure~\ref{f.vsys} shows the continuum-subtracted spectra of the lines toward the three nuclei, 
plotted against the LSRK velocities and normalized with the peak intensities. The line names are in the figure legend.
The spectra of NGC 4418 are from our 0\farcs35 data sampled in a 1\arcsec-diameter aperture.
The line profiles of Arp 220 were taken from our companion spectral scan reduced in the same way 
since it covers a wider frequency range and contains more lines though at lower angular resolution. 
It is evident in the figure that line shapes vary much among the lines for each nucleus, reflecting at least 
the blending of nearby lines and variation of absorption and excitation conditions.
We therefore averaged the normalized spectra for each nucleus to reduce the line-shape distortion. 
The mean line profiles are also shown in Fig.~\ref{f.vsys} with black lines.
We measured, for each nucleus, the central velocity of the mean line profile at 20, 30, 40, and 50 \% of its peak.
The measured central velocities are shown as vertical line segments in Fig.~\ref{f.vsys}.
These intermediate levels help avoid the self-absorption evident at higher levels toward the Arp 220 nuclei. 
In addition, the mean profiles at these levels should be less affected by adjacent lines 
and asymmetric line wings, if any, than at lower levels.
The central velocities agree well among the four levels, and therefore, the fiducial velocities decided from 
them through averaging and rounding should be precise to about 10 \kms. 
A systemic error is possible if a nucleus has a lopsided distribution of gas or gas properties, although
we used larger apertures than our data resolution allows to reduce the effect of any small-scale asymmetries.
For better systemic velocities, one needs to properly de-blend lines and model any intrinsic asymmetries
and radial motion of the molecular gas around the individual nuclei.\footnote{
It would be rather complex. For example, Arp 220W has double-peaked profiles in many lines but the brighter peak
is the redshifted one in CO and the blueshifted one  in HCN and CS while HNC shows single-peaked profiles. 
Such a nucleus requires a model more complicated than having an $m=2$ symmetry of a single gas parameter.}

For comparison, \citet{Sakamoto13} estimated for NGC 4418
the central velocity of 2100 \kms\ with CO 
and 2088 \kms\ with several other dense gas tracers,
noting that the latter may be closer to the true systemic velocity of the nucleus.
\citet{Martin16} estimated the systemic velocities of the eastern and western nuclei of Arp 220 to be
5454 and 5342 \kms, respectively, from \HCOplus(3--2) line profiles toward the nuclei.
Both are about 50 \kms\ larger than what we adopt here.
We excluded \HCOplus\ in our velocity analysis above 
because it blends with bright HCN($v_2=1$, $l=1f$) on the redshifted side 
and has deep absorption on the blueshifted side.

\subsection{Initial Line Identification}
\label{s.Vsys_lineID.lineID}
We identified lines through a comparison between the observed and synthesized spectra.
Rather than individually identifying lines with their frequencies alone, 
we modeled all transitions of each species together to add the lines to our model. 
Their relative line intensities were calculated with assumptions of the local thermodynamic equilibrium (LTE) and 
that all lines originate from the same slab of isothermal gas.
We employed for this analysis the MADCUBA/SLIM software \citep{MADCUBA} that allows iterative line identification
and spectral fitting through the successive generation of synthetic spectra using all identified and fitted species.
We used our NGC 4418 spectrum at 0\farcs35 resolution from Bands 6 and 7 
because this nucleus has a much narrower linewidth than the Arp 220 nuclei and hence suffers less from line-blending.
The Band 9 spectrum was not used for simultaneous fitting because it has less accurate flux calibration, 
a small number of lines, and more continuum absorption.
Acknowledging that the isothermal slab is not a good model for our target nuclei, 
we still show our line identifications hoping it to be an instructive starting point. 
Appendix \ref{ap.spid} lists lines and species in our initial model for NGC 4418 and compares the observed
and model spectra.

Figures~\ref{f.triNuc_spec.B9.350mas}, \ref{f.triNuc_spec.B7}, 
and \ref{f.triNuc_spec.B6.350mas} show our continuum-subtracted spectra 
with labels for major identified lines as well as some lines of interest.
Many lines are observed with similar relative strengths among the three nuclei,
though some show notable differences. 
As we intended in our target selection, NGC 4418 serves as a good reference to disentangle 
the heavily blended lines in Arp 220.
For comparison, the previous interferometric spectral scans at lower resolution
by \citet{Costagliola15} for NGC~4418 and \citet{Martin11} for Arp 220 also provided line identifications.
Comparing their spectra and ours, 
as well as our 0\farcs2 and 0\farcs35 spectra in Band 7 (Fig.~\ref{f.triNuc_spec.B7}), 
it is evident that relative line strengths depend on spatial resolution.
So does the degree of self-absorption and continuum absorption.  
This resolution dependence indicates a radial variation of ISM conditions in these nuclei, and we will see it again in Section \ref{s.lineIm.size.observations}.

\section{Spectral Line Analysis}
\label{s.lineIm}
We made spectral line images using the visibility data from which our continuum models constructed in \citest{Paper1}
had been subtracted.
Figures \ref{f.chmap.n4418.B9} through \ref{f.chmap.a220.B6} show the channel maps of major lines in our data
for a $50$ \kms\ resolution and {\tt robust} = 0 for the imaging.
Table \ref{t.chmaps} has their parameters, including the peak brightness temperatures (peak \Tb) and information on major nearby lines that may be blended.
The tabulated peak \Tb\ may be biased low because of the continuum subtraction; we explain it in Section \ref{s.lineIm.cautions}.
These peak \Tb\ measured in the image domain suffer from beam dilution, unlike the continuum peak \Tb\ from visibility fitting.

Figures \ref{f.sqash.n4418} and \ref{f.sqash.a220} show integrated intensity images
of NGC 4418 and Arp 220, respectively.
For NGC 4418, we integrated channels from 1900 to 2300 \kms\ for a total $\Delta V$ = 450 \kms, 
except for the two lines noted in the caption.
For Arp 220, we integrated channels from 4850 to 5850 \kms\ for a total $\Delta V$ = 1050 \kms.
Figure \ref{f.sqash.COlarge} also shows the CO(2--1) and (3--2) integrated-intensity images of the two galaxies 
in larger areas.
Figures~\ref{f.mom1.n4418} and \ref{f.mom1.a220} show the mean velocity images of selected lines.\footnote{
We define the mean velocity at each position as 
\begin{equation} 
	\langle v \rangle = \frac{\sum_i | I_i | v_i}{\sum_i | I_i |}   ,
\end{equation}
where $I_i$ is the intensity in the $i$-th channel at velocity $v_i$.
We use the absolute value of intensity here because both emission and absorption are in our continuum-subtracted image cubes.
The summations are over channels with significant detection of either emission or absorption.
The detection criterion was that the absolute value of intensity is above a threshold in the reference data cube 
that was made by smoothing the original data cube to about 1.5 to 2 times the original spatial resolution.
We used the masking threshold of $2.5\sigma$, where $\sigma$ is the rms noise in each original cube.
}
We describe below our observations in these data.

\subsection{Cautions about Continuum Subtraction and Line Blending}
\label{s.lineIm.cautions}
We start from cautions about continuum subtraction and line blending.
These have become acute issues in extragalactic spectral scans 
with the angular resolution and sensitivity of ALMA.

While we subtracted our modeled continuum emission from our line data,
the necessity for the subtraction itself is a complex issue. 
The continuum subtraction is undoubtedly necessary for any channels 
where the continuum nucleus and line-emitting gas have no overlap on the sky.
The necessity is not evident for the channels where the continuum nucleus is covered by
a foreground line-emitting gas whose line emission is optically thick. 
In this situation, the source has its line photosphere outside its continuum photosphere.
If the continuum emission from outside the line photosphere is negligible,
then we only detect line photons (i.e., those from molecules) and detect no continuum photons that need subtraction.
There are also intermediate situations, e.g., partial coverage of the continuum source and optically thin line emission
on top of optically thin dust-continuum emission from the same interstellar cloud.
For simplicity, we subtracted the continuum from all the channels using our Gaussian source models 
with power-law spectra \citesp{Paper1}.
The continuum models have no correction for an occultation of the continuum emission by line-emitting gas. 
Therefore, a line-emitting gas that completely covers the continuum source would have zero (or a negative) intensity 
if the line is optically thick and has the same (or a lower) brightness temperature compared to the background continuum. 
Therefore, the lack of positive emission toward a continuum nucleus in our continuum-subtracted line image does not necessarily 
mean a lack of line emission (and molecular gas) on the sightline to the nucleus.
For analytic expressions of this issue, see \citet[\S 13.4.1]{ToRA6} and \citet{MADCUBA}.

\begin{figure}[t]
\fig{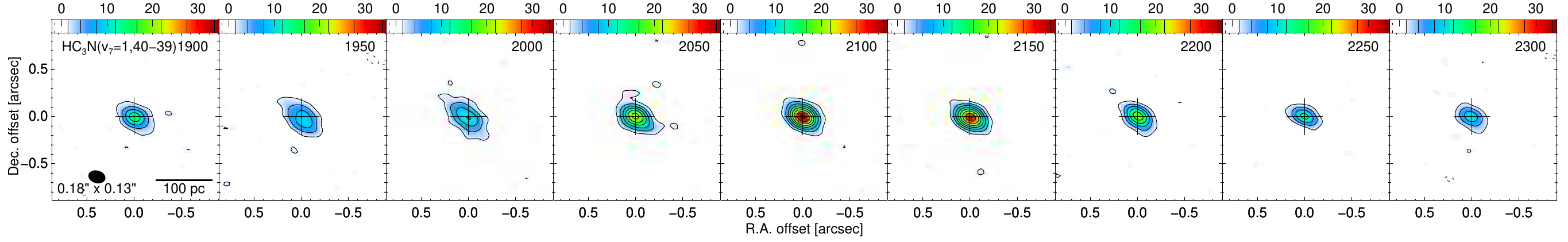}{1.0\textwidth}{(a) \HCthreeN($v_7=1$, 40--39, $l=1f$)}
\fig{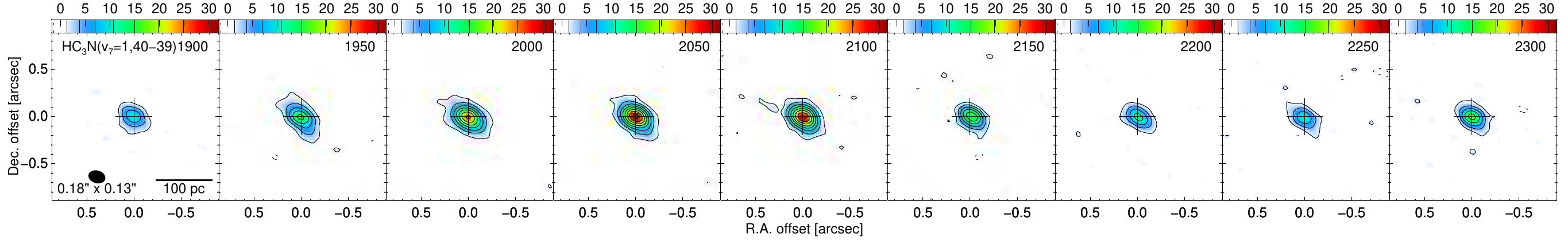}{1.0\textwidth}{(b) \HCthreeN($v_7=1$, 40--39, $l=1e$)}
\fig{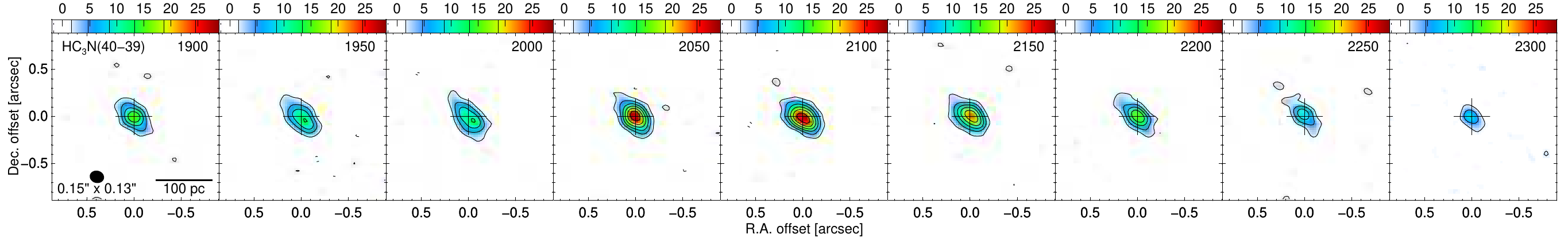}{1.0\textwidth}{(c) \HCthreeN(40--39)}
\fig{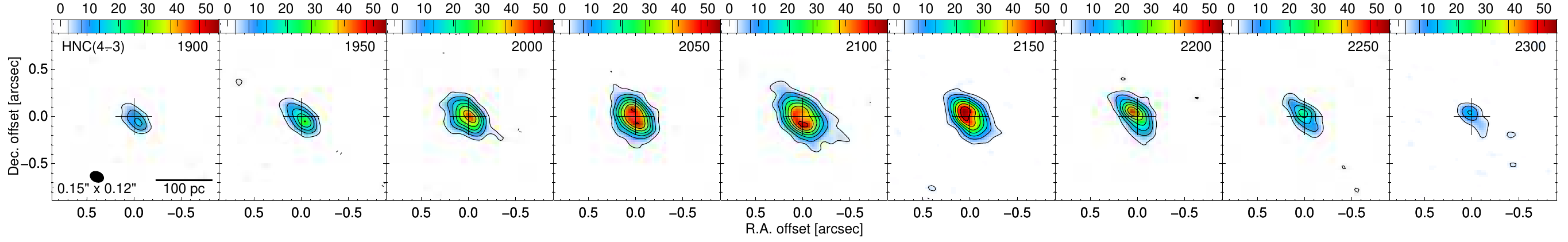}{1.0\textwidth}{(d) HNC(4--3)}
\fig{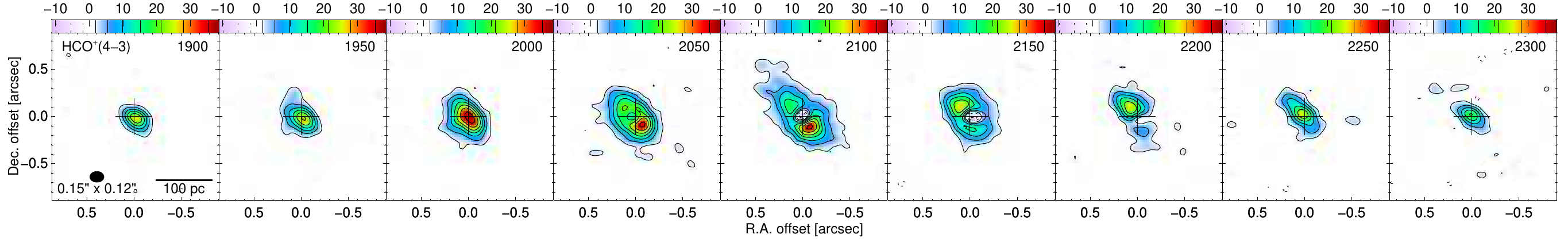}{1.0\textwidth}{(e) \HCOplus(4--3)}
\fig{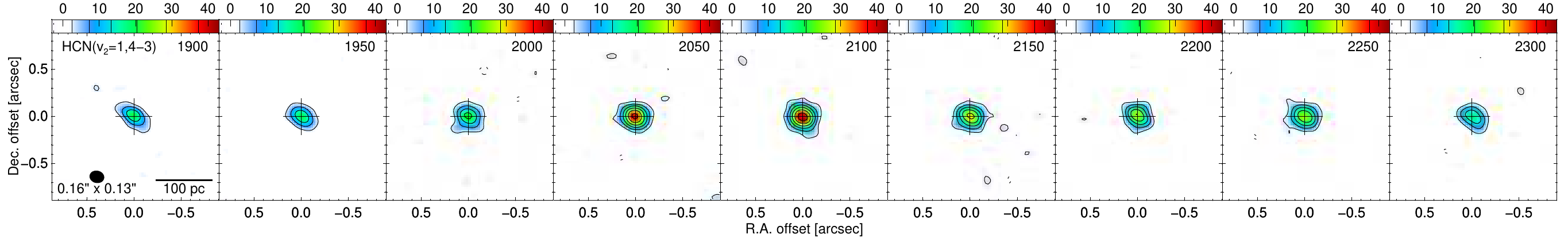}{1.0\textwidth}{(f) HCN($v_2=1$, 4--3, $l=1e$)}
\caption{ 
NGC 4418 channel maps of 50 \kms\ width for major lines in the ALMA Band 7.
Contours are at $\pm3n^p\sigma$ for $n=1,2,3,\ldots$;
the index $p$ and the rms noise per 50 \kms\ channel $\sigma$ are in Table \ref{t.chmaps}.
Negative contours are dashed.
Velocity (LSRK, radio) in \kms\ is in the upper-right corner of each channel map.
The crosses are at the continuum positions of the nuclei.
The synthesized beam, labeled with its FWHM, is shown in the bottom-left corner of the first channel.
The intensity color wedges are labeled with brightness temperature in Kelvin.
}
\label{f.chmap.n4418.B7}
\end{figure}

Lines are blended not only in our spectra (Figs.~\ref{f.triNuc_spec.B7} and \ref{f.triNuc_spec.B6.350mas})
but also in our channel and integrated-intensity maps. 
The line-blending is more severe in Arp 220 than in NGC 4418 because the lines are broader in the former.
For example, \HthirteenCN(4--3) is evident in the CO(3--2) channel maps  in Fig.~\ref{f.chmap.n4418.B7}(g).
Much of the emission within 0\farcs5 of the western nucleus  in the 5700 \kms\ channel (and adjacent channels)
must be \HthirteenCN(4--3). 
Indeed, that CO velocity corresponds to 5311 \kms\ for \HthirteenCN(4--3).
Moreover, the emission shape around the W nucleus agrees very well with those in the 5300--5350 \kms\ channels of 
other dense gas tracers such as HCN(4--3) and CS(7--6) in Figs.~\ref{f.chmap.n4418.B7}(d) and (h).
What is worse, CO(3--2) probably has a minor but non-zero contribution to the (circum)nuclear emission at 5700 \kms\ 
(and in adjacent channels) around the western nucleus judging from CO(2--1) channel maps in Fig.~\ref{f.chmap.n4418.B6}(a).
Thus CO(3--2) and \HthirteenCN(4--3) must coexist around this location in the
observed position-position-frequency space.
Therefore, we cannot separate the two by simply applying a three-dimensional mask made from the CO(2--1) information.
Major cases of line-blending are listed in Table \ref{t.chmaps}, and the line list in the Appendix suggests more.
As we see below multiple times, consideration for line blending is often crucial for data interpretation 
at the level of line density in our data.

\subsection{Spatial Distribution around the Nuclei --- Multi-line Comparison}
\label{s.lineIm.size}
We find that the size of the emitting region varies among lines with certain trends.
This finding has implications on the radial structure in the nuclei,
the analysis of multi-line data, 
and our choice of lines to probe these nuclei.

\subsubsection{Size variation of line-emitting regions}
\label{s.lineIm.size.observations}
Molecular line emission in our images tends to be sharply peaking toward the three continuum nuclei,
with some lines showing decrement or absorption on the continuum peaks.  
The size of the line emitting region varies among the lines.
In both galaxies, the line-detection areas are by far the largest for CO.
It is consistent with our SMA analysis that CO emission is more extended than other lines in these galaxies
\citep{Sakamoto09,Sakamoto13}.
Among the non-CO lines, the emission lines from vibrationally excited molecules, such as HCN($v_2$=1), 
have much more compact distributions than the other ordinary lines (i.e., transitions within the vibrational ground state).
In NGC 4418, the compactness of the vibrationally-excited line emission is seen in both the channel maps 
and the integrated intensity images.
In Arp 220, the compactness of HCN($v_2$=1) emission is better seen in the channel maps 
at $V \gtrsim 5300$ \kms, where the line is not blended with the adjacent \HCOplus($v=0$) line.
In short, the size of the line-emitting area has the following apparent trend: 
CO $>$ other lines without vibrational excitation $>$ lines from vibrationally excited states.

\begin{figure}[t]
\includegraphics[width=0.233\textwidth]{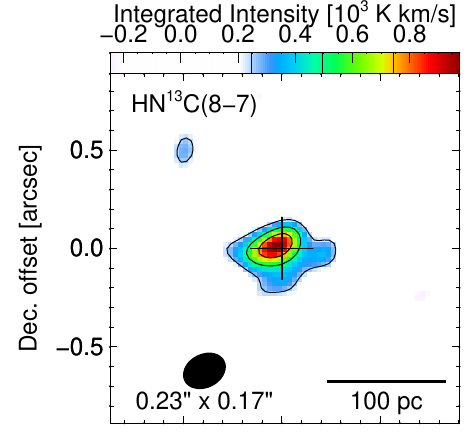} 
\includegraphics[width=0.185\textwidth]{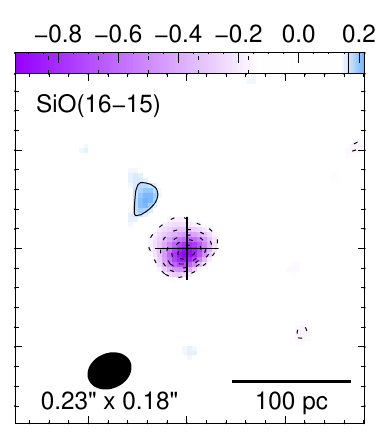} 
\includegraphics[width=0.185\textwidth]{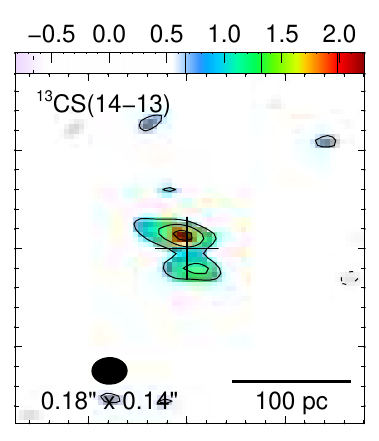} 
\includegraphics[width=0.185\textwidth]{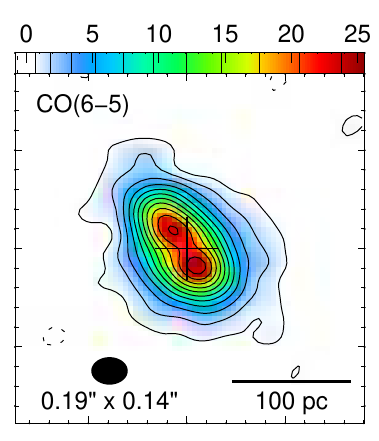} 
\includegraphics[width=0.185\textwidth]{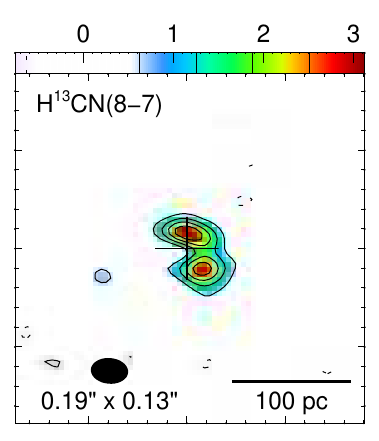} 
\\
\includegraphics[width=0.233\textwidth]{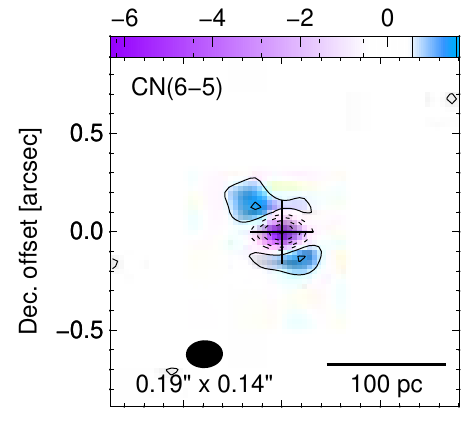} 
\includegraphics[width=0.185\textwidth]{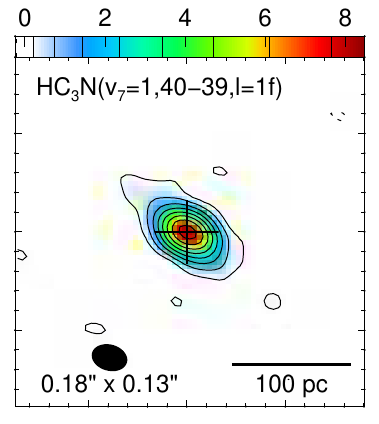} 
\includegraphics[width=0.185\textwidth]{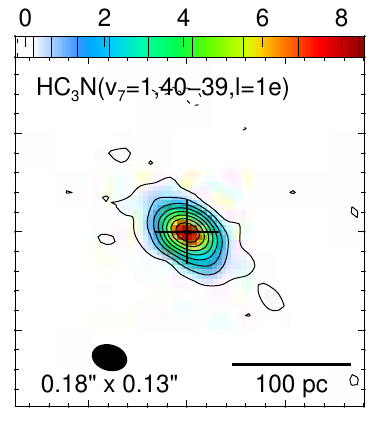} 
\includegraphics[width=0.185\textwidth]{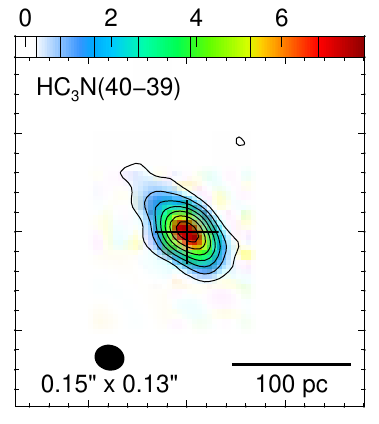} 
\includegraphics[width=0.185\textwidth]{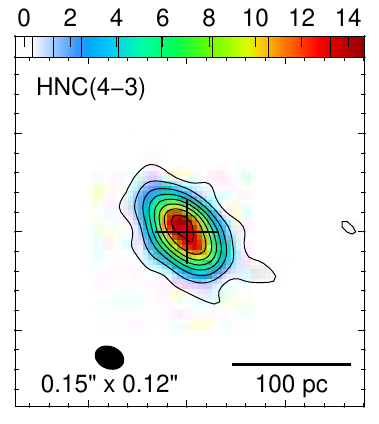} 
\\
\includegraphics[width=0.233\textwidth]{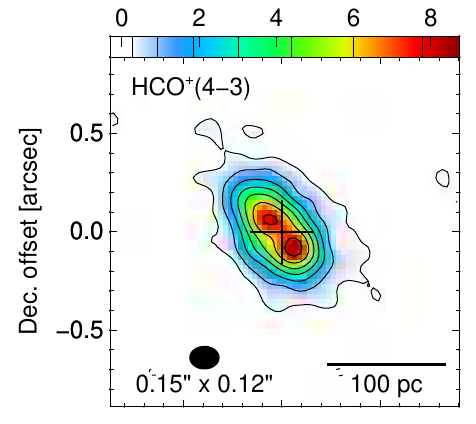} 
\includegraphics[width=0.185\textwidth]{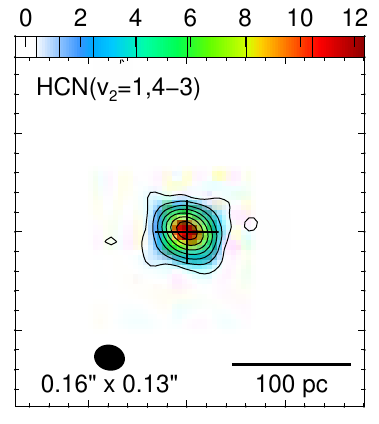} 
\includegraphics[width=0.185\textwidth]{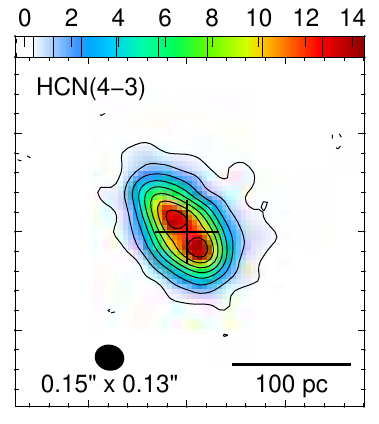} 
\includegraphics[width=0.185\textwidth]{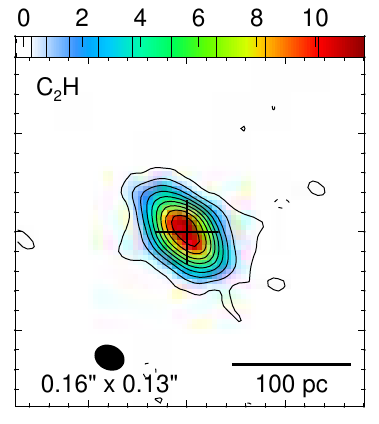} 
\includegraphics[width=0.185\textwidth]{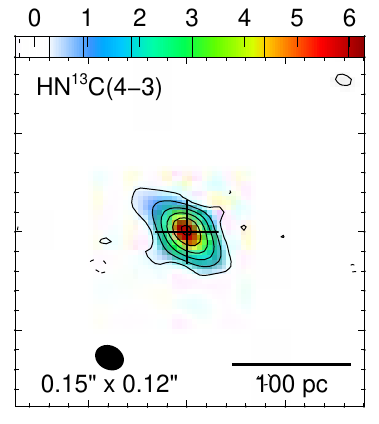} 
\\
\includegraphics[width=0.233\textwidth]{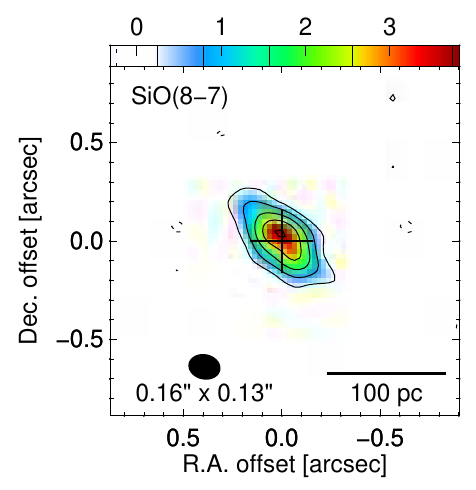} 
\includegraphics[width=0.185\textwidth]{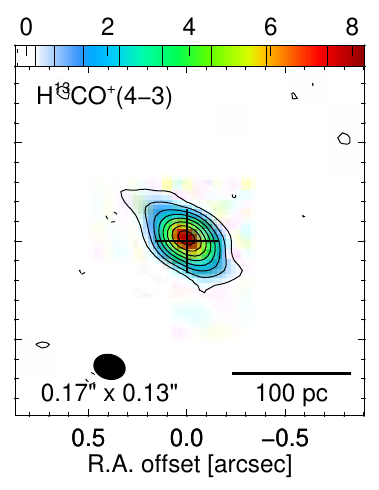} 
\includegraphics[width=0.185\textwidth]{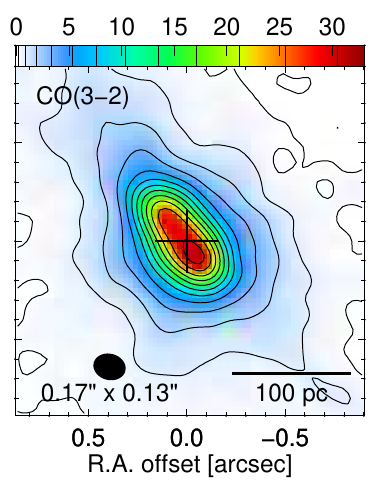} 
\includegraphics[width=0.185\textwidth]{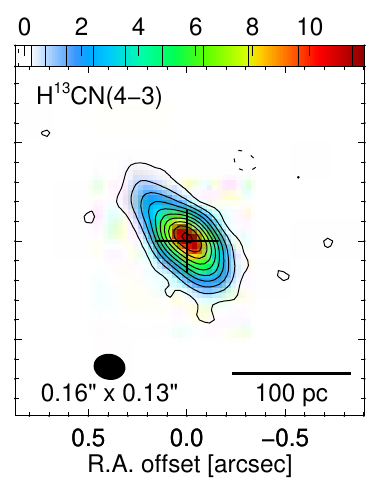} 
\includegraphics[width=0.185\textwidth]{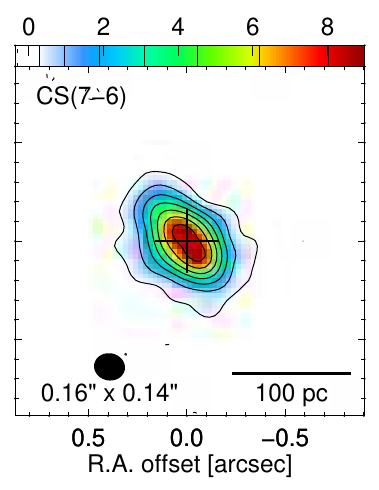} 
\caption{ 
Integrated intensity maps of NGC 4418 for major lines in our ALMA observations.
Integration is from the 1900 \kms\ channel to the 2300 \kms\ channel 
except the following; SiO(16--15) integration is from 1900 to 2150 \kms to reduce contamination by \HthirteenCOplus(8--7),
\thirteenCS(14--13) is integrated over the 1950--2300 \kms\ channels, and \HthirteenCN(8--7) is integrated over
the 1950--2250 \kms\ channels.
Contours are at $\pm3n^p\sigma$ for $n=1,2,3,\ldots$, where $p$ is in Table \ref{t.chmaps}
and $\sigma$ is calculated from the noise in the channel maps in Table \ref{t.chmaps} and the number of integrated channels.
Negative contours are dashed.
In each panel, the crosse is at the continuum position of the nucleus.
The synthesized beam, labeled with its FWHM, is shown in the bottom-left corner.
}
\label{f.sqash.n4418}
\end{figure}

\begin{figure}[t]
\epsscale{0.9}
\plottwo{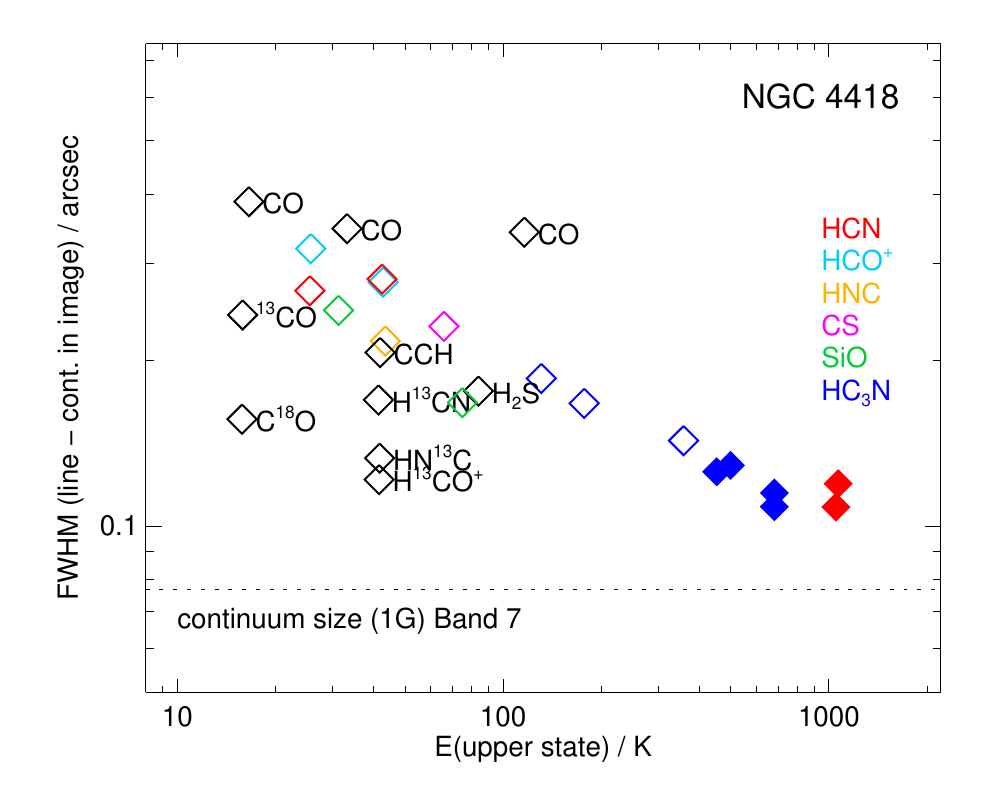}{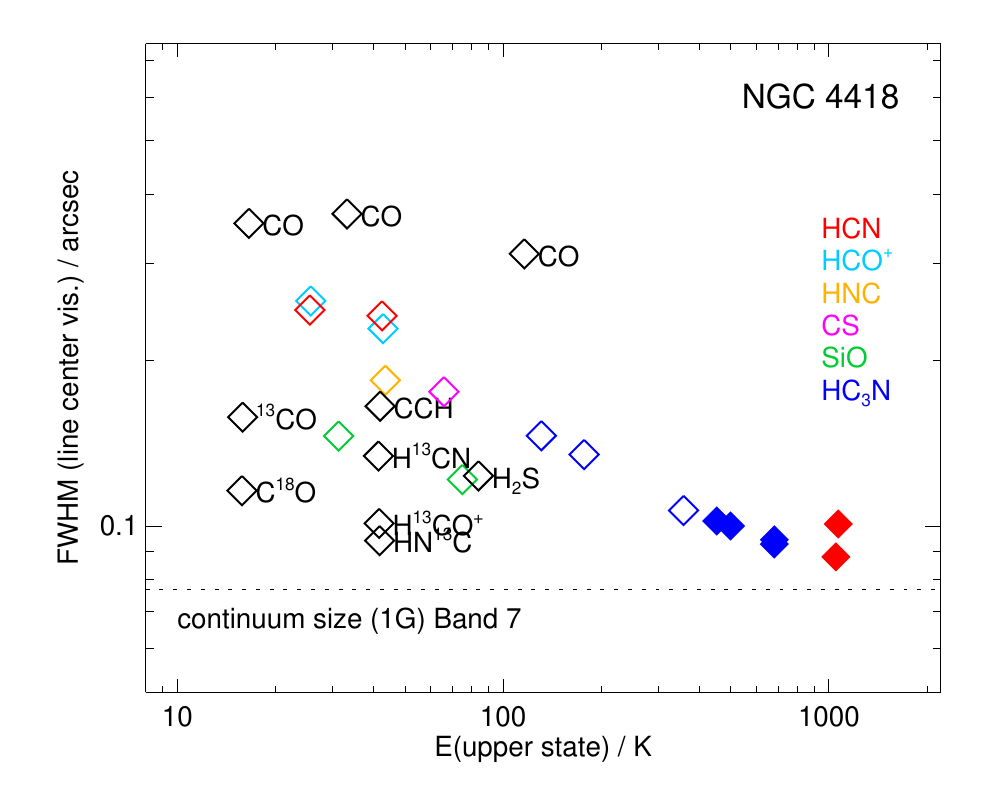} \\
\plottwo{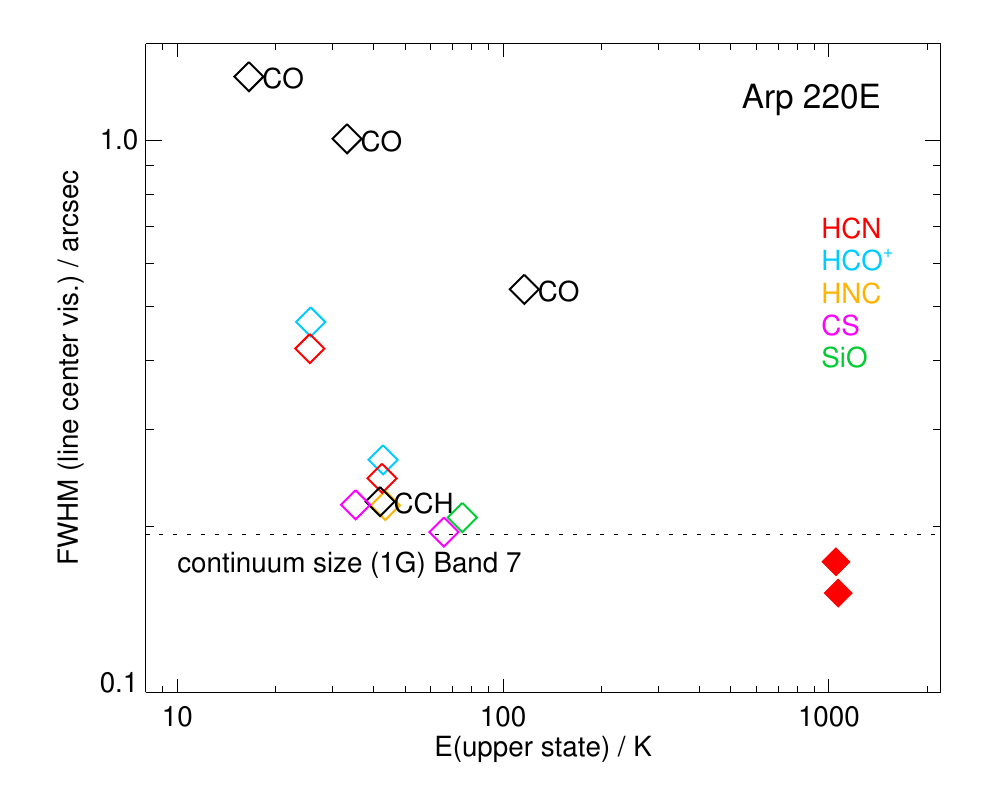}{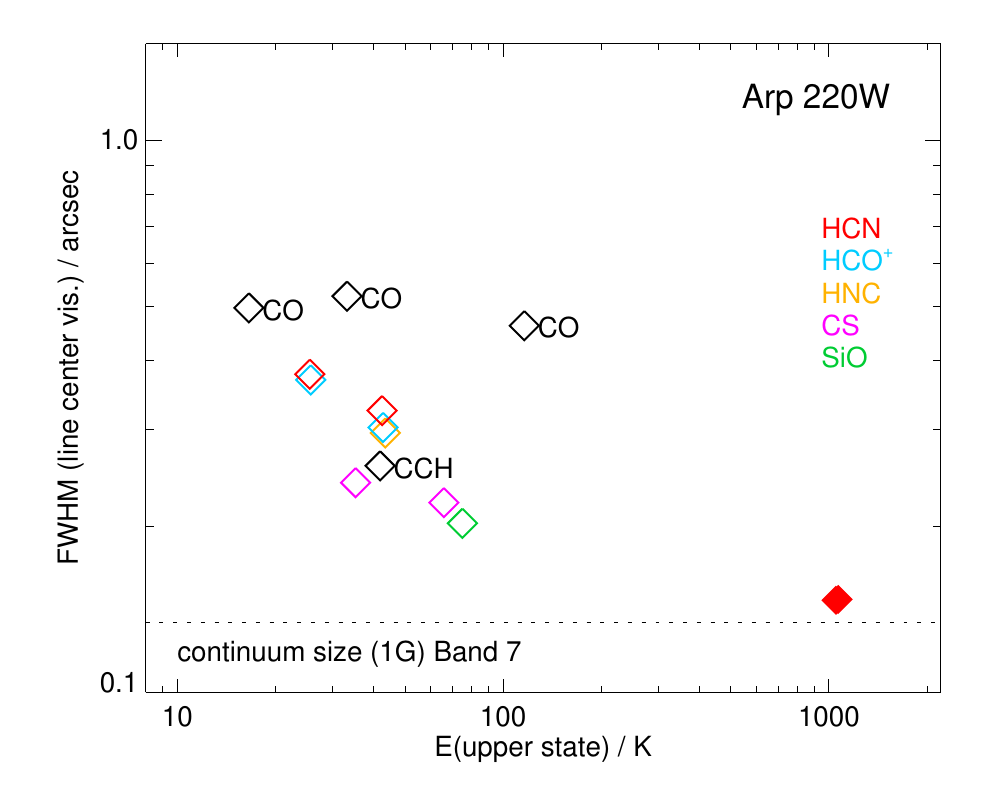}
\caption{ \label{f.line_extent.n4418_a220}
The extent of line emission in the nuclei of NGC 4418 and Arp 220 plotted against the upper-state energy of the individual lines.
The lines are those in Bands 6 and 7 and CO(6--5).
The top-left panel shows image-domain FWHM measured from the line integrated-intensity images in Figure \ref{f.sqash.n4418}.
The top-right and bottom panels show FWHM measured from the visibility fitting without continuum subtraction in \citest{Paper1}; channels within 100 \kms\ of the systemic velocities were averaged.
All the FWHM sizes are the geometrical means of the deconvolved major- and minor-axis FWHM.
Their formal errors are about the size of the symbols.
Filled symbols are for lines in vibrationally excited states.
Dotted horizontal lines are for the 1G (single-Gaussian) continuum size in Band 7 \citesp{Paper1}.
Note that each nucleus has a continuum structure that is more sharply peaked than the 1G fit. 
For example, the core component of the continuum emission of Arp 220E has an extent of \about0\farcs07.
}
\end{figure}

Figure \ref{f.line_extent.n4418_a220} plots the extent of the emitting regions in FWHM for major lines 
in the nuclei of NGC 4418 and Arp 220.
The deconvolved source sizes are plotted against the upper-state energies of the lines.
We use the FWHM to characterize source size to avoid the bias that a faint source appears smaller than 
a bright source of the same size and shape if evaluated at the same absolute intensity (e.g., at the $3\sigma$ contours). 
Even the weaker lines in the plot have high signal-to-noise ratios; S/N = 47, 66, 97, 110, and 115 respectively for 
SiO(8--7), \HNthirteenC(4--3), HCN(\vtwo=1,4--3), \CeighteenO(2--1), and \HthirteenCOplus(4--3) 
in the integrated intensity maps of NGC 4418.
Hence their FWHM have only small errors due to thermal noise.
The formal errors of the plotted FWHM are at most the size of the plotting symbols.
We measured the source FHWM in two ways for NGC 4418 and found them generally consistent.
For Fig.~\ref{f.line_extent.n4418_a220}a, we used Gaussian fitting of the continuum-subtracted integrated intensity images 
(Fig.~\ref{f.sqash.n4418}). 
For Fig.~\ref{f.line_extent.n4418_a220}b, we used visibility fitting with a Gaussian source model for the line$+$continuum data 
within 100 \kms\ of the systemic velocity; we already fitted our entire spectral scan in visibilities in \citest{Paper1}.
We used the latter method for the nuclei of Arp 220 in Fig.~\ref{f.line_extent.n4418_a220}c and d.
It mitigates the effect of line-blending by only using channels close to \Vsys\ 
and avoids absorption and ring-like structure by not subtracting continuum.

Figure \ref{f.line_extent.n4418_a220} confirms our visual observation that emission 
from vibrationally excited molecules is compact.
It also reveals two general trends. 
First, the line-emitting regions are smaller for lines whose upper-level energies are higher;
lines from vibrationally excited levels are extreme cases in this trend.
Second, the line-emitting regions are also smaller for lines of less abundant molecules for similar upper-level energies;
rarer isotopologues and \twelveCO\ lines are at the opposite ends of this trend.
All three nuclei show these trends, except that the lines of rare isotopologues in Arp 220
are too heavily blended with nearby brighter lines to be reliably measured for their emitting regions.

The two trends in the size of the line-emitting regions must be because the excitation conditions 
for the transitions from higher energies and for less abundant species are achieved 
only at locations (or on sight-lines passing) closer to the center of each nucleus.
The bright continuum emission sharply peaking toward the center of each nucleus also
suggests such radial variation in physical conditions \citep{Sakamoto17, Paper1}.
Parameters that are very likely to have radial variations include the following;
gas density, gas temperature, and radiation field for radiative excitation (both 
through a vibrationally-excited state by IR radiation 
and 
within the vibrational ground level by millimeter-submillimeter radiation).
The latter two are most likely to be higher toward the center \citep[e.g.,][]{GA12, Scoville15, GS19, Wheeler20}. 
Radial variations are also expected in chemical abundance \citep{Harada10}.

\subsubsection{Lines to Probe the Nuclei}
\label{s.lineIm.size.implications}
Our observations on the size variation of line-emitting regions tell us about the choice of lines and line combinations
to better probe galaxy nuclei similar to our three targets.
The size of the line-emitting areas approaches from above (i.e., decreases) to  the size of the continuum emission 
as the upper-energy level increases or the molecular abundance decreases.
This tendency makes lines from very high energy levels and rare isotopologues better probes of the innermost structure of the nucleus,
down to the radius at which either the continuum or line emission becomes optically thick.
In particular, the lines from vibrationally excited molecules,  having the highest upper-level energies in our data, 
should be the most suitable of lines to probe the warm compact nuclei. We will discuss their usage in \S\ref{s.lineIm.vibLines}.

The size variation of the line-emitting regions also cautions that line ratios require careful interpretation for these galactic nuclei. 
The simplest one-zone modeling has problems in the following assumptions, 
1) that the compared lines are from the same volume and hence have the same beam-filling factors,
and 
2) that the physical conditions are uniform across each nucleus. 
This complexity is a major obstacle to spectral modeling and line identification. 
It is the reason why we regard our line identification in this paper as an initial attempt.

\subsection{Velocity Structures around the Nuclei}
\label{s.lineIm.velstructure}
The channel maps and the mean velocity maps show rich velocity structures in and around the nuclei.
The most notable are the following.

\subsubsection{NGC 4418 --- Counter-rotating nuclear disk and turbulent gas around it}
\label{s.lineIm.velstructure.n4418}
\begin{figure}[t]
\fig{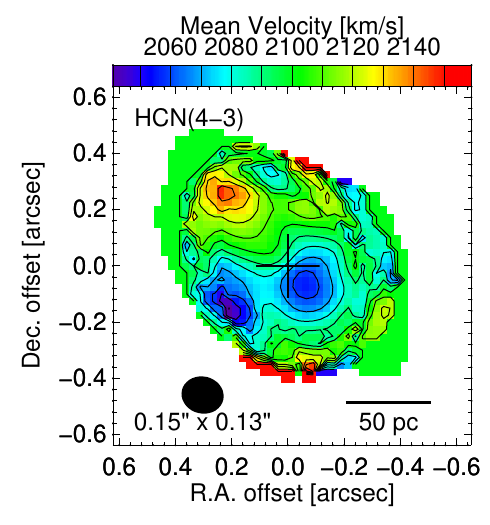}{0.26\textwidth}{}
\fig{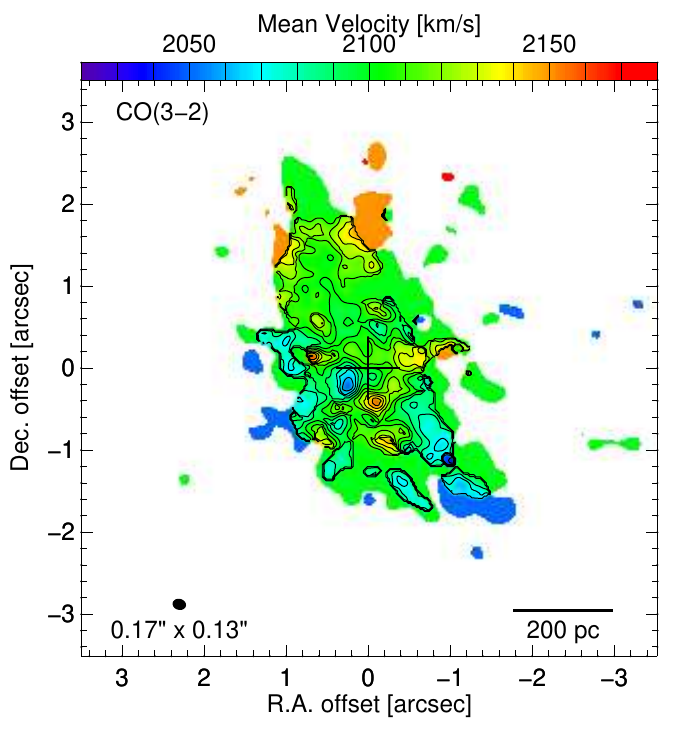}{0.36\textwidth}{}
\fig{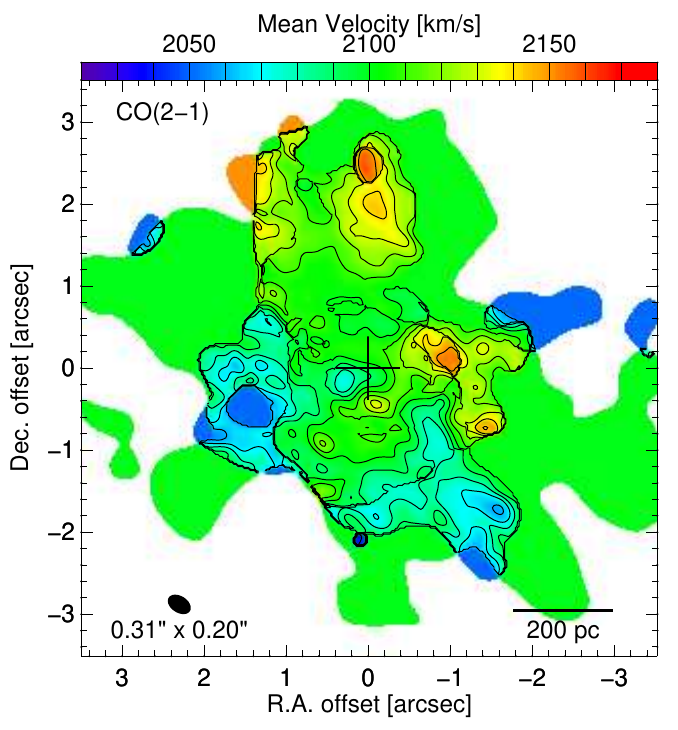}{0.36\textwidth}{}
\caption{ 
Mean velocity maps of NGC 4418 in HCN(4--3), CO(3--2), and CO(2--1).
The cross in each map marks the position of the continuum nucleus.
}
\label{f.mom1.n4418}
\end{figure}

A general velocity structure in the nucleus of NGC 4418, 
seen in the channel maps of, for example, \HCOplus(4--3), HCN(4--3), CO(3--2), and CS(7--6), is
a southwest-northeast velocity gradient in the nucleus.
The brightest line emission is offset (by $\lesssim$ 0\farcs3) from the continuum position 
to the southwest in the low-velocity channels up to 2100 \kms\ and the northeast at higher velocities. 
The offset is smaller or not visible in the channels furthest from the systemic velocity.
This velocity structure becomes less visible at lower angular resolution ($\gtrsim$ 0\farcs25) and in lines with smaller spatial extents.
The velocity gradient is in the direction of the elongation of the integrated intensity for the same lines. 
It is confined to the central 100 pc (i.e., $r \lesssim 50$ pc) of the nucleus, 
as seen in the channel maps and the mean velocity image of HCN(4--3) in Fig.~\ref{f.mom1.n4418}.

The circumnuclear velocity gradient was first reported in \citet{Sakamoto13} and 
was attributed to gas rotation around the nucleus.
They inferred a rotation curve in the nuclear gas disk to rise toward smaller radii 
from the observation that lines with higher critical densities having larger line widths.
Although that alone can also be explained by the combination of higher velocity dispersion 
and higher gas densities toward the center,
our new observation of larger spatial offsets closer to the systemic velocity supports
the rising rotation curve toward the center.
The rotation curve at the very center should be better studied in the lines that most strongly peak toward the center, 
such as lines from vibrationally excited molecules. 
The non-detection of velocity offsets in these lines in our current channel maps is probably because the line-emitting 
regions are too small compared to our angular resolution and presumably also because of high continuum opacity toward the nucleus.

Although the observed southwest-northeast velocity gradient in the close vicinity of
the nucleus is roughly parallel to the major axis of the galaxy \citep[position angle = 60\arcdeg ;][]{Jarrett00},
the direction of velocity gradient is opposite to that of the stars and atomic gas in the disk of the galaxy.
The galactic disk rotates in such a way that its northeast side is blueshifted 
and the southwest side redshifted \citep{Ohyama19}. 
By confirming the nuclear velocity gradient of molecular gas in \citet{Sakamoto13}, our new data
strengthen the surprising finding of \citet{Ohyama19} 
that the kpc-scale galactic disk and the nuclear gas disk counterrotate with each other.

\begin{figure}[t]
\fig{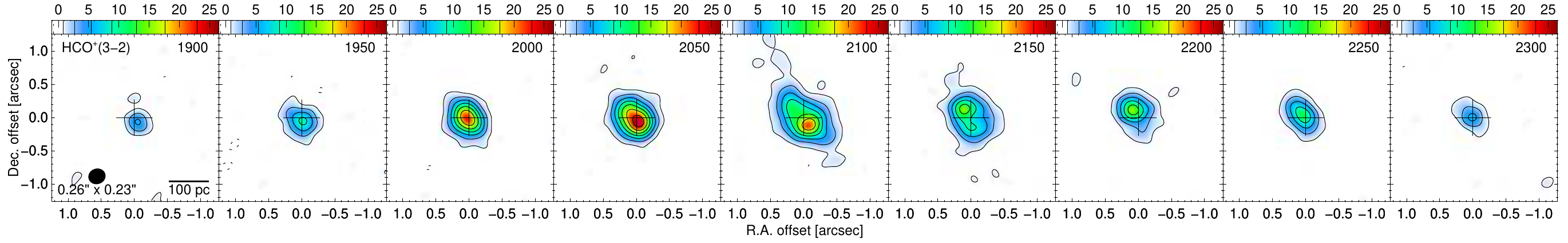}{1.0\textwidth}{(a) \HCOplus(3--2) }
\fig{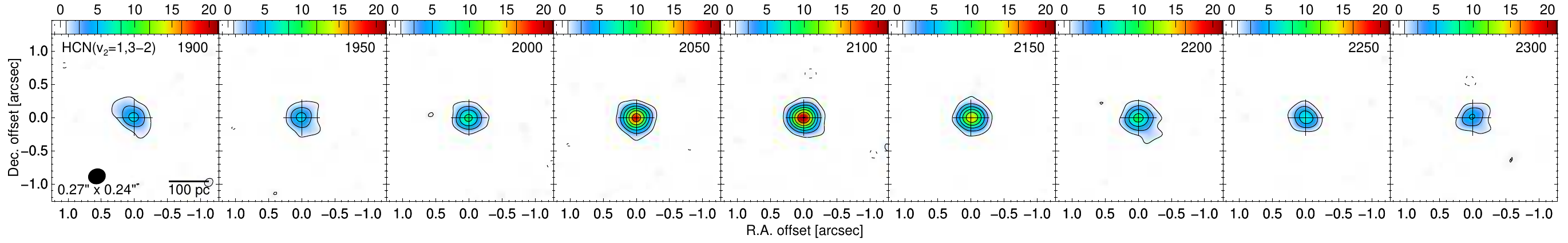}{1.0\textwidth}{(b) HCN($v_2=1$, 3--2, $l=1f$) }
\fig{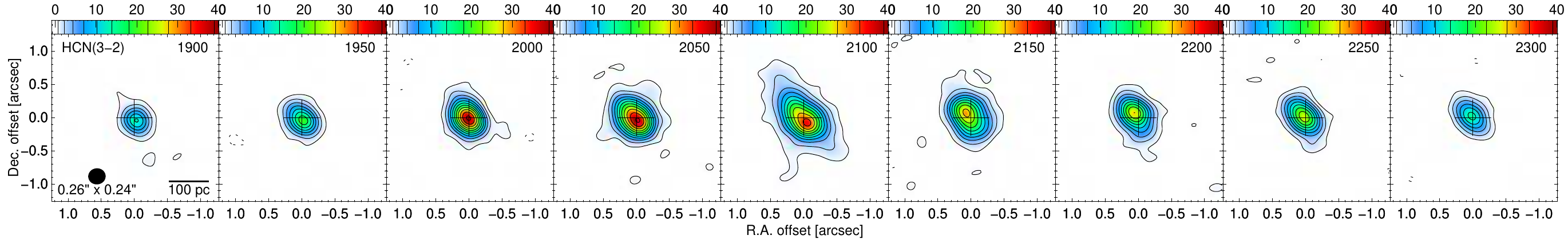}{1.0\textwidth}{(c) HCN(3--2) }
\fig{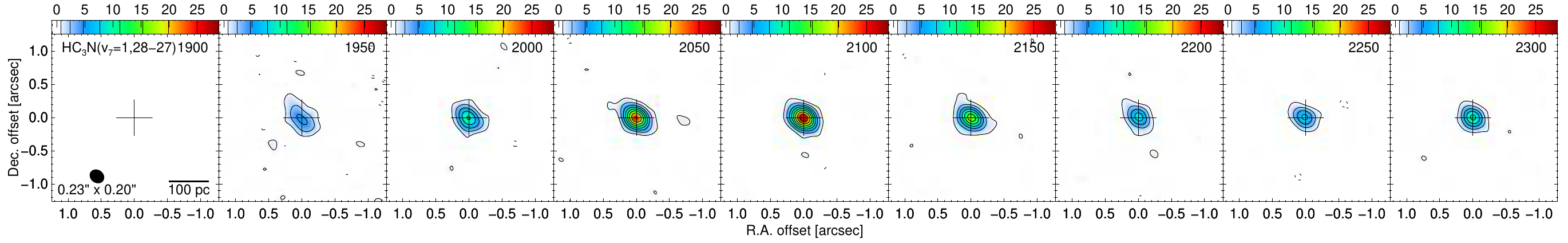}{1.0\textwidth}{(d) \HCthreeN($v_7=1$, 28--27, $l=1e$) }
\fig{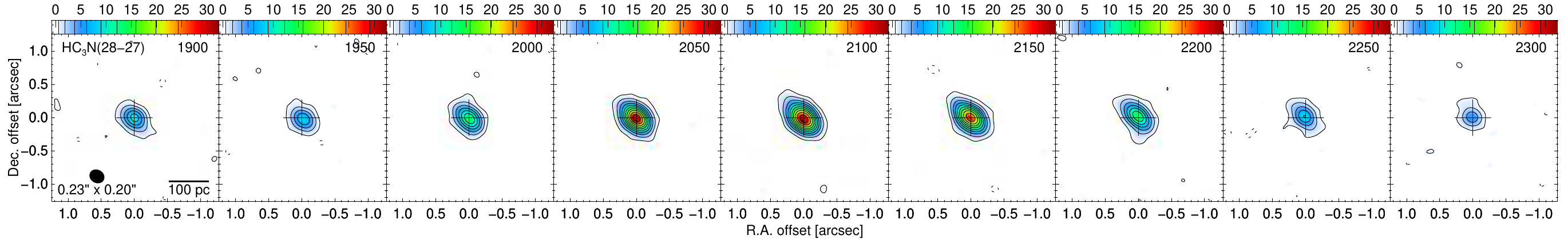}{1.0\textwidth}{(e) \HCthreeN(28--27) }
\fig{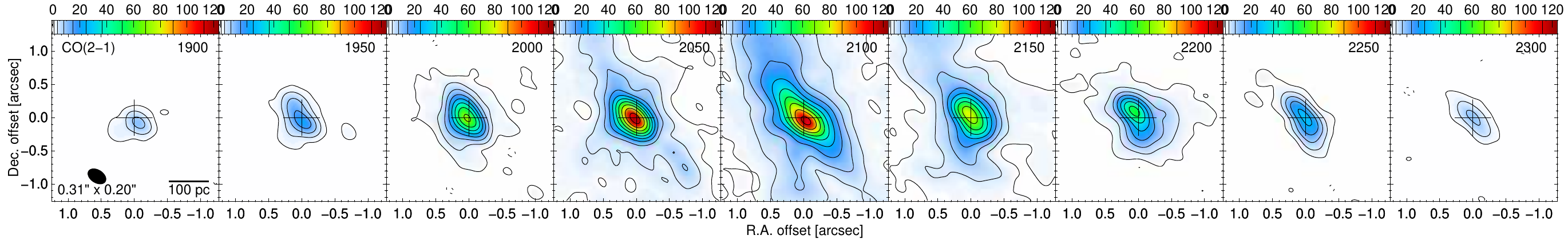}{1.0\textwidth}{(f) CO(2--1) }
\caption{ 
NGC 4418 channel maps of 50 \kms\ width for major lines in the ALMA Band 6.
Other descriptions are the same as in Fig. \ref{f.chmap.n4418.B7}.
The first channel of panel (d) is at the edge of our bandpass and is flagged.
}
\label{f.chmap.n4418.B6}
\end{figure}

The velocity field of molecular gas outside the nuclear disk is highly disturbed without
a regular butterfly pattern of a rotating disk in Figs.~\ref{f.chmap.n4418.B7}l, \ref{f.chmap.n4418.B6}f,
and the CO mean velocity maps in Fig.~\ref{f.mom1.n4418}.
The kpc-scale bipolar outflow from the galactic center \citep{Sakamoto13,Varenius17,Ohyama19} is
a likely source of this disturbance, but a simple bipolar pattern of high-velocity emission is not evident in our CO images.
An outflow has also been inferred from CO(2--1) and (1--0) line wings in the central \about5\arcsec -- 1\farcs4 \citep{Fluetsch19, Lutz20}.
While the CO wing emission is not evident in our spectral decomposition of the central 0\farcs35 (Fig.~\ref{f.specfit} in the Appendix),
dense gas tracers such as HCN, \HCOplus, HNC, and CCH show some broad emission in the residual spectrum, consistent with their 
larger line widths than CO noticed in SMA observations \citep{Sakamoto13}.
The large gas velocities can be due to an outflow, turbulence, or rotation near the nucleus.
If it is due to an outflow, its current extent in these dense-gas tracers must be small 
since no extended outflow is visible in our channel maps.
Other likely sources for the disturbed velocity field at 0.1--1kpc scale are possible gas infall from a companion galaxy \citep{Varenius17} 
and gas supply from the outer disk of NGC 4418 through tidal torquing by the passing companion galaxy \citep{Boettcher20}.
We will note in \S \ref {s.lineIm.absline.n4418} signs of radial gas motion toward the nucleus. 
Combining all these observations, NGC 4418 has surprisingly disturbed kinematics of molecular gas in its central kpc 
despite its undisturbed optical appearance at first sight.

\subsubsection{Arp 220 --- Two nuclear disks and a prominent bipolar outflow from the western nucleus}
\label{s.lineIm.velstructure.a220}
\begin{figure}[t]
\epsscale{0.7}
\plottwo{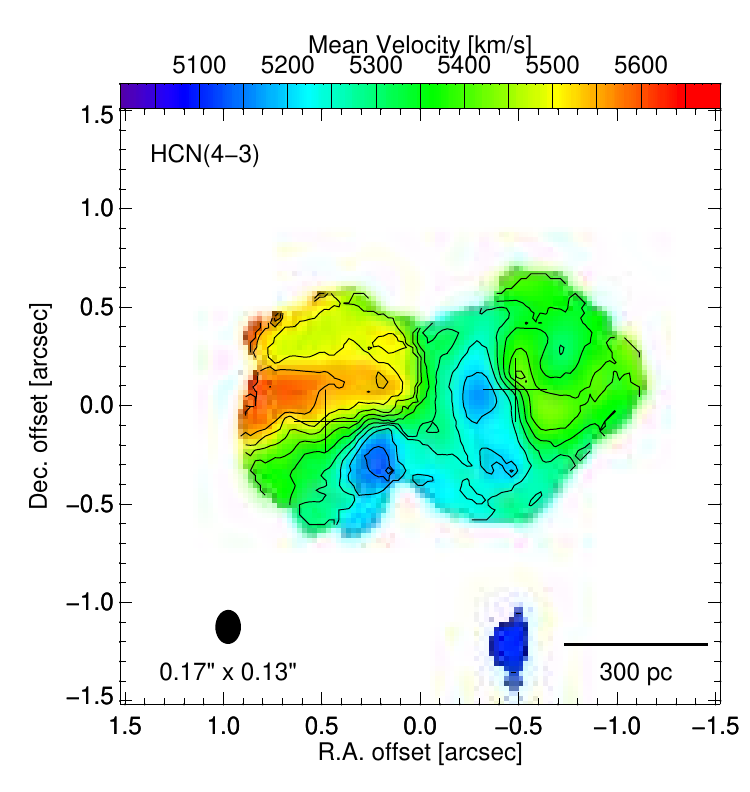}{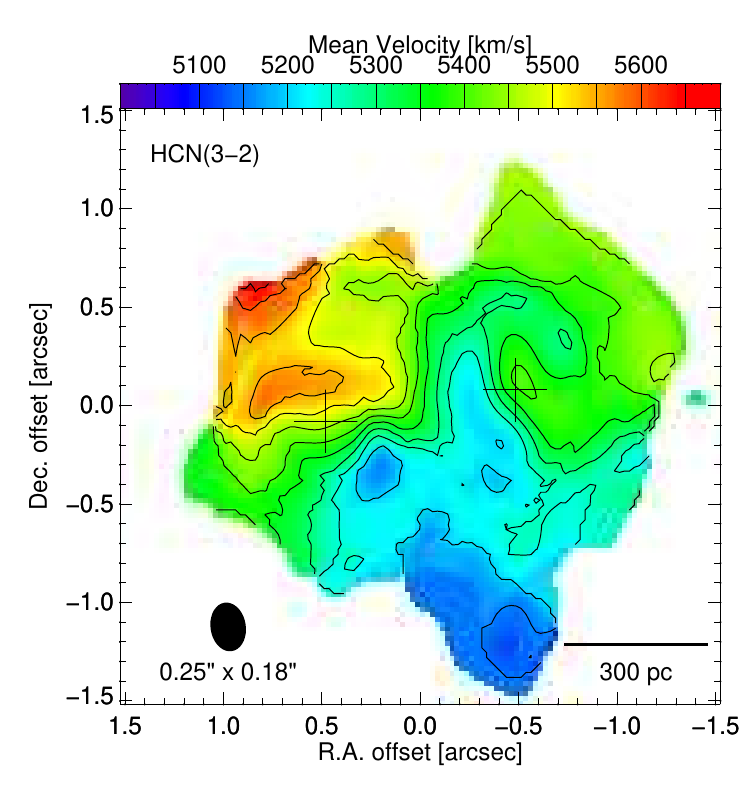}
\\
\epsscale{0.8}
\plottwo{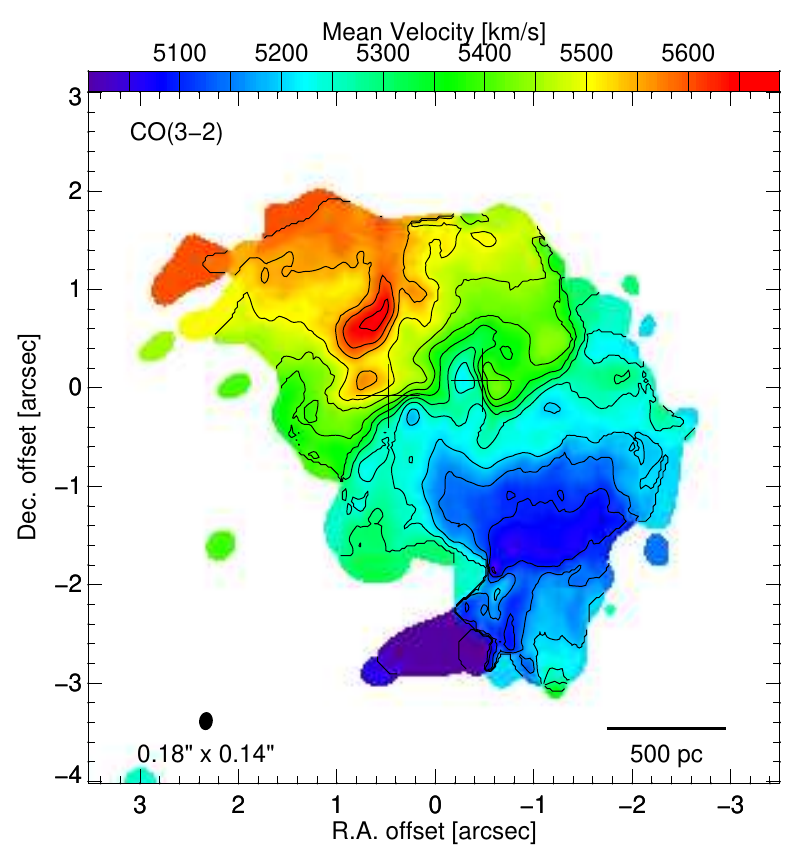}{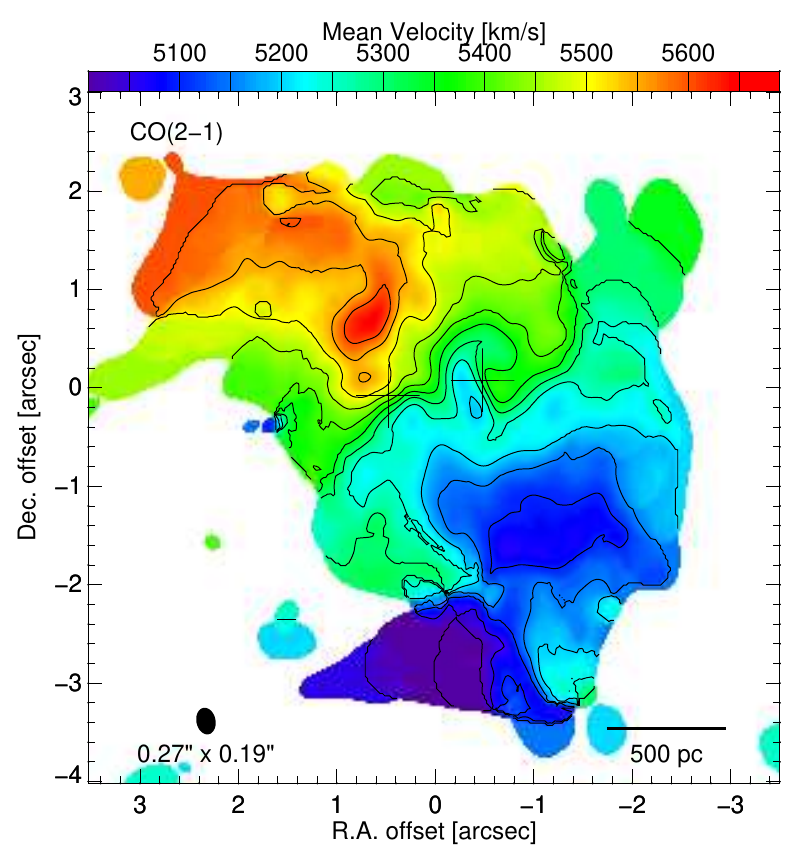}
\caption{ 
Mean velocity maps of Arp 220 in HCN(4--3), HCN(3--2), CO(3--2), and CO(2--1). 
The crosses mark the two nuclei.
}
\label{f.mom1.a220}
\end{figure}

In the channel maps of Arp 220, we see a gradual spatial shift with velocity in line emission around each nucleus.
The shift from lower to higher velocities is from southwest to northeast around \arpE,
while it is from east to west around \arpW.
The mean velocity maps in Fig.~\ref{f.mom1.a220} also show the velocity gradients across the individual nuclei
in CO and HCN.
This velocity structure has been attributed to two rotating nuclear disks in the individual galactic nuclei with 
misaligned rotational axes and apparent counter-rotation \citep{Sakamoto99,Scoville17}.
The velocity structure around the western nucleus appears to be also present in the vibrationally excited HCN lines
in their redshifted halves, 
while their blueshifted halves are severely contaminated by the blending \HCOplus.

\begin{figure}[t]
\includegraphics[width=0.280\textwidth]{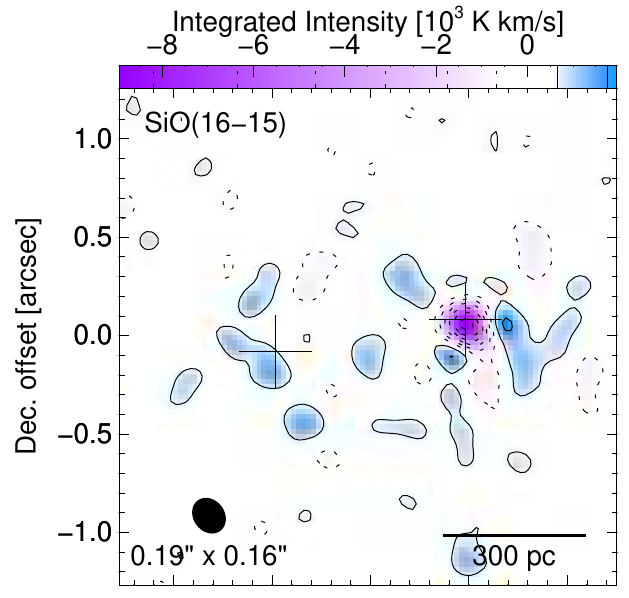} 
\includegraphics[width=0.230\textwidth]{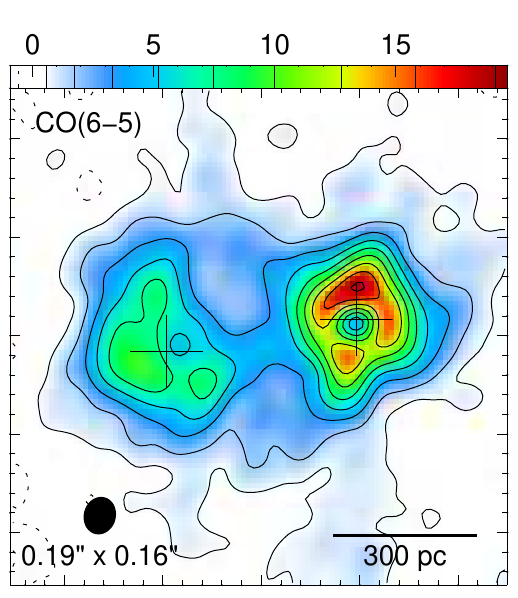} 
\includegraphics[width=0.230\textwidth]{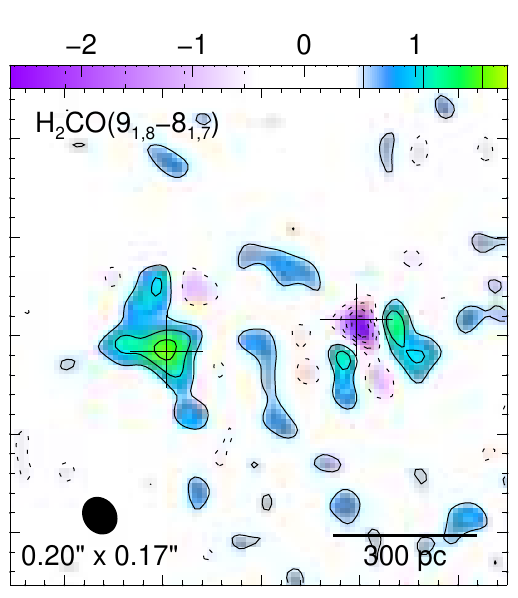} 
\includegraphics[width=0.230\textwidth]{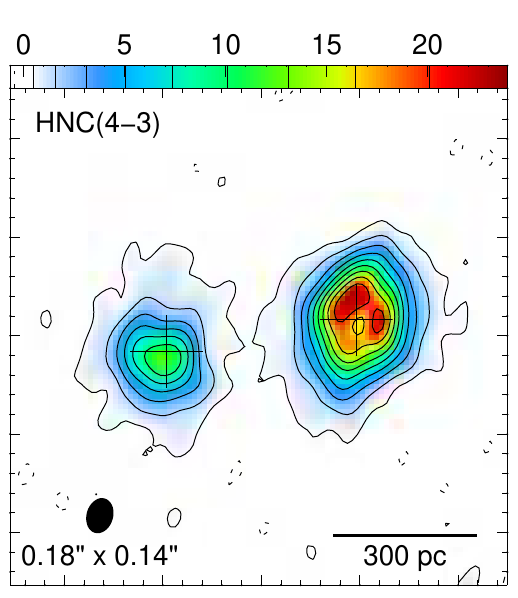} 
\\
\includegraphics[width=0.280\textwidth]{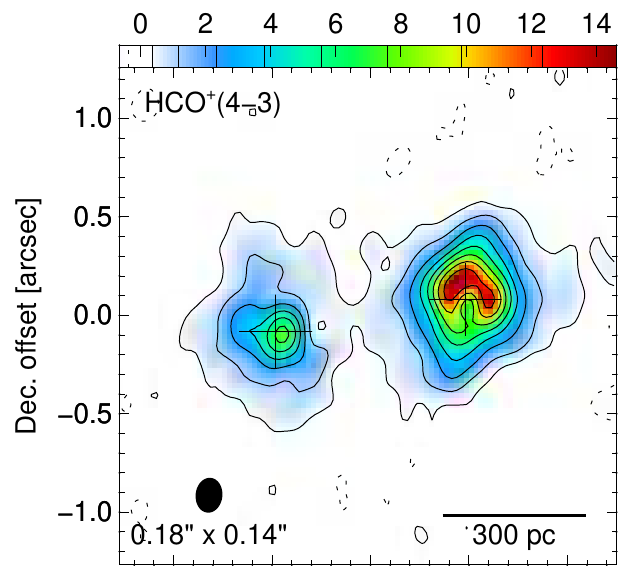} 
\includegraphics[width=0.230\textwidth]{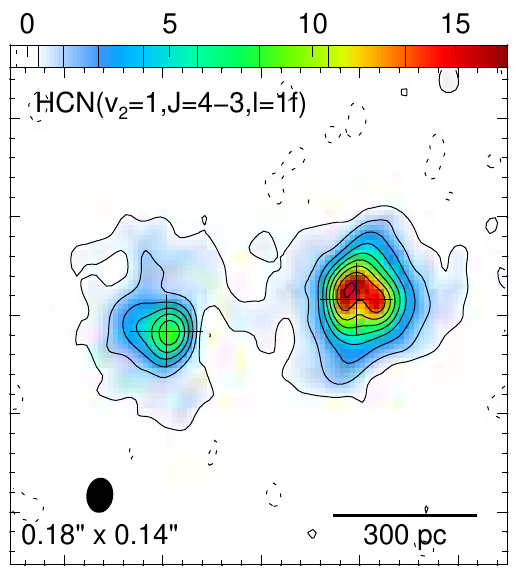} 
\includegraphics[width=0.230\textwidth]{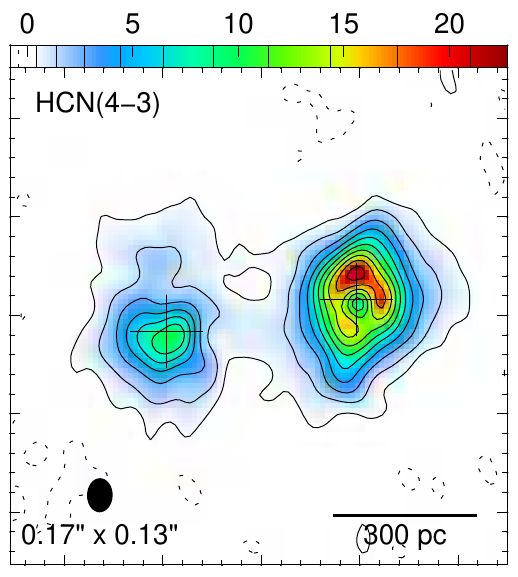} 
\includegraphics[width=0.230\textwidth]{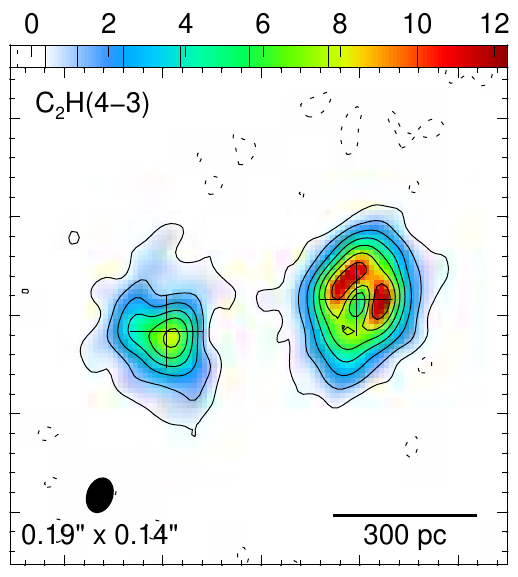} 
\\
\includegraphics[width=0.280\textwidth]{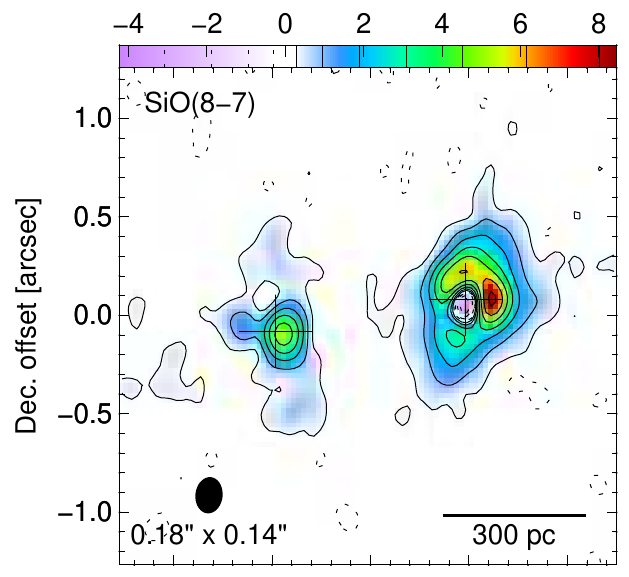} 
\includegraphics[width=0.230\textwidth]{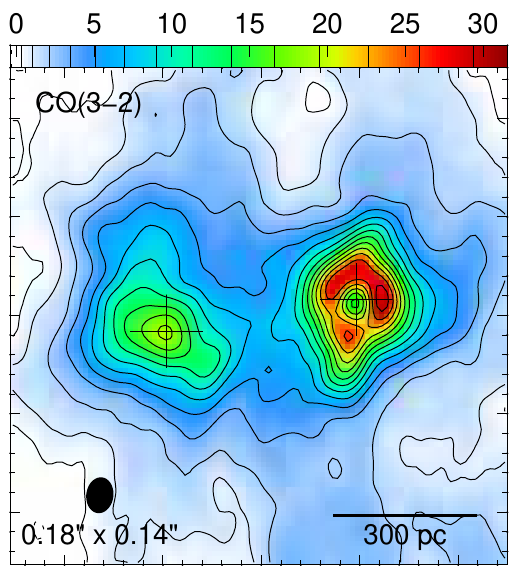} 
\includegraphics[width=0.230\textwidth]{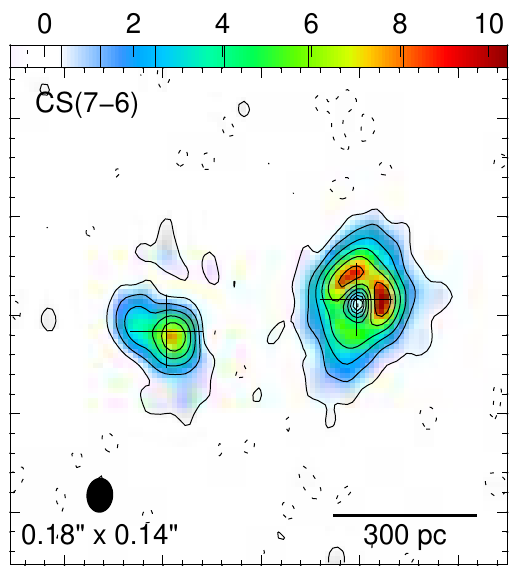} 
\includegraphics[width=0.230\textwidth]{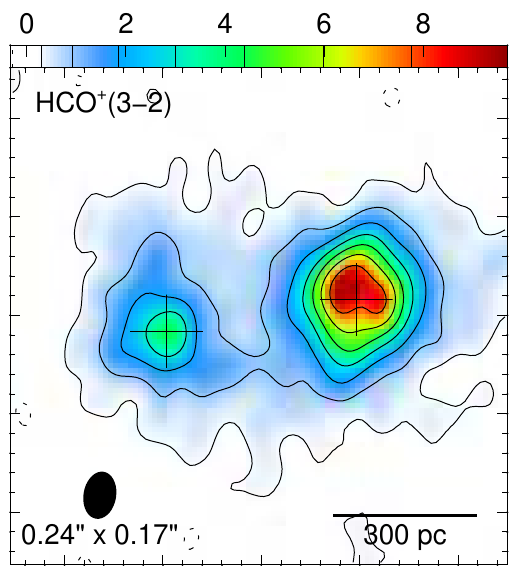} 
\\
\includegraphics[width=0.280\textwidth]{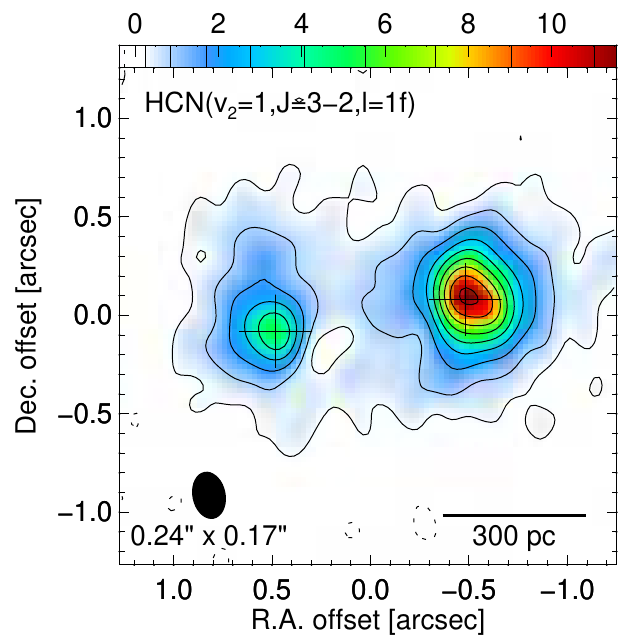} 
\includegraphics[width=0.230\textwidth]{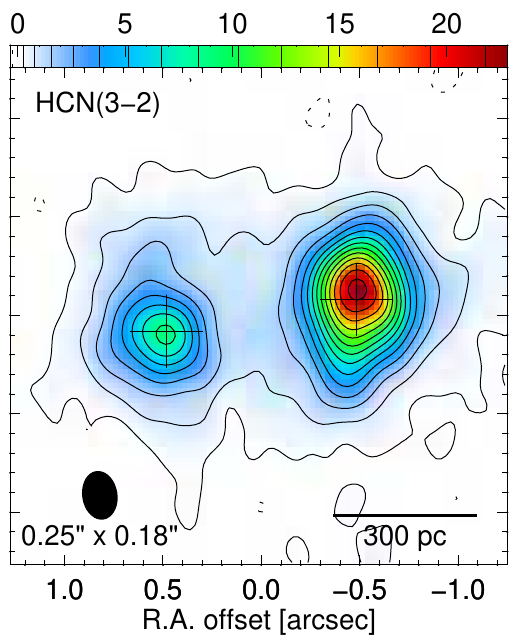} 
\includegraphics[width=0.230\textwidth]{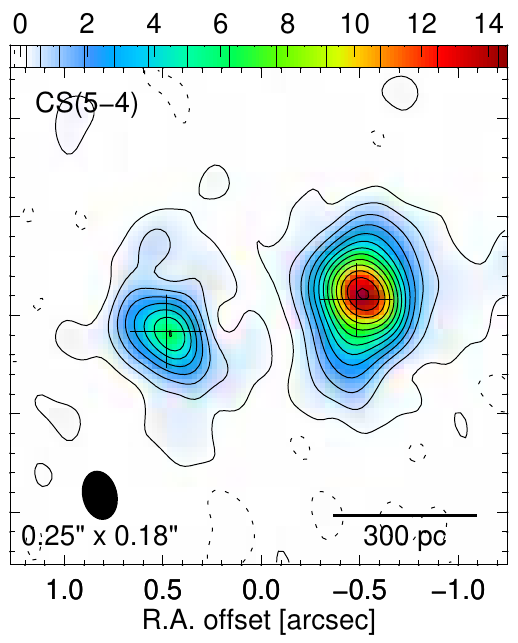} 
\includegraphics[width=0.230\textwidth]{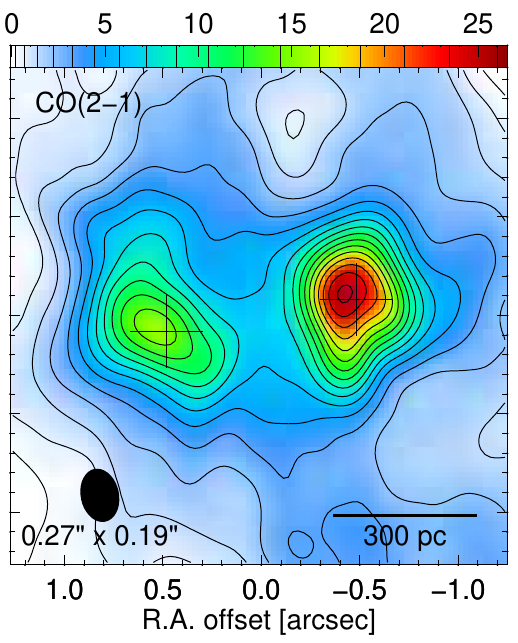} 
\caption{ 
Integrated intensity maps of Arp 220 for major lines in our ALMA observations.
Integration is over the 4850--5850 \kms\ channels, i.e., twenty-one 50 \kms channels. 
(See Table \ref{t.chmaps} for some contaminating lines in this velocity range).
Contours are at $\pm3n^p\sigma$ for $n=1,2,3,\ldots$, where $p$ is in Table \ref{t.chmaps}
and $\sigma$ is calculated from the number of integrated channels (21) and $\sigma_{\rm 50\,km/s}$ in Table \ref{t.chmaps}.
Negative contours are dashed and the correction for the primary beam pattern is applied.
In each panel, the crosses mark the continuum nuclei.
The synthesized beam, labeled with its FWHM, is shown in the bottom-left corner.
}
\label{f.sqash.a220}
\end{figure}

The western nucleus also shows emission elongated to the south (north) at the velocities about 200--450 \kms\ lower
(higher) from the systemic velocity of 5300 \kms. For example, see the channel maps of HCN(4--3) and CO(2--1).
This feature corresponds to the bipolar outflow from the western nuclear disk imaged at 0\farcs1 resolution in
both 3 mm continuum and line emission by \citet{Sakamoto17} and \citet{Barcos-Munoz18},
at 0\farcs2 in CO(3--2) by \citet{Wheeler20}, and at 0\farcs5 in 150 MHz continuum emission by  \citet{Varenius16}.
Figure \ref{f.redblue.a220w} presents this outflow by plotting high-velocity emission in HCN(4--3) and CO(2--1), (3--2), and (6--5). 
The extent of the outflow in these maps is about 0\farcs5 (200 pc) on the sky in each direction.
The HCN plot is for a lower velocity offset from the systemic, 
and it shows a mixture of the nuclear disk component with the east--west velocity gradient
and the outflow component with the north--south velocity gradient. 
The CO(3--2) and (6--5) data are plotted only for the blueshifted emission because the redshifted emission is contaminated by
\HthirteenCN.
These high-velocity emissions are aligned to the outflow axis that we measured to be p.a. = $-15\degr$
in the $\lambda=3$ and 0.45 mm continuum emission \citesp{Paper1}.
The outflow is warm and dense enough to be visible in HCN(4--3) and CO(6--5).
This outflow also appears as the diamond shapes of molecular emission around the western nucleus
in most of the integrated intensity maps in Fig.~\ref{f.sqash.a220}.

We leave our discussion on the molecular outflow from the eastern nucleus to Section \ref{s.lineIm.absline.outflows}.
Although the outflows from the eastern and western nuclei were found using the same technique, 
i.e., blueshifted absorption in \HCOplus \citep{Sakamoto09}, 
the one from the eastern nucleus is not as evident in line emission as the one from the western nucleus.

\begin{figure}[t]
\epsscale{0.85}
\plottwo{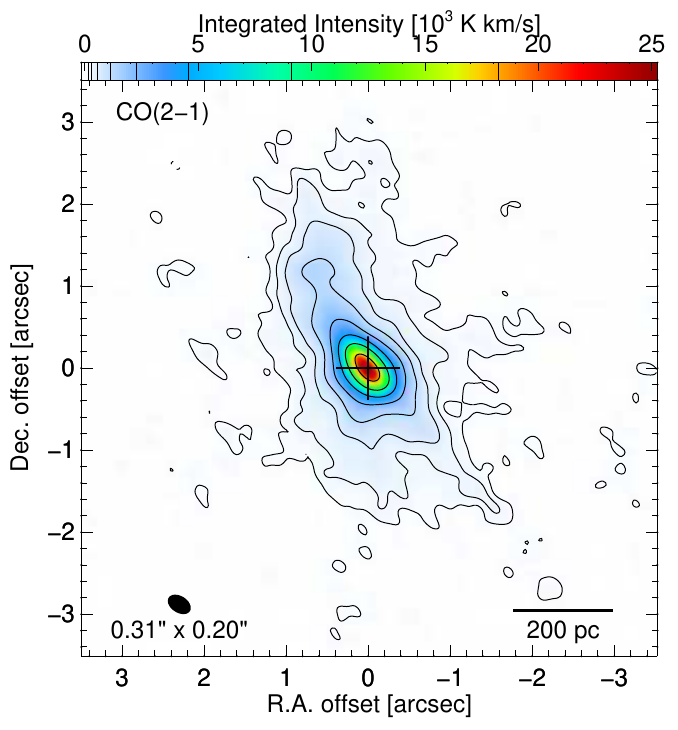}{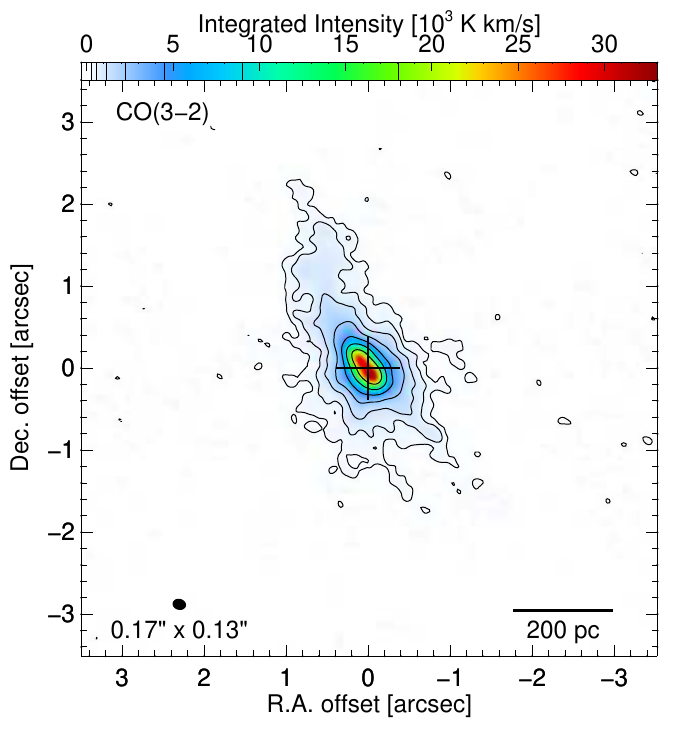}
\plottwo{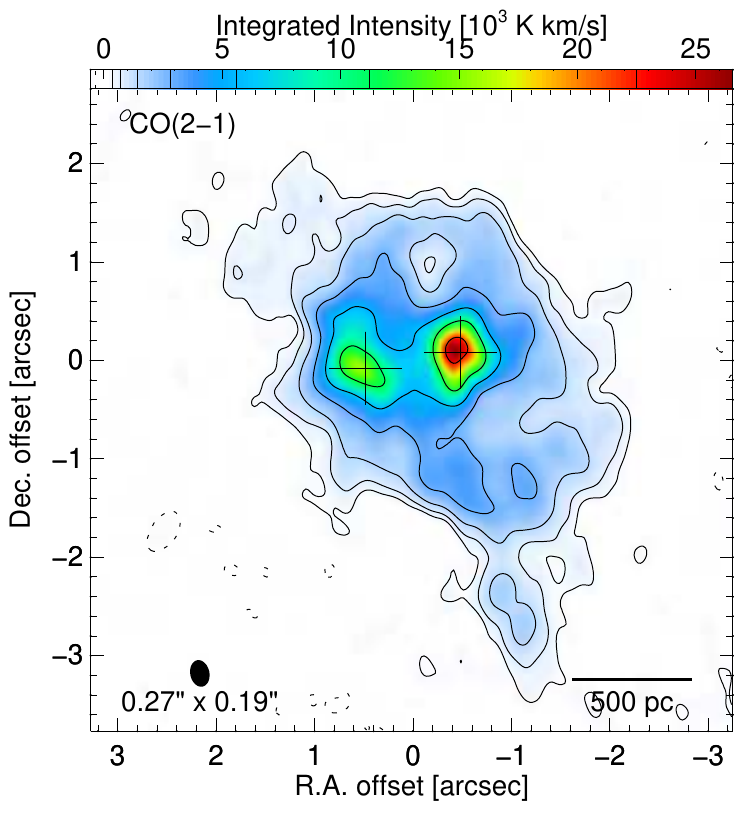}{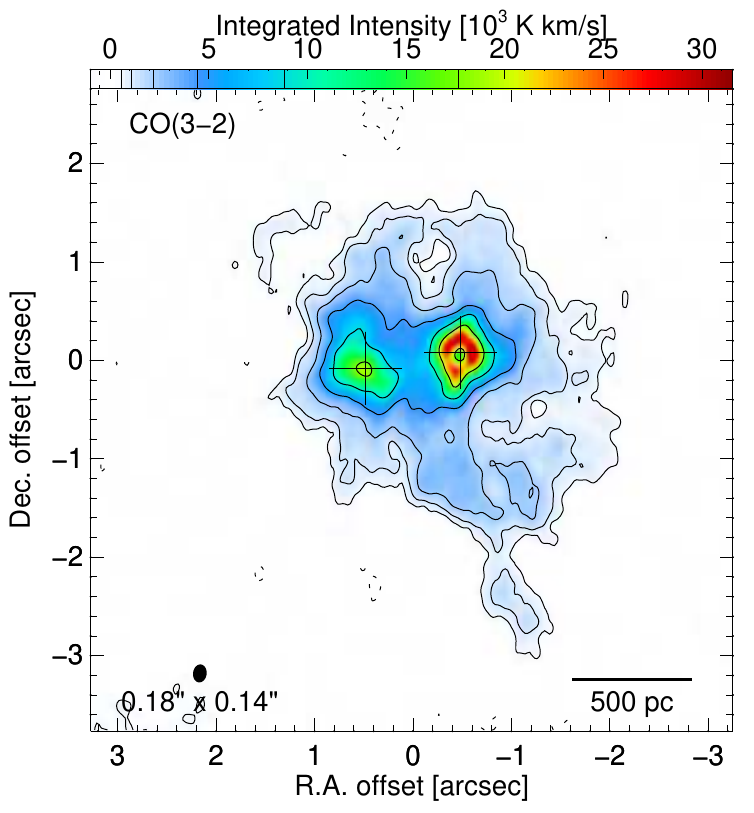}
\caption{ 
Integrated intensity images of NGC 4418 (top) and Arp 220 (bottom) in CO(2--1) and CO(3--2);
the small-area images in Figs.~\ref{f.sqash.n4418} and \ref{f.sqash.a220} are plotted in a wider area. 
Attenuation by the primary beam pattern is corrected.
Contours are at $\pm3 n^2\sigma$ for $n=1,2,3,\ldots$,
where
$\sigma = \sigma_{\rm 50 km/s} \times (50\, \kms) \times \sqrt{N_{\rm ch}}$,
$\sigma_{\rm 50 km/s}$ is in Table \ref{t.chmaps},
and $N_{\rm ch}$ is 9 for NGC 4418 and 21 for Arp 220.
Negative contours are dashed.
Each nucleus is marked with a cross at its continuum position.
The synthesized beam is shown in the bottom-left corner and labeled with its FWHM.
}
\label{f.sqash.COlarge}
\end{figure}

Around the two nuclei is a larger-scale structure shown in Fig.~\ref{f.sqash.COlarge} and referred to as
the outer disk. 
The CO mean velocity maps  in Fig.~\ref{f.mom1.a220} show that while the outer disk has an overall velocity gradient 
in the northeast--southwest direction, its velocity field is disturbed.
This disturbance must be partly due to the nuclear disks and the molecular outflows.
In addition, the outer disk itself is expected to be disturbed in the merging galaxy.

\subsection{Absorption Lines and Skewed Line Profiles}
\label{s.lineIm.absline}
The three nuclei have many absorption features that tell us about the gas along the sightlines to the nuclei.
Most notably, asymmetric absorption in frequency (i.e., in the spectral line profiles) and space 
(i.e., absorption offset from the central continuum peak)
indicates gas motion that is otherwise hard to detect.
As noted in Section \ref{s.lineIm.cautions}, the continuum we subtracted from our line data has simple power-law
spectra and constant shapes over wide frequency ranges. 
They do not artificially cause frequency-dependent features that vary within a single line.

\subsubsection{NGC 4418 --- Skewed Line Profiles and Absorption toward the Nucleus}
\label{s.lineIm.absline.n4418}
\begin{figure}[t]
\epsscale{0.4}
\plotone{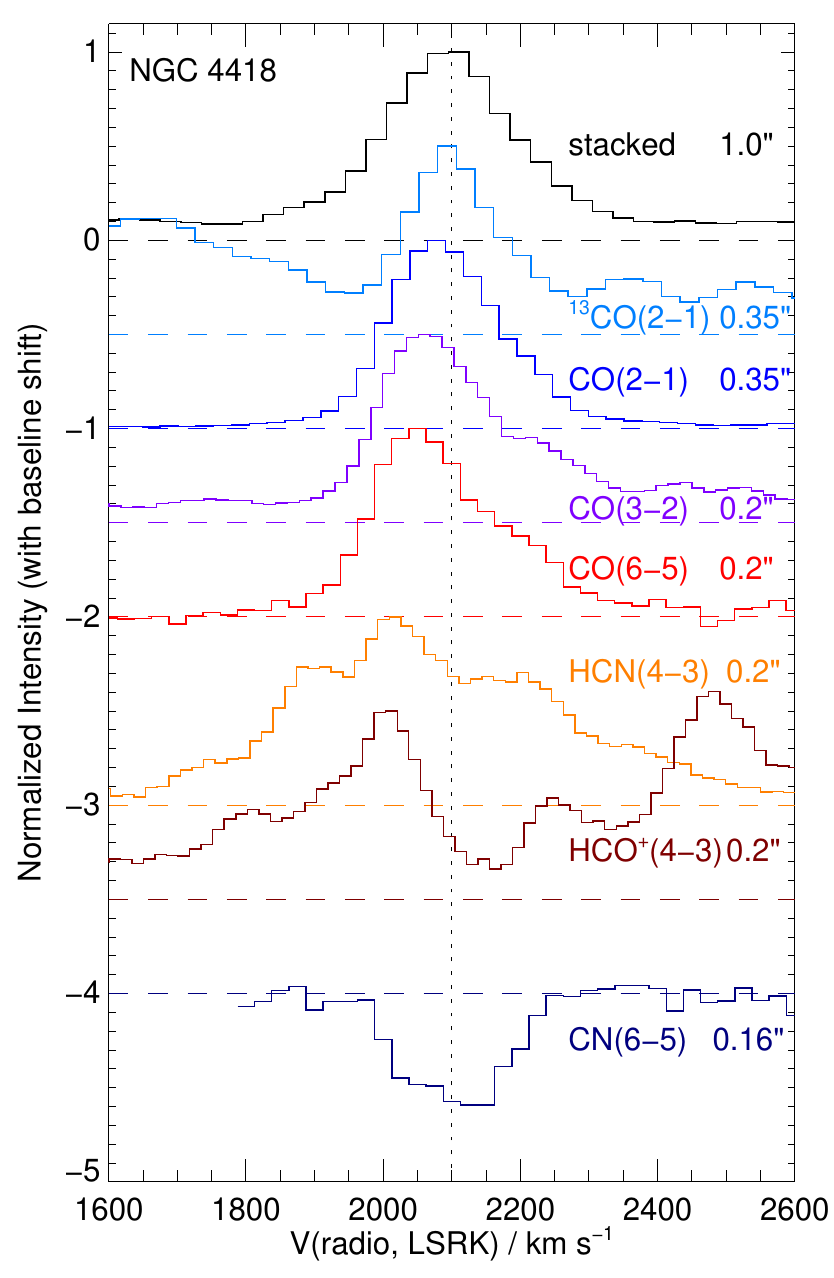} 
\caption{ \label{f.n4418.absorption}
Continuum-subtracted line profiles toward the NGC 4418 nucleus.
The top one is the normalized and stacked line profile in a 1\arcsec\ aperture and was used to decide \Vsys\ in Fig.~\ref{f.vsys}.
The line profiles below, except for the one at the bottom, 
are from 0\farcs35 (Band 6) and 0\farcs20 (Band 7 and Band 9) beams.
They are skewed in various degrees.
Each spectrum, except for CN(6--5) at the bottom, is normalized by the peak intensity within 300 \kms\ from \Vsys = 2100 \kms.
The CN(6--5) spectrum is normalized by the continuum intensity. The continuum is about 60\% absorbed
at the minimum.
Baseline offsets in multiples of $0.5$ are applied to reduce overlap between the spectra.
}
\end{figure}

Figure \ref{f.n4418.absorption} provides an updated view of the skewed line profile in the center of NGC 4418 
reported in \citet{Sakamoto13}. 
At the top and in black is the normalized-and-stacked line profile in the central 1\arcsec. 
It is the profile that we used to estimate the systemic velocity. 
It is symmetric about the adopted \Vsys\ of 2100 \kms. 
The normalized line profiles below (and in color) are from individual lines and smaller apertures.
They are skewed more in lower profiles. 
The second from the top is \thirteenCO(2--1).  
Although the profile is nearly symmetric about \Vsys,
it peaks at a slightly lower velocity and has a longer tail in higher velocities. 
The CO(2--1), (3--2), and (6--5) profiles below are progressively more skewed in the same manner.
The HCN(4--3) and  \HCOplus(4--3) profiles below have their peaks at \about2000 \kms.  
The former is flat between 2100 and 2200 \kms.  
The latter shows an absorption-like minimum at around 2150 \kms.
This \HCOplus(4--3) profile reminds us that its continuum absorption is the deepest among the absorption features 
toward Arp 220E and W in Bands 6 and 7 (Figs.~\ref{f.triNuc_spec.B7}, \ref{f.triNuc_spec.B6.350mas}).
At the bottom of Fig.~\ref{f.n4418.absorption} is the very deep CN(6--5) absorption near \frest=680 GHz.\footnote{
It is a blended feature of N=6--5 doublet of J=$\frac{11}{2}$--$\frac{9}{2}$ and $\frac{13}{2}$--$\frac{11}{2}$, each of which containins six 
hyperfine features, spanning 126 \kms\ in total. 
We used for our velocity labeling the weighted-mean frequency of the twelve transitions using $g_u A$ for the weighting,
where $g_u$ is the statistical weights of the upper state and $A$ is the Einstein $A$ coefficient.}
Similar to the case of \HCOplus(4--3), 
the deepest absorption is between 2100 and 2150 \kms\ and is hence slightly redshifted from \Vsys. 
This spectrum is normalized by the continuum intensity, and as much as 60\% of the continuum emission is absorbed.
The CN optical depth would be 0.9 for a foreground absorber having a covering factor of unity.
The CN absorption is broad, covering both sides of \Vsys\ over at least 250 \kms, 
although this is partly because this CN feature consists of twelve transitions spanning 126 \kms.
Our Band 9 data on Arp 220 do not cover CN(6--5), but deep CN absorption has been observed toward both nuclei 
in N=3--2 and 1--0 \citep{Scoville15,Sakamoto17}.

We attribute, as in \citet{Sakamoto13}, the skewed line profiles to absorption.
It is probably a combination of self-absorption of the lines and line absorption of the background continuum.
The near-symmetric \thirteenCO\ line profile strengthens this interpretation.
Our observations do not favor an alternative model for the skewed CO(2--1) profile to introduce
two emission components at different velocities \citep{Costagliola13}.
The profiles of other liens should be similar to CO(2--1) if there were two emission components and no absorption. 
In addition, the redshifted CN absorption has negative intensities in the continuum-subtracted spectra.
They cannot be explained either by two gas velocities without absorption or by line self-absorption only.
The redshifted absorption is also in \ion{O}{1}, OH, and \ion{H}{1} toward NGC 4418 \citep{GA12,Costagliola13}.

The redshifted absorption is most easily interpreted as an inward gas motion toward the nucleus along our sightline
\citep{GA12, Sakamoto13}. 
It does not necessarily mean a net gas inflow to the nucleus since the circumnuclear gas may
have outward motions in directions other than our sightline.
We point out in \S\ref{s.lineIm.absline.outflows} that the redshifted absorption can also be due to an outflow 
when the background continuum is not a point source but a disk.

We also detected, for the first time for NGC 4418, SiO absorption around the systemic velocity.
The continuum-subtracted SiO(8--7) intensity goes below zero at the center of the nucleus at $\lesssim0\farcs2$ resolution
in Figs.~\ref{f.chmap.n4418.B7}j and \ref{f.triNuc_spec.B7}. 
SiO(16--15) also shows sub-continuum absorption. It has a broad wing in blueshifted velocities
(Figs.~\ref{f.triNuc_spec.B9.350mas} and \ref{f.chmap.n4418.B9}).
The minima of the SiO spectra are around the fiducial systemic velocity or slightly blueshifted from it in the two transitions.
The absorption velocities may be, however,  affected by the nearby \HthirteenCOplus\ lines in redshifted velocities.
The large width of more than 100 \kms\ in the SiO(16--15) absorption is consistent with the broad CN(6--5) absorption.
Judging from the large width and the need for high excitation, it is more likely that the absorbing gas is inside the nucleus 
rather than a gas that happens to be on our sightline to the nucleus but is far from it.
The SiO(16--15) absorptions in NGC 4418 and Arp 220 are discussed  in Section \ref{s.lineIm.absline.others}.

\subsubsection{Arp 220 --- Absorption-line Systems toward the Nuclei}
\label{s.lineIm.absline.a220abssystem}
\begin{figure}[t]
\centering
\includegraphics[height=87mm]{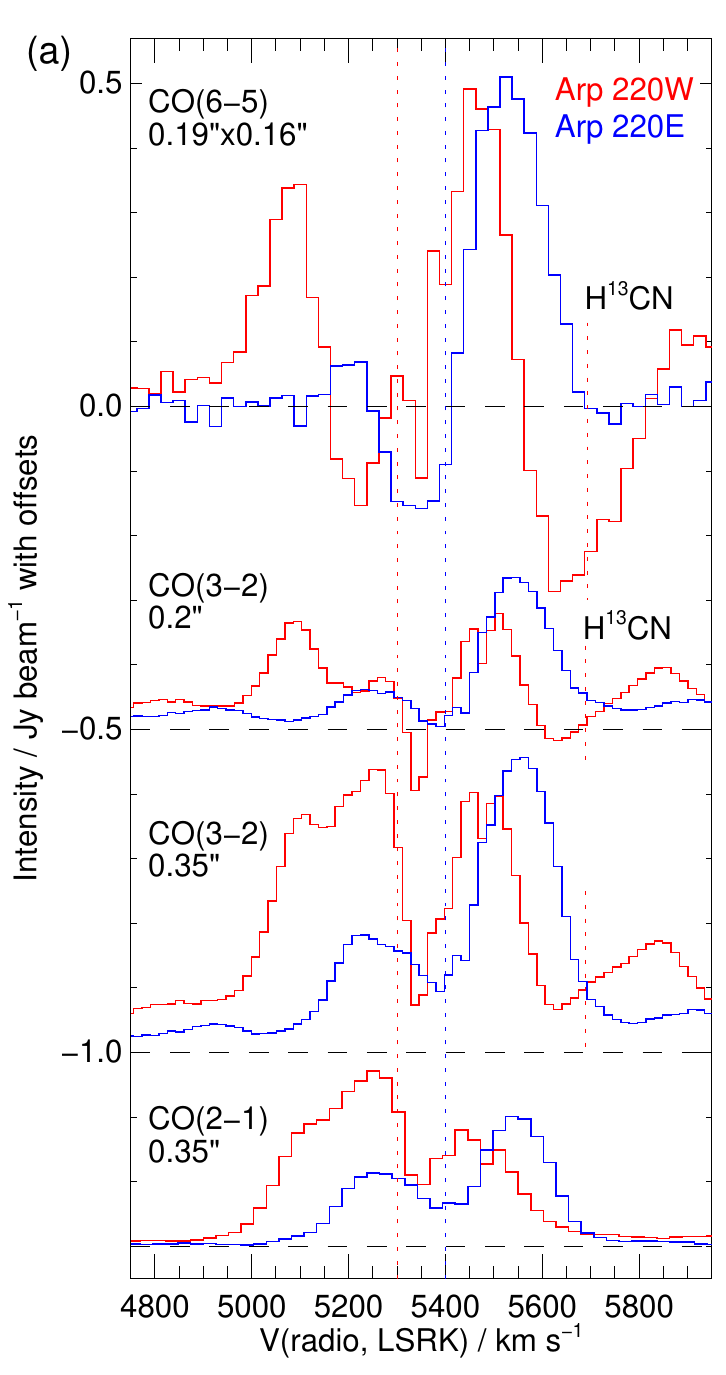}
\includegraphics[height=87mm]{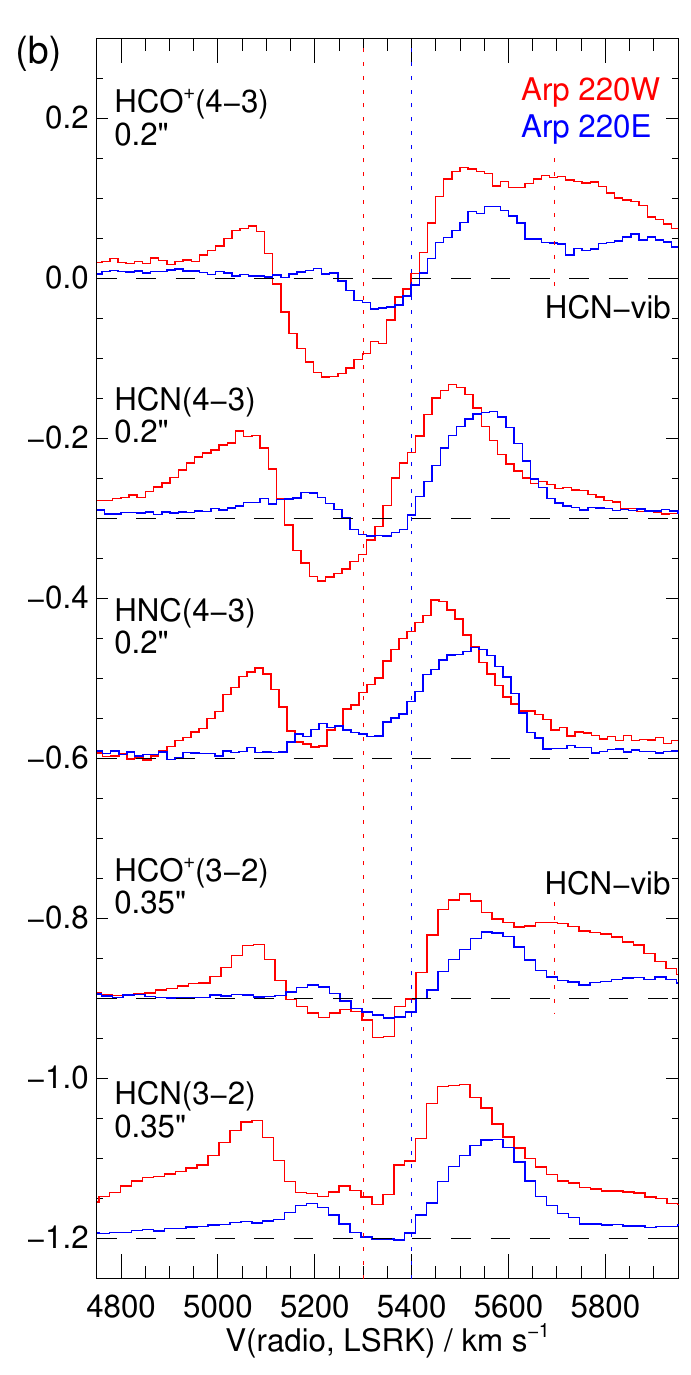}
\includegraphics[height=87mm]{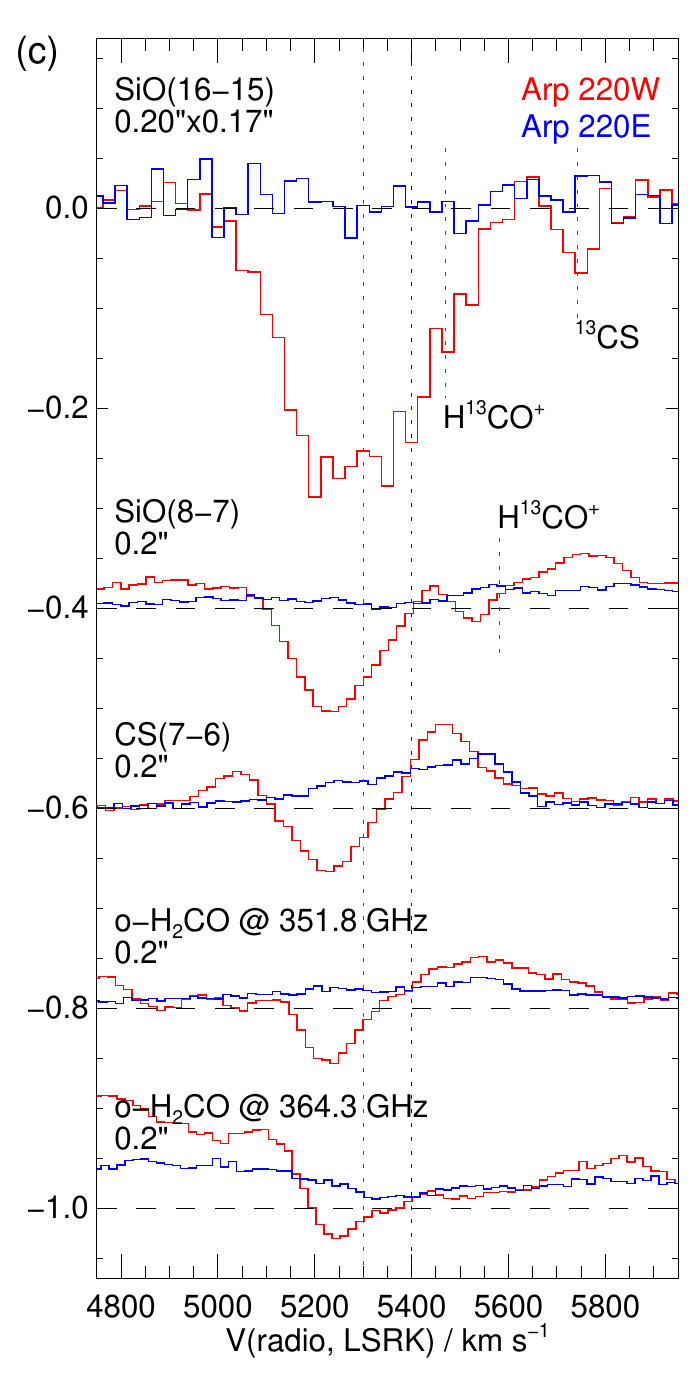}
\includegraphics[height=87mm]{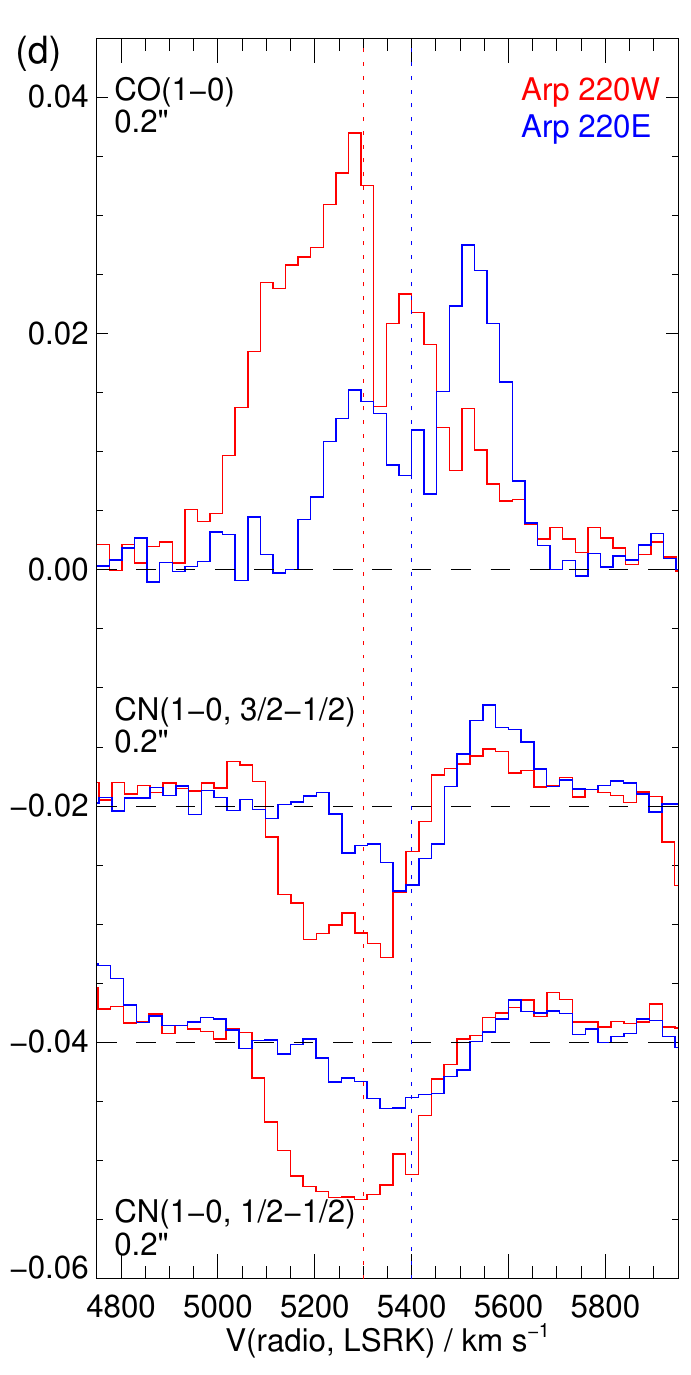}
\caption{ \label{f.a220.absorption}
Continuum-subtracted line profiles toward the nuclei of Arp 220. 
The W nucleus  is plotted in red and E in blue.
Their fiducial systemic velocities at 5300 and 5400 \kms\ are the vertical dotted lines in the same color scheme.
The spectra are plotted with vertical shifts to reduce overlaps along with black dashed lines at the zero level.
To the left of the individual spectra are the line names and the data resolution.
In panel (c), ortho-\HtwoCO\ at 351.8 GHz is the ($5_{1,5}$--$4_{1,4}$) transition, 
and that around 364.3 GHz is a mixture of the ($5_{3,2}$--$4_{3,1}$) and ($5_{3,3}$--$4_{3,2}$) transitions.
Major blended lines are marked with vertical dotted line segments in red at the systemic velocity of the W nucleus. 
CO(6--5) is blended with \HthirteenCN(8--7),
CO(3--2) with \HthirteenCN(4--3), 
\HCOplus\ lines with HCN($v_2$=1, $l$=1f),
SiO(16--15) with \HthirteenCOplus(8--7) and \thirteenCS(15--14),
 and SiO(8--7) with \HthirteenCOplus(4--3).
The spectra in panel (d) are from the 3 mm data in \citet{Sakamoto17}.
}
\end{figure}

Figure~\ref{f.a220.absorption} presents absorption features in the continuum-subtracted line profiles of the two Arp 220 nuclei.
We selected lines showing significant absorption of the continuum (i.e., negative intensities in the continuum-subtracted data) 
toward at least one nucleus and supplemented them in panel (d) with spectra of some 
\about3 mm lines from \citet{Sakamoto17}.
Both nuclei show sub-continuum absorption in blueshifted velocities, confirming
our first report \citep{Sakamoto09} and companion ALMA observations \citep{Martin16}.
It is evident in, e.g., CO(6--5), \HCOplus(4--3), and HCN(4--3), 
The blueshifted absorption has also been seen toward both nuclei in molecular lines at $\lambda \sim 1$ cm \citep{Zschaechner16}.

\paragraph{\arpW:}
On closer inspection of the new high-quality spectra, we see toward \arpW\ absorption features of distinct properties 
at two velocities.
One is a narrow absorption feature seen in CO at around 5330 \kms. 
It is most evident in the CO(3--2) spectra in panel (a).
As in this example, we denote an absorption feature with the velocity of its minima, \Vmin.
We caution that \Vmin, in general, differs from the central velocity of the absorber,  \Vabs.
For the latter, one needs to decompose an observed spectrum to emission and absorption features, 
e.g., by assuming a Gaussian profile for each component \citep{Martin16}.\footnote{
If a line profile is the sum of a Gaussian emission centered 
at the systemic velocity and a Gaussian absorption feature, 
then the absorption minimum in the observed line profile is further from the systemic velocity than 
the central velocity of the absorbing gas; $|\Vmin - \Vsys| \geq |\Vabs - \Vsys|$.
It is because the emission component acts like a non-flat spectral baseline to the absorption line profile.
}
However, this 5330 \kms\ feature is so narrow that the velocity-shifting effect by other spectral components must be small,
hence \Vmin \about \Vabs \about 5330 \kms.
Therefore, this feature is close to but about 30 \kms\ redshifted from our fiducial velocity of the western nucleus.
This feature has negative intensities (i.e., continuum absorption) in our CO(6--5) and 0\farcs2 CO(3--2) data.
This feature is also discernible as narrow troughs in our 0\farcs35 CO(3--2) spectrum and CO(2--1) and (1--0) profiles,
although the minimum intensities are still positive. 
Given the agreement in \Vmin\ and similar widths, we regard these to be the same `5330 \kms\ absorption feature' toward \arpW. 
We note that the apparent absorption depth depends not only on the line opacity but also on the angular resolution and observing frequency.
It is because deep absorption against a compact continuum source can be filled with off-nuclear emission when both are in the same beam and also because the continuum brightness and source size vary with frequency. 
The effect of the angular resolution is evident when comparing the 0\farcs2 and 0\farcs35 spectra of CO(3--2) in panel (a).
This 5330 \kms\ absorption feature is not evident in other species, except possibly in \HCOplus(3--2) and CN(1--0, 3/2--1/2).

Another absorption feature toward \arpW\ is the broad absorption having \Vmin \about 5200--5250 \kms,
most clearly seen in \HCOplus(4--3) in panel (b).
We call it the 5230 \kms\ feature for brevity. 
This feature is unambiguously blueshifted compared with the W nucleus by about 130 \kms\
and is much broader than the \about5330 \kms\ feature.
The 5230 \kms\ feature shows negative intensities (i.e., continuum absorption) in Fig.~\ref{f.a220.absorption}
in CO(6--5), \HthirteenCN, \HCOplus, HCN(4--3), SiO, CS, \HtwoCO, and CN. 
Note in particular the \HthirteenCN(8--7) and (4--3) absorption lines. 
They have $\Vmin\ \sim \Vsys({\rm W}) - 100$ \kms\ as other 5330 \kms\ features do,
and hence the absorption features are not due to redshifted CO at  $\Vmin \sim \Vsys + 300 $ \kms.
The 5230 \kms\ feature is also visible as troughs at positive intensities in CO(3--2) at 0\farcs2, HNC, and HCN(3--2). 
As noted above, the positive intensities must be partly due to the data resolution. 
Indeed, the HNC(4--3) channel maps in Fig.~\ref{f.chmap.a220.B7}(a) show negative intensity on the W nucleus in the 5200 \kms\ channel. 
The 5230 \kms\ absorption feature has an apparent FWHM of \about200 \kms\ in \HCOplus(4--3) in panel (b).
Because of that, a part of this feature extends to the redshifted velocities of the W nucleus. 
The same is seen at least in \HthirteenCN(8--7), \HCOplus, HCN(4--3), SiO, CS, and CN(1--0).

The western nucleus, therefore, has an absorption line system of two components.
One is the narrow feature at \Vmin \about 5230 \kms\ seen in CO. 
The other is the broad feature at \Vmin \about 5330 \kms\ seen in CO and many other lines. 
Although at a much lower signal-to-noise ratio, we already recognized both absorption features in our SMA observations \citep{Sakamoto09}.
The only feature we do not confirm is their 4890 \kms\ feature.
In our new ALMA observations with much higher data quality and wider frequency coverage, 
the relative strength of the two absorption features significantly varies with molecular species, 
line transitions, observing frequencies, and angular resolution.

\paragraph{\arpE:}
The eastern nucleus of Arp 220 has a broad feature of negative intensity with minima at \Vmin \about 5350 ($\pm50$) \kms\
in CO(6--5), \HCOplus, HCN(4--3), and CN in Fig.~\ref{f.a220.absorption}.
In addition, troughs of positive intensities are seen at similar \Vmin\ and have comparable widths in such lines as 
CO(3--2) and (2--1), HNC, and HCN(3--2). 
The presence of absorption is unambiguous in the former group of lines. 
Therefore, it is more sensible to attribute the troughs of positive intensities in the latter group of lines 
to absorption (line self-absorption and or continuum absorption) 
than to uneven distribution of gas parameters such as lower gas temperature in blueshifted velocities. 
Consequently, we regard these features as the `$\Vmin \sim 5350$ \kms\ absorption' toward \arpE.
This absorption was also first found with the SMA \citep{Sakamoto09}.

The \Vmin\ of all sub-continuum absorptions are lower than the \Vsys\ of the nucleus,
indicating $\Vabs < \Vsys$.
The line profiles having troughs of positive intensities at \about5350 ($\pm$50) \kms\ are generally asymmetric 
in the sense that they have significantly higher peaks in redshifted velocities (i.e., $V > \Vsys$) 
than in blueshifted velocities, also implying a blueshift of the absorption.
The width of the \about5350 \kms\ absorption feature is comparable to that of the `broad' 5230 \kms\ feature 
toward the western nucleus.
As in the case of the \about5230 \kms\ feature toward the W nucleus, 
negative intensities of the broad \about5350 \kms\ absorption feature toward the E nucleus are partly in 
redshifted velocities, e.g., in the CN spectra in Fig.~\ref{f.a220.absorption}. 
A redshifted wing of absorption profile can also be inferred by decomposing some line profiles (e.g., CO, HCN) into
a Gaussian emission and a Gaussian absorption.

Similar to our observations toward \arpW,
there might be an additional narrow absorption feature in the CO line profiles at a slightly more redshifted velocity 
than the broader component and the systemic velocity of the nucleus.
It may be why the nominal minimum in the trough of the CO(1--0) profile
is slightly redshifted compared to the systemic velocity of the nucleus.

Therefore, the eastern nucleus has an absorption line system of at least one and maybe two components.
This system is similar in many ways to that toward the W nucleus. 
A notable difference is the lack of deep sub-continuum absorption in SiO and \HtwoCO\ toward the E nucleus.

\paragraph{Two outflows:}
Both nuclei have a broad absorption feature that is blueshifted relative to the nucleus; 
$\Vmin(\rm{W}) \sim 5230\; \kms \sim \Vsys(\rm{W}) - 70$~\kms\ 
and
$\Vmin(\rm{E}) \sim 5350\; \kms \sim \Vsys(\rm{E}) - 50$~\kms. 
These blueshifted absorptions have been attributed to molecular outflows from the individual nuclei \citep{Sakamoto09}.
Our ALMA observations support this interpretation in three ways.
First, the absorbing gas must have different line-of-sight velocities (\Vabs) toward the two nuclei 
after correcting their apparent \Vmin.
It is because for $\Vmin < \Vsys$ the corrected \Vabs\ will be at $\Vmin < \Vabs < \Vsys$, 
and hence one obtains $\Vmin(\rm{W}) < \Vabs(\rm{W}) < \Vsys(\rm{W}) < \Vmin(\rm{E}) < \Vabs(\rm{E}) < \Vsys(\rm{E})$ 
for Arp 220.
Second, the two broad-line absorbers have distinctively different degrees of absorption between the two sightlines 
in CS(7--6), SiO(8--7), and o-\HtwoCO\ transitions.
Third, the absorbing gas must be close to each nucleus rather than in a common envelope of the two nuclei.
For these broad absorptions to occur, the absorbing gas must have 
a high enough excitation yielding a significant population of the lower levels of the observed absorption features,
such as J=3 for HCN(4--3) absorption.
We saw that such high excitation was limited to the immediate vicinity of each
nucleus for species other than \twelveCO\ (Section \ref{s.lineIm.size.observations}).
For these reasons, we maintain that these blueshifted, broad-line absorption features are due to two molecular outflows, 
one from each nucleus.

\paragraph{Narrow absorption:}
The narrow-line absorption feature toward \arpW, as well as the possible one toward \arpE,
may well be due to gas that is farther out from each nucleus than the broad-line absorbers.
Their small velocity widths (i.e., less turbulence) and lower excitation implied by their detection mostly in CO are 
consistent with this interpretation. 
The gas causing the narrow line absorption can be either in the outskirts of the individual nuclear disks
or outside the nuclear disks and in a large gas structure in the merger, such as the outer disk to encompass the two nuclei.
In either case, the small offset of the narrow features from \Vsys\ of each nucleus may only represent 
disturbed gas motion (e.g., in the outskirts of the nuclear disks) 
and not necessarily a coherent motion of gas on large scales.

\subsubsection{Outflow-Disk Configurations through Spatially Lopsided Line-absorption}
\label{s.lineIm.absline.outflows}
\begin{figure}[t]
\epsscale{0.6}
\plotone{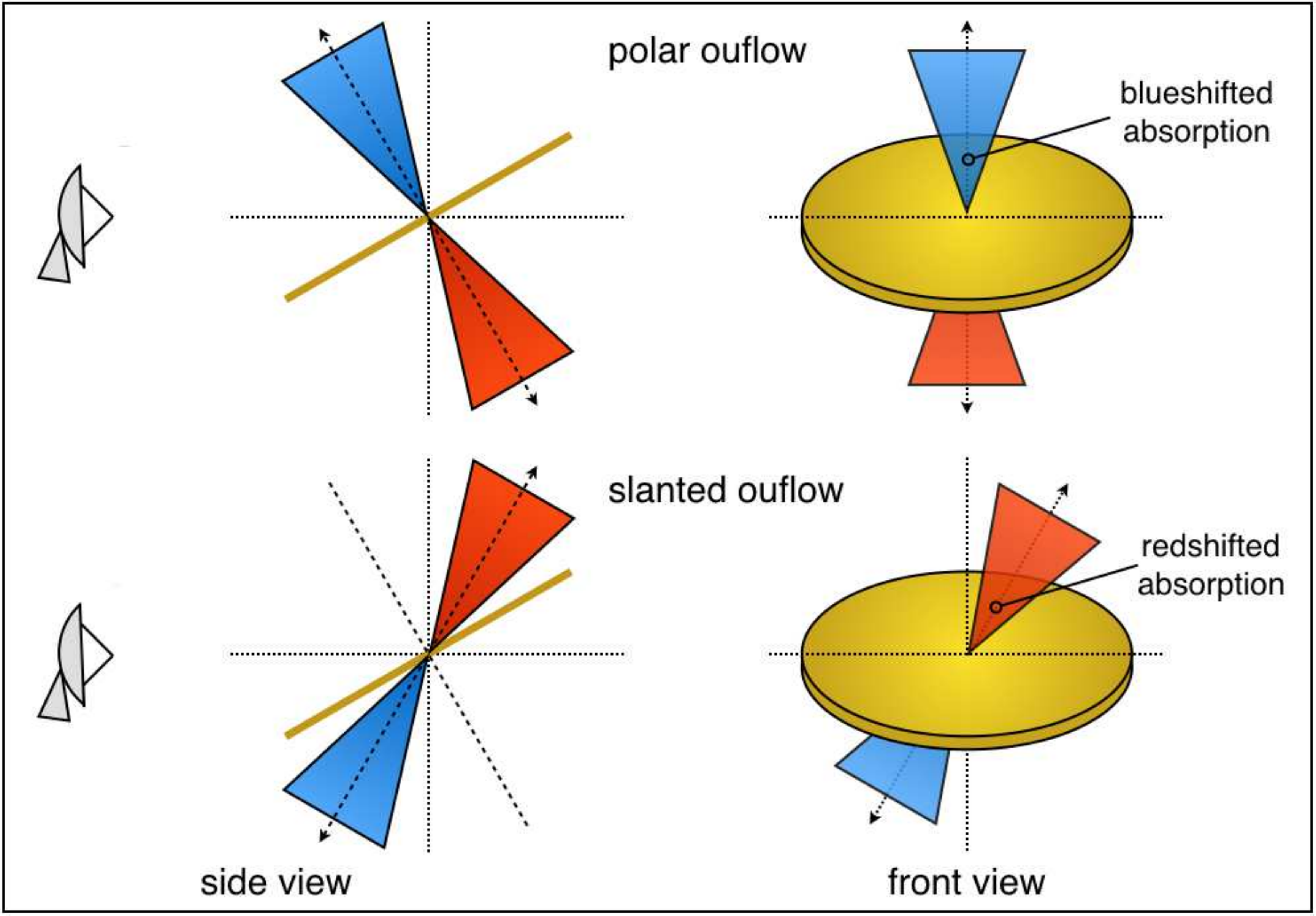} 
\caption{ \label{f.outflow_configs}
Configurations of a disk--outflow system and their effects on line absorption. 
To the left are the side views of two configurations of a bipolar outflow and a circumnuclear disk.
To the right are their sky-plane projections. 
The outflow is along the polar axis of the disk in the top row while it is slanted in the bottom row.
The outflow perpendicular to the disk has its approaching gas in front of the disk, and hence the outflow absorbs the disk light
at blueshifted velocities. 
(If the disk is close to edge-on or the outflow opens wide enough, then 
a minor part of the foreground outflow can be redshifted while the majority is blueshifted.)
For a slanted outflow that is not perpendicular to the disk, both blueshift and redshift are possible 
for the central velocity of the absorbing gas.
In this illustration, the receding (i.e., redshifted) part of the slanted outflow is in front of the disk,  and hence the
line absorption by the outflow is redshifted. 
}
\end{figure}

If a disk is the continuum source and the absorbing gas is due to a bipolar outflow, 
then the spatial distribution of the line absorption tells us about the outflow-disk configuration.
In the top row of Figure \ref{f.outflow_configs}, a bipolar outflow of molecular gas is perpendicular to a continuum-emitting
nuclear disk.
The outflow should cause line absorption preferentially around the far-side semi-minor axis of the disk
in blueshifted velocities.
The outflow may also be visible in emission further from the center around the minor axis,
either on the near-side of the disk in redshifted velocities
or on the far-side in blueshifted velocities.
Thus, it is possible to use
a spatially-lopsided system of absorption and emission around the center 
to detect a bipolar outflow normal to a bright and opaque nuclear disk.
This method is parallel to detecting an outflow through a P-Cygni line profile, 
i.e., an anti-symmetric system of absorption and emission in the velocity space.
Figure \ref{f.outflow_configs} also shows in the bottom row 
that a slanted bipolar outflow that is not perpendicular to the nuclear disk can have lopsided and redshifted absorption.
The absorption occurs on the far side of the disk but not necessarily on the minor axis of the disk.
From a spatially-lopsided intensity profile and the velocity information of the line absorption, 
one can tell the direction of the bipolar outflow and near and far sides of the inclined disk 
and constrain whether the outflow is normal to the disk.\footnote{
We consider a collimated {\em inflow} toward a nuclear disk much less likely and disregard it without new evidence, 
although it could also explain the spatially-lopsided absorption in blue- and red-shifted velocities. 
}

\begin{figure}[t]
\epsscale{0.35}
\plotone{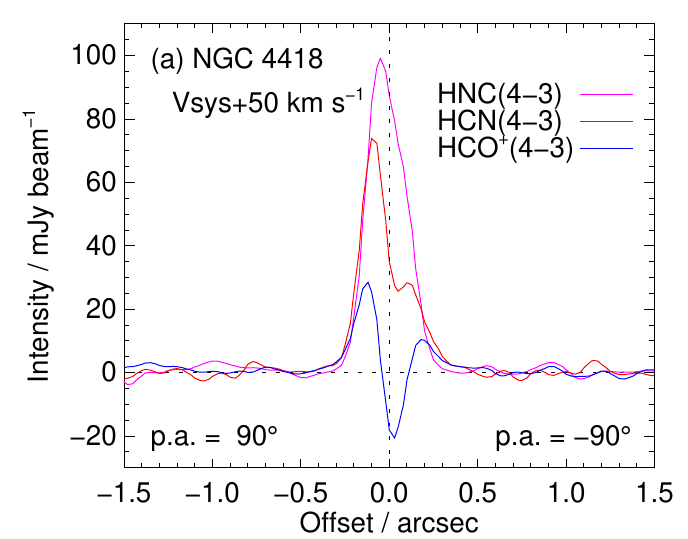} \\
\plotone{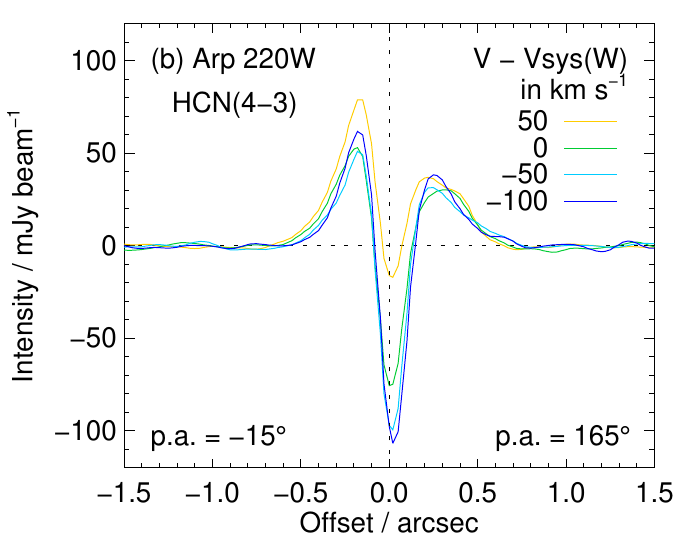} \\
\plotone{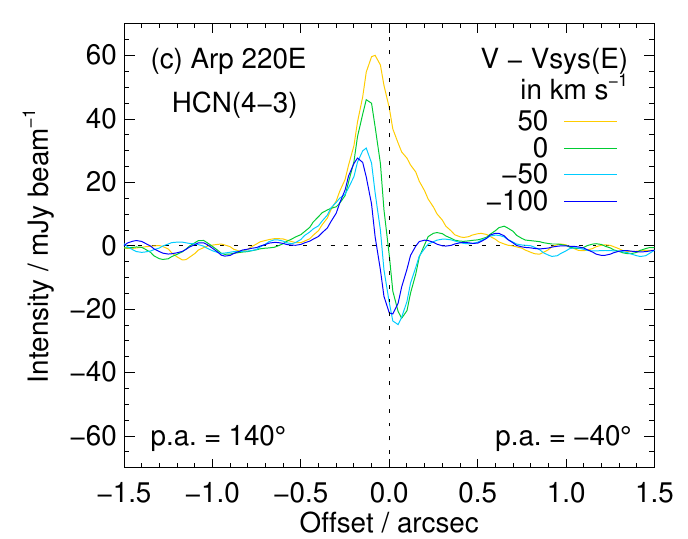} \\
\epsscale{1.0}
\caption{ \label{f.1Dslice}
Lopsided line-intensity profiles across the three nuclei.
The origin of the offset coordinate is at the continuum nucleus in each panel.
Panel (a) is for NGC 4418 and shows the intensity profiles of three lines along p.a. = 90\degr.
The profiles are from the channel maps at $\Vsys + 50$ \kms.
Panel (b) is for Arp 220W and (c) for Arp 220E. 
Both show the HCN(4--3) intensity profiles along the minor axes of the two nuclear disks.
The profiles are taken from the channel maps at $(+50, 0, -50, -100)$ \kms\ from the systemic velocity of each nucleus.
}
\end{figure}

\paragraph{NGC 4418:}
This nucleus has redshifted absorption in many lines (Section \ref{s.lineIm.absline.n4418}).
Figure~\ref{f.1Dslice}(a) shows intensity profiles of three lines along the east--west axis across the continuum peak. 
The profiles are from the channel maps at $\Vsys + 50$ \kms, where the redshifted absorption is most significant.
The profiles show that redshifted absorption is slightly shifted to the west (p.a. = \minus90\degr) relative to the continuum peak.
(We adopted the E--W slicing based on the offset seen in the \HCOplus(4--3) channel map.)
If this offset is due to an outflow, then the far-side of the nuclear disk, whose major axis is at p.a. $\approx$ 47\degr,
is in the northwest, which is also the far-side of the large-scale galactic disk (and hence the two disks are nearly coplanar).
In this outflow model, the redshifted velocity of the absorption means that the outflow direction is oblique to the nuclear disk, 
similar to the lower row in Fig.~\ref{f.outflow_configs}.
Such slanted configurations are often found in AGN-driven outflows, e.g., in NGC 1068 \citep{GarciaBurillo14}.
The slanted outflow is not immediately visible in emission, except that the HCN(4--3) mean velocity map in Fig.~\ref{f.mom1.n4418}
shows a blueshifted component about 0\farcs2 east of the nucleus. 
We present this slanted mini outflow as a possible model for the redshifted absorption in NGC 4418.
This model still needs verification because the spatial offset of the absorption is small (since the continuum nucleus is compact) 
and because the lopsidedness in Fig.~\ref{f.1Dslice}(a) may be due to line emission in the nuclear disk.
Another model for the redshifted absorption is the inward-moving gas along our sightline, explained 
in Section \ref{s.lineIm.absline.n4418}.
We add that the SiO(16--15) absorption at the systemic velocity is at the continuum peak position without an offset. 
A possible reason for the lack of lopsided absorption and circumnuclear emission is 
that the SiO excitation to J \about\ 15 owes much to sub/millimeter continuum radiation from the nucleus
(see Section \ref{s.lineIm.absline.others}).

\paragraph{\arpW:} 
The blueshifted absorption toward \arpW\ was first found spatially lopsided by \citet{Tunnard15},
who linked it to a lopsided occultation of the nuclear disk by the outflow.
The absorption offset has been confirmed in subsequent observations \citep{Rangwala15,Martin16, Wheeler20}, 
and it is also in earlier SMA observations \citep{Sakamoto08,Sakamoto09} as a small offset 
between the peak of continuum emission and that of CO integrated intensity.
This lopsidedness is very clear in many of our channel maps. 
Specifically, the line emission around the continuum peak of the nucleus 
is stronger to the north and weaker to the south in velocities around 5300 \kms; see
HNC(4--3), \HCOplus(4--3) and (3--2), HCN(4--3) and (3--2), \CtwoH(4--3), SiO(8--7), and CS(7--6)
in Figs.~\ref{f.chmap.a220.B7} and \ref{f.chmap.a220.B6}.
Notable exceptions, SiO(16--15) and \HtwoCO(9$_{1,8}$--8$_{1,7}$) in Band 9, are addressed in Section \ref{s.lineIm.absline.others}.
Figure~\ref{f.1Dslice}(b) shows the lopsidedness in HCN(4--3) using the intensity profiles along the outflow axis
of p.a. = $-15$\degr\ that is within 10\degr\ from the minor axis of the nuclear disk \citesp{Paper1}.
The lopsided absorption is evident, with the absorption slightly shifted to the south of the continuum nucleus.
We attribute this spatial lopsidedness to the absorption of the photons from the background nuclear disk. 
Then, the absorbing gas must cover more of the southern side of the western nuclear disk. 
Therefore, the southern side of the nuclear disk must be the far side of the disk, whether the outflow is normal to the nuclear disk or not.

\begin{figure}[t]
\centering
\includegraphics[height=50mm]{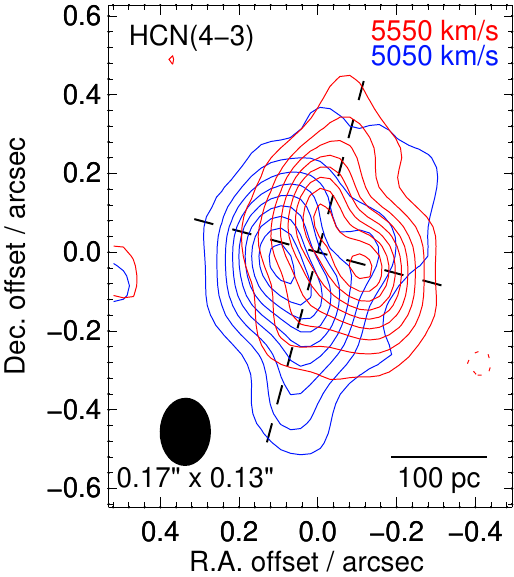}
\includegraphics[height=50mm]{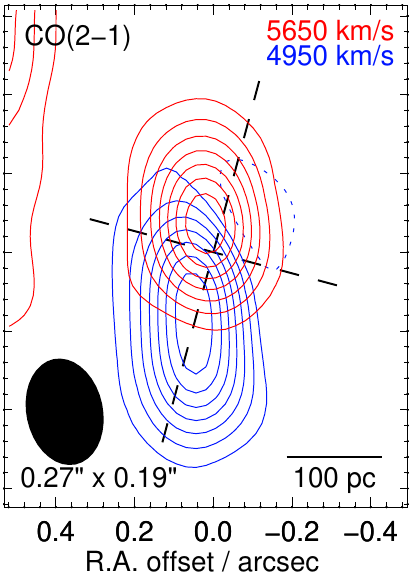}
\includegraphics[height=50mm]{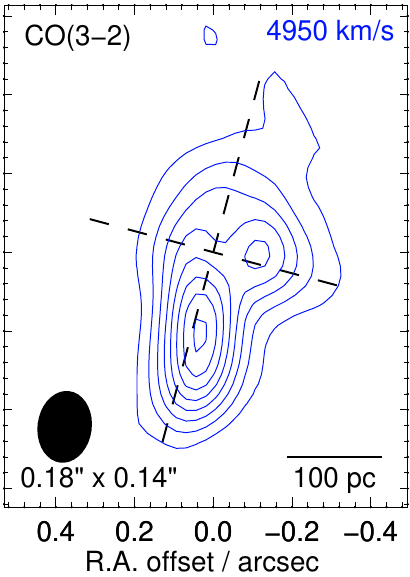}
\includegraphics[height=50mm]{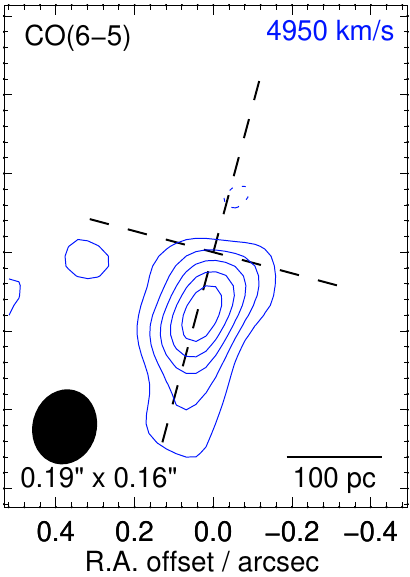}
\caption{ 
High-velocity gas around Arp 220W showing the bipolar outflow.
The HCN(4--3) and CO(2--1) maps show emission at $\pm 250$ and $\pm 350$ \kms, respectively, 
from the fiducial velocity of the western nucleus, 5300 \kms.
CO(3--2) and (6--5) data are only shown for $\Vsys - 350$ \kms\ 
because of the contamination by \HthirteenCN\ at $\Vsys + 390$ \kms.
Each channel map integrates 50 \kms.
Contours are at $\pm Cn^{p} \sigma_{\rm 50\, km/s}$ with 
$(C, p) = (3, 1.5)$ for HCN, (3, 1.3) for CO(2--1) and (3--2), and (3, 1) for CO(6--5).
Negative contours dashed.
The rms noise, $\sigma_{\rm 50\, km/s}$, is in Table \ref{t.chmaps} for each line.
The black dashed cross marks the continuum position of the western nucleus, which is the origin of the offset coordinates,
and the major and minor axes of the continuum outflow measured at $\lambda = 3$ and 0.45 mm 
\citep[major axis p.a. = $-15\degr$;][]{Paper1}.
}
\label{f.redblue.a220w}
\end{figure}

\begin{figure}[t]
\epsscale{0.45}
\plotone{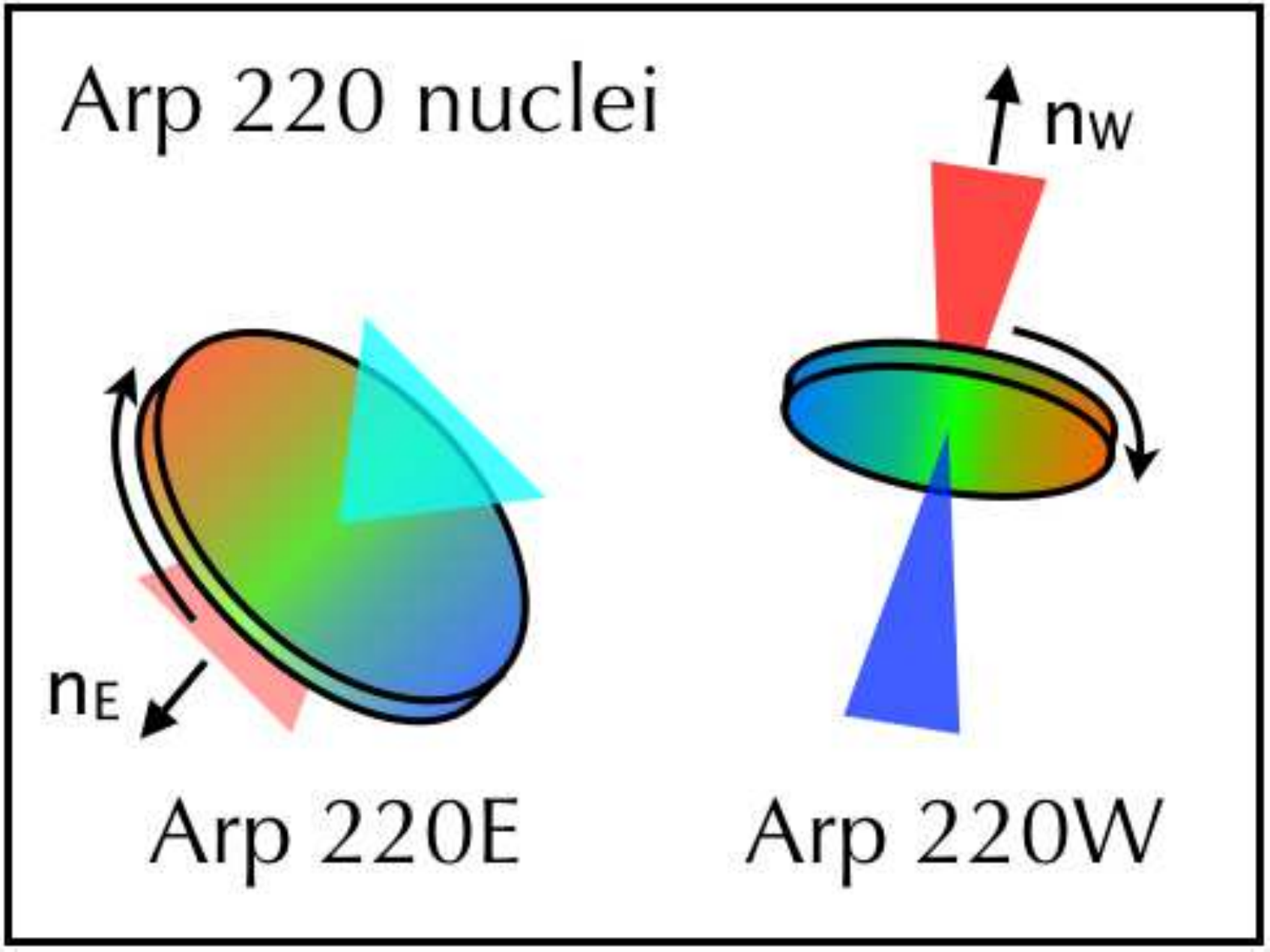} 
\caption{ \label{f.a220_illust}
A schematic diagram of Arp 220 for the configuration of its nuclear disks and outflows.
The color coding is blue(red) for blue(red)shifted gas 
and green for the gas near the systemic velocities of the individual nuclei.
(Arp 220E is about 100 \kms\ more redshifted than Arp 220W.)
The disk major axes and axial ratios are from our fitting of supernova distributions and 
3 and 0.4 mm continuum emission \citep{Paper1}; both nuclei have inclinations $\gtrsim 60\degr$.
The near- and far-sides of the nuclear disks are determined from the spatially-lopsided line absorption explained in the main text.
The normalized angular-momentum vectors of the disks are shown as $\bm{n}_{\rm E}$ and $\bm{n}_{\rm W}$.
The nuclear disks are in counter-rotation.
In Arp 220E, only the blueshifted outflow has been detected; the redshifted lobe here is our prediction.
We caution that this illustration is highly simplified for clarity.
For example, the actual nuclear disks may well be thicker, distorted, and interacting with the surrounding outer disk.
}
\end{figure}

The direction of the outflow can be determined from gas velocities.
In \arpW, blueshift is evident in the broad absorption  (Section \ref{s.lineIm.absline.a220abssystem}).
Moreover, the blueshifted absorption is slightly shifted to the south from the centroid of the continuum nucleus in Fig.~\ref{f.1Dslice}(b); 
i.e., the spatial lopsidedness is clearer in blueshifted velocities.
Furthermore, blueshifted gas is seen in emission south of the nuclear disk in the channel maps at velocities $\lesssim 5000$ \kms.  
There is also redshifted emission elongated to the north of the nucleus at $V \gtrsim 5600$ \kms\ 
(Section \ref{s.lineIm.velstructure.a220} and Fig.~\ref{f.redblue.a220w}).
Therefore, the outflow is bipolar. 
Its southern side is approaching us, and the northern side is receding from us.
Figure~\ref{f.a220_illust} illustrates this outflow configuration.

The configuration of the nuclear disk and the orientation of the bipolar outflow determined above are consistent 
with the outflow axis agreeing (to a few 10 degrees in the three-dimensional space) with the rotational axis of the nuclear disk.
While it is also possible that the two axes overlap only in their sky-plane projection,
our observations disfavor some large angles between them, e.g., the outflow to the far-side of the nuclear disk cannot be receding.
There remains a puzzle \citep{Tunnard15,Sakamoto17} that the configuration of the
nuclear disk (i.e., the southern side being the far side) is opposite to what was suggested by the near-IR light distribution 
\citep{Scoville98}. 
A likely key to this puzzle is that  
the nuclear disk may well be warped, flared, or non-axisymmetric given the tidal interaction with the eastern nucleus and 
the high luminosity activities inside.

The overall outflow-nuclear disk configuration of \arpW\ is  summarized in Figure~\ref{f.a220_illust}.
The major axis of the western nuclear disk was measured to be at p.a. $\approx 83$\degr\ in our fitting
of 3 and 0.4 mm continuum emission and the distribution of supernovae \citesp{Paper1}. It is consistent
with the well-known east-west velocity gradient in molecular lines, seen abundantly in our data.
The bipolar outflow at p.a. $\approx -15\degr$, seen in both continuum and line emission, 
is almost along the minor axis of the nuclear disk.
The continuum analysis in \citest{Paper1} also found the minor-to-major axis ratio of 0.47 for the western nuclear disk.
Thus the disk inclination must be $\gtrsim60\degr$; that is, \about60\degr\ for a thin disk and edge-on for 
an oblate spheroid whose diameter is twice its full height.
The presence of spatially-lopsided line absorption disfavors a completely edge-on configuration. 
The blue- and red-shifted lobes of line emission from the bipolar outflow also disfavor the edge-on nuclear disk
if the outflow is perpendicular to it.

\paragraph{\arpE:}
We found the blueshifted absorption toward the eastern nucleus to be spatially lopsided too.
The absorption is slightly shifted from the continuum peak in the channel maps of HCN and \HCOplus\ lines
in both the (4--3) and (3--2) transitions. 
At and below the fiducial systemic velocity of this nucleus, 5400 \kms, 
the lines show absorption northwest of the continuum peak and emission in the southwest.
Figure \ref{f.1Dslice}(c) shows this shift using HCN(4--3) intensity profiles along the minor axis of the nuclear disk,
whose major axis was measured to be at p.a. $\approx 51$\degr\ in our 3 and 0.4 mm continuum analysis 
and the distribution of supernovae \citesp{Paper1}.

The anti-symmetric intensity pattern around the continuum nucleus suggests that an out-of-plane gas is
in front of the eastern nuclear disk to the northwest of its center.
The bulk of this foreground gas must be moving toward us since the anti-symmetric pattern is clearer in blueshifted velocities.
This configuration is consistent with a polar outflow from the eastern nuclear disk, whose northwest side is the far side.
We can, however, less stringently constrain the angle between the outflow and the nuclear disk than for the W nucleus. 
By analogy with many galactic outflows, 
we assume the outflow to be bipolar, having an undetected redshifted lobe in the southeast direction.

Figure \ref{f.a220_illust} contains our current best estimate of the disk and outflow configuration
in the E nucleus. 
Our data suggest that the disk is inclined by around 70\degr\ and that the E outflow is smaller than the W outflow.
The minor-to-major axial ratio of the eastern nuclear disk is 0.45 in the continuum analysis in \citest{Paper1},
suggesting the disk inclination to be $\gtrsim 63$\degr.
A nearly edge-on configuration is disfavored from the presence of the lopsided absorption unless
the outflow axis forms a large angle with the axis of disk rotation.
Under the simplest assumptions that both E and W nuclear disks are oblate spheroids and have similar axial ratios 
(or disks having similar thickness-to-radius ratios), they should have similar inclinations.
If we further assume that both outflows are in the polar direction of the nuclear disks, then the outflow from the eastern nucleus should be of a smaller scale (e.g., in mass, velocity, and extent) than the outflow from the western nucleus.
It is because, unlike the W nucleus, the E nucleus does not show symmetric line emission along the minor axis 
at large blue- and red-shift velocities.
This scale difference can be because the E outflow is slower, has a smaller extent, or is closer to pole-on
than the W outflow. 
The non-detection of this E outflow in continuum emission at 3 mm \citep{Sakamoto17} and 0.4 mm \citesp{Paper1}
also implies it to have less emissivity (e.g., less massive) or a smaller extent on the sky.

The configuration we estimated above for the outflow from \arpE\ does not match some previous studies.
\citet{Varenius16} found in the 150 MHz continuum that \arpE\ shows  a north-south elongation possibly due to a plasma outflow.
This position angle does not agree with that of the spatially-lopsided line absorption.
\citet{Wheeler20} claimed CO(3--2) detection of the \arpE\ outflow in a different disk configuration.
Our comment on it is in Section \ref{s.followup.Wheeler20}.

\paragraph{Counter-rotating nuclear disks:}
Finally, we note that the two nuclear disks of Arp 220 in their updated configuration (see Fig.~\ref{f.a220_illust}) 
are still counter-rotating with each other as proposed in \citet{Sakamoto99}. 
This counter-rotation is despite the revision of the near and far sides of the western nuclear disk to the ones 
opposite from those previously adopted from \citet{Scoville98}, 
who used the pattern of near-IR extinction to decide the disk orientation.
Using the coordinate system where the $x$-axis is to the east, the $y$-axis is to the north, and the $z$-axis is along our sightline directed from 
us to the target, the direction vector of the angular momentum of a rotating disk is
\begin{equation}
	\bm{n} 
	= 
	\left(
	\begin{array}{r}
		  \cos \theta \sin i \\
		- \sin \theta \sin i \\
		   \cos i  
	\end{array}	
	\right),
\end{equation}
where $i$ is the disk inclination that is the angle between the angular momentum vector and the positive direction of the $z$-axis
and $\theta$ is the position angle of the receding major axis of the disk measured from the north to east on the sky plane.
Using $\theta_{\rm E} = 51\degr$ and $\theta_{\rm W} = 263\degr$ from \citest{Paper1} and assuming
$i_{\rm E} = i_{\rm W} = 70\degr$, the angle between the two rotation vectors 
is $\cos^{-1} (\bm{n}_{\rm E} \cdot \bm{n}_{\rm W}) = 129\degr$. 
(This angle is 150\degr\ for $i_{\rm W} = 110\degr$, i.e., 
the previous choice of near and far sides of the western nuclear disk.)
An angle larger than 90\degr, i.e., $\bm{n}_{\rm E} \cdot \bm{n}_{\rm W} < 0$, means counter-rotation.
Using conservative errors of $\pm5\degr$ for $\theta_{\rm E}$ and $\theta_{\rm W}$
and $\pm10\degr$ for $i_{\rm E}$ and $i_{\rm W} $, we obtained $128\degr \pm11\degr$ from $10^3$ simulations. 
Therefore, the counter-rotation of the two nuclear disks in Arp 220 is a robust conclusion.

It is worth noting that the western nuclear disk also counter-rotates with the kpc-scale gas structure 
around the two nuclei of Arp 220, referred to as the outer disk in \citet{Sakamoto09}.
The gas in the western nuclear disk can efficiently lose angular momentum by interacting with the outer disk. 
Interestingly, the nuclear disk of NGC 4418 also counter-rotates with the gas and stars in the surrounding 
environment, i.e., the kpc-scale galactic disk. 
It is plausible that the counter-rotation of the nuclear disks of \arpW\ and NGC 4418 relative to their
surroundings helped the two nuclei to have extreme central gas concentrations of $\NH \sim 10^{26}$ \persquarecm\ 
and their luminous nuclear activities.

\subsubsection{Other Absorption-related Issues}
\label{s.lineIm.absline.others}
\begin{figure*}[!hbt]
\centering
\includegraphics{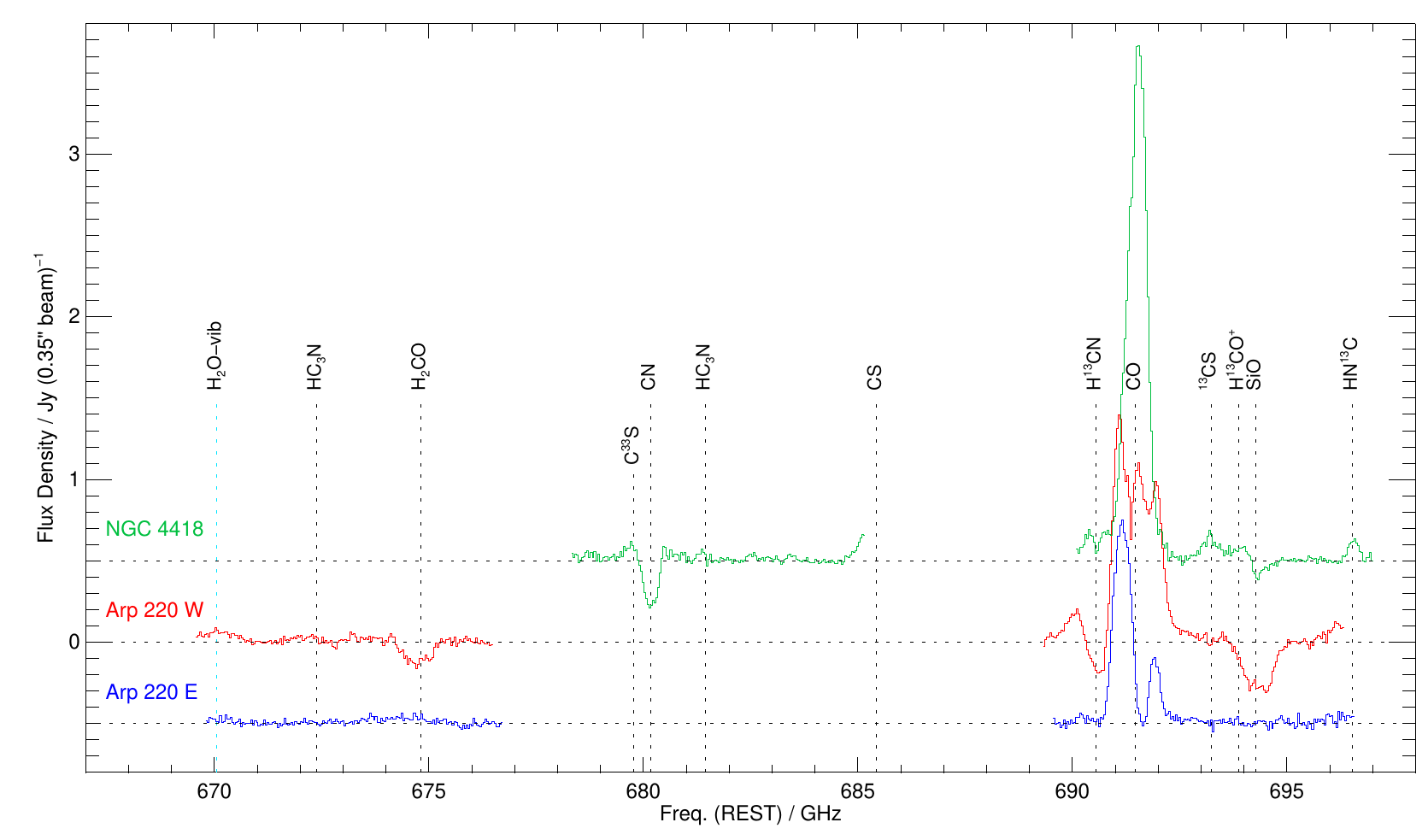}
\caption{ \label{f.triNuc_spec.B9.350mas}
Spectra of the three nuclei in NGC 4418 and Arp 220 
in the rest-frequency range of 670--697 GHz 
sampled in a 0\farcs35 beam at 40 MHz resolution.
Continuum has been subtracted and constant offsets are added for clarity.
Major lines and lines of interest are marked with the species. 
Among them,  \HCthreeN\ is not firmly detected unlike in lower bands 
and vibrationally excited \HtwoO\ is only a possible identification.
CN consists of a dozen transitions spanning 0.29 GHz; it is plotted at their intensity-weighted mean frequency.
}
\end{figure*}

\begin{figure}[t]
\fig{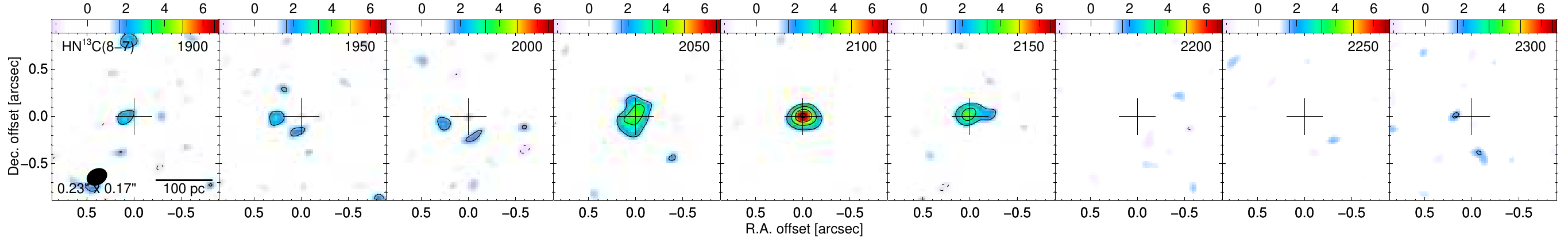}{1.0\textwidth}{(a) \HNthirteenC(8--7)}
\fig{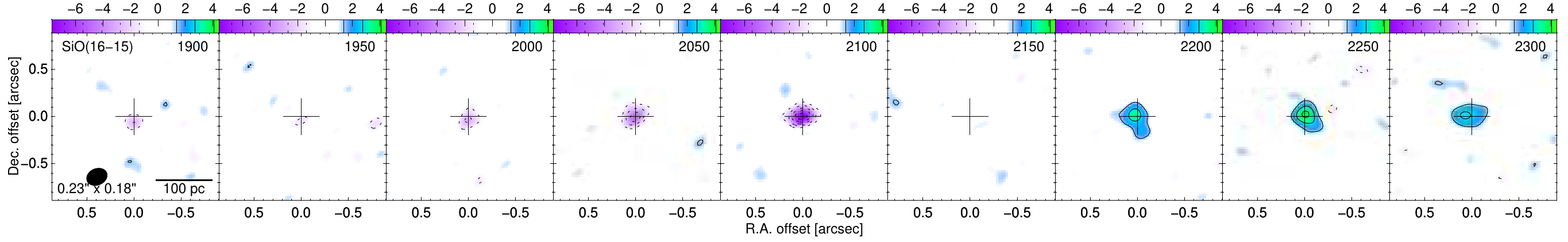}{1.0\textwidth}{(b) SiO(16--15) [blended with \HthirteenCOplus(8--7)]}
\fig{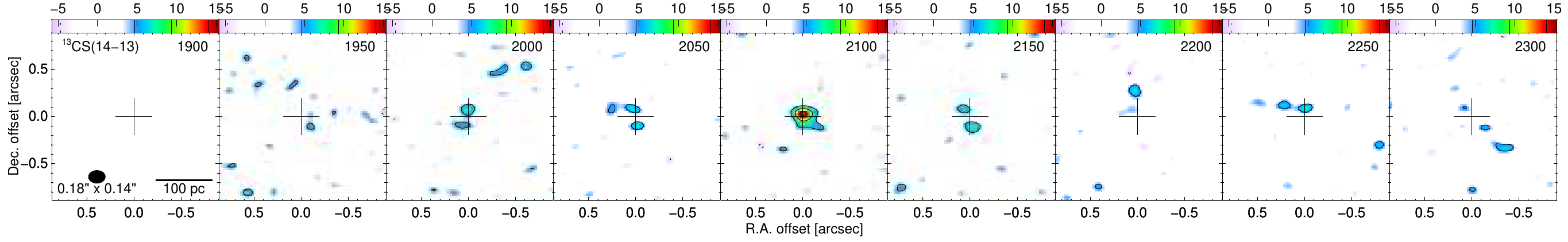}{1.0\textwidth}{(c) \thirteenCS(14--13)}
\fig{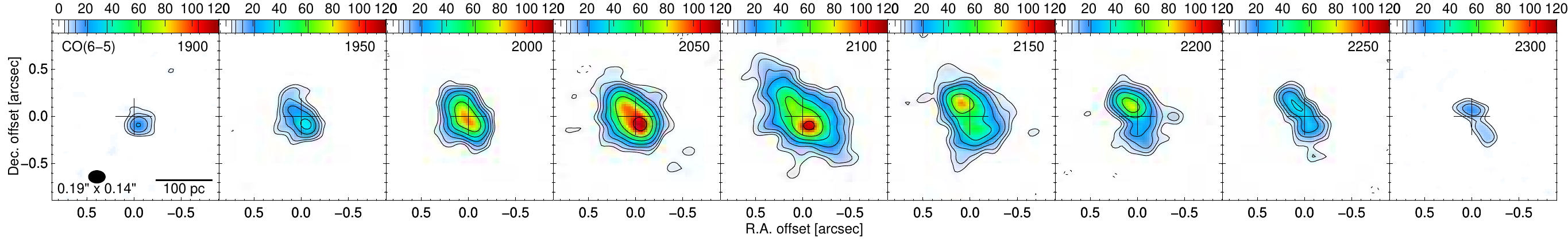}{1.0\textwidth}{(d) CO(6--5)}
\fig{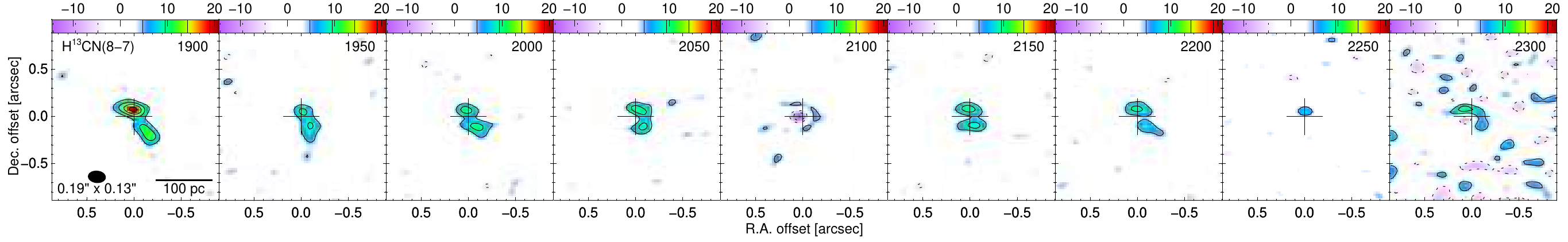}{1.0\textwidth}{(e) \HthirteenCN(8--7) [blended with CO(6--5)]}
\fig{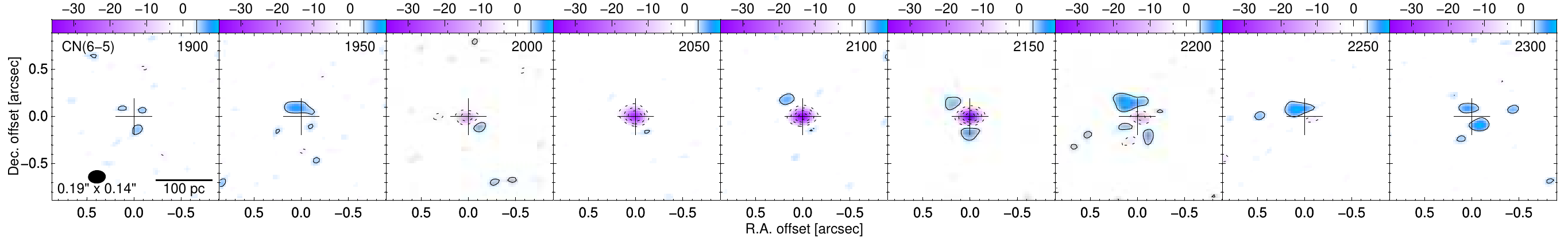}{1.0\textwidth}{(f) CN(6--5)}
\caption{ 
NGC 4418 channel maps of 50 \kms\ width for major lines in the ALMA Band 9.
Contours are at $\pm3n^p\sigma$ for $n=1,2,3,\ldots$;
the index $p$ and the rms noise per 50 \kms\ channel $\sigma$ are in Table \ref{t.chmaps}.
Negative contours are dashed.
Velocity (LSRK, radio) in \kms\ is in the upper-right corner of each channel map.
The crosses are at the continuum positions of the nuclei.
The synthesized beam, labeled with its FWHM, is shown in the bottom-left corner of the first channel.
The intensity color wedges are labeled with brightness temperature in Kelvin.
}
\label{f.chmap.n4418.B9}
\end{figure}

\begin{figure}[t]
\fig{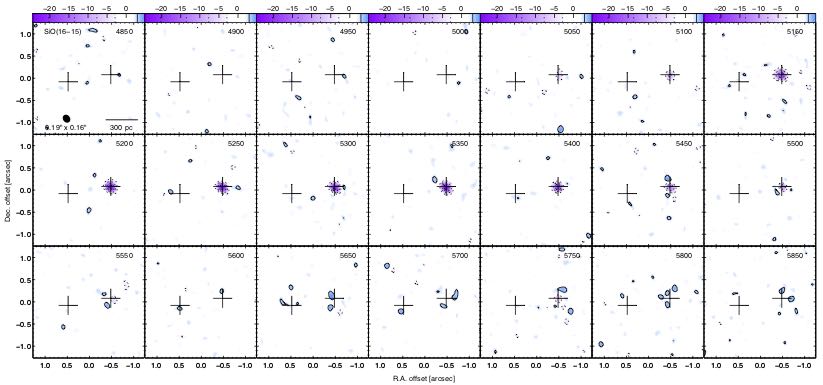}{1.0\textwidth}{(a) SiO(16--15)}
\fig{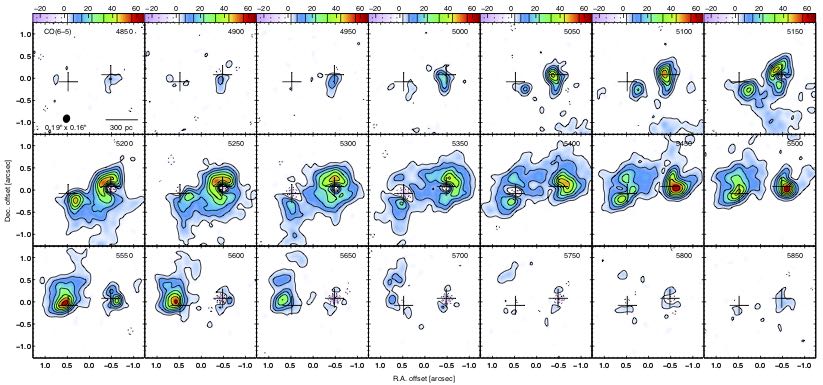}{1.0\textwidth}{(b) CO(6--5) [blended with \HthirteenCN(8--7)]}
\caption{ 
Arp 220 channel maps for major lines in Band 9.
Contours are at $\pm3n^p\sigma$ for $n=1,2,3,\ldots$, 
where the index $p$ and the rms noise per 50 \kms\ channel $\sigma$ are in Table \ref{t.chmaps}.
Negative contours are dashed.
Velocity (LSRK, radio) in \kms\ is in the upper-right corner of each channel map.
The crosses are at the continuum positions of the nuclei.
The synthesized beam, labeled with its FWHM, is shown in the bottom-left corner of the first channel.
The intensity color wedges are labeled with brightness temperature in Kelvin.
Major line-blendings are indicated in sub-panel titles and more information is in Table \ref{t.chmaps}.
}
\label{f.chmap.a220.B9}
\end{figure}

\begin{figure}[t]
\fig{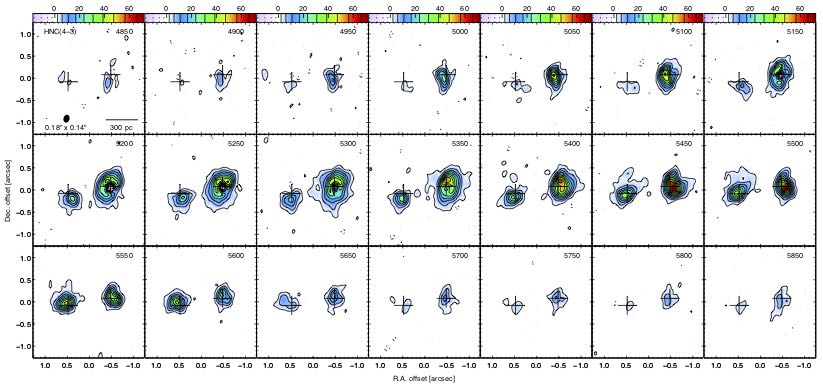}{1.0\textwidth}{(a) HNC(4--3)}
\fig{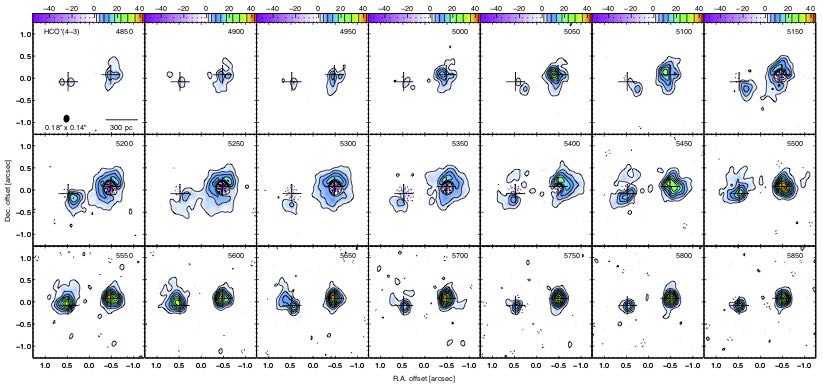}{1.0\textwidth}{(b) \HCOplus(4--3) [blended with HCN($v_2=1$)]}
\caption{ 
Arp 220 channel maps for major lines in Band 7.
Contours are at $\pm3n^p\sigma$ for $n=1,2,3,\ldots$, 
where the index $p$ and the rms noise per 50 \kms\ channel $\sigma$ are in Table \ref{t.chmaps}.
Negative contours are dashed.
Velocity (LSRK, radio) in \kms\ is in the upper-right corner of each channel map.
The crosses are at the continuum positions of the nuclei.
The synthesized beam, labeled with its FWHM, is shown in the bottom-left corner of the first channel.
The intensity color wedges are labeled with brightness temperature in Kelvin.
Major line-blendings are indicated in sub-panel titles and more information is in Table \ref{t.chmaps}.
}
\label{f.chmap.a220.B7}
\end{figure}

\begin{figure}[t]
\fig{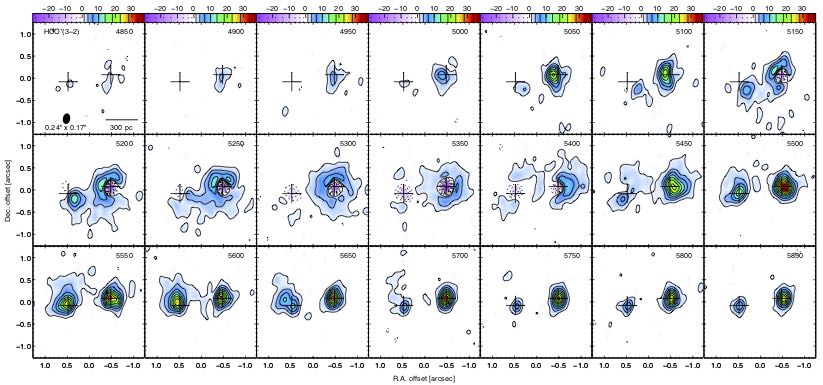}{1.0\textwidth}{(a) \HCOplus(3--2) [blended with HCN($v_2=1$)]}
\fig{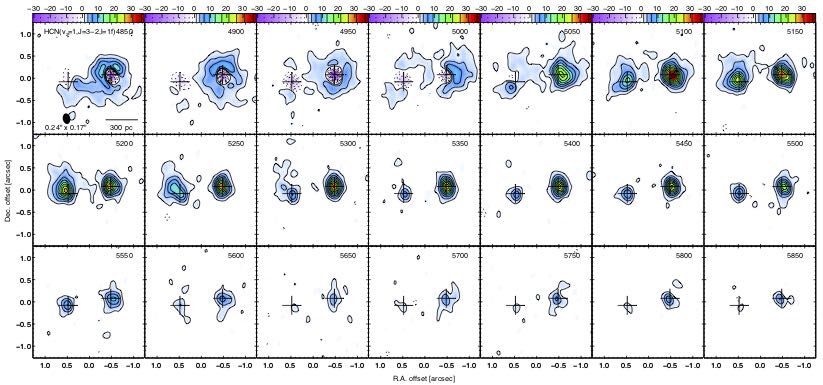}{1.0\textwidth}{(b) HCN($v_2=1$, 3--2, $l=1f$) [blended with \HCOplus(3--2)]}
\caption{ 
Arp 220 channel maps of 50 \kms\ width for major lines in Band 6.
Other descriptions are the same as in Fig.~\ref{f.chmap.a220.B7}.
}
\label{f.chmap.a220.B6}
\end{figure}

\paragraph{SiO:}
We have seen a striking variation of SiO in its absorption and emission among transitions and the three nuclei
(see Figs.~\ref{f.triNuc_spec.B7}, \ref{f.triNuc_spec.B9.350mas}, \ref{f.chmap.a220.B9}, and \ref{f.chmap.a220.B7}).
SiO(8--7) is detected in absorption toward all three nuclei; the absorption is deepest in \arpW, next in NGC 4418,
and least in \arpE.
The contrast between the two nuclei of Arp 220 is consistent with previous SiO observations 
\citep{Tunnard15, Rangwala15, Wheeler20}.
SiO(8--7) is also detected in emission around the compact absorption in the three nuclei. 
In contrast, SiO(16--15) is detected only toward NGC 4418 and \arpW\ and only (or mostly) in absorption.
The emission in redshifted velocities in the SiO(16--15) channel maps of NGC 4418 (Fig.~\ref{f.chmap.n4418.B9}b) 
is probably \HthirteenCOplus(8--7), judging from the line frequency. 
Similarly, the faint circumnuclear emission in the SiO(16--15) channel maps of Arp 220 is probably
the same line blending because it is only in redshifted velocities.
The SiO(16--15) absorption is compact and, unlike the SiO(8--7) absorption, has its centroid on the continuum peak
in both NGC 4418 and \arpW.

\citet{Tunnard15} deduced that the two nuclei of Arp 220 are chemically distinct 
from a striking difference in SiO(6--5) absorption.
On the other hand, the overall pattern of emission seen in the remaining majority of lines in our spectra 
does not appear drastically different between the two nuclei.
Therefore, it may be that the drastic difference in SiO may be partly due to this particular species.
For example, the more prominent molecular outflow in \arpW\ may have caused the difference 
\citep{Tunnard15,Wheeler20} because the SiO abundance can be enhanced by shocks in an outflow \citep{MP92}.
The same explanation may apply to the difference of \arpW\ and NGC 4418 in their
SiO(8--7) absorption.

Another possible reason for the variation of SiO absorption is the different degrees of luminosity concentration
and SiO excitation by radiation.
\arpW\ and NGC 4418 are comparably bright, and the two are brighter than \arpE, in the 200-700 GHz continuum \citesp{Paper1}.
In addition, the former two are more highly obscured than the latter \citesp{Paper1}.
These differences will result in different SiO excitation by both IR radiation (through vibrational SiO excitation) and
submillimeter radiation.
The larger column densities of \arpW\ and NGC 4418, 
on the order of $\NH \sim 10^{26}$ \persquarecm\ \citesp{Paper1},
are more favorable for vibrational SiO excitation by 8.1 \micron\ IR radiation.
It is because the trapping of continuum photons works more to increase the inner temperature of the nuclei through the greenhouse
effect \citep{GS19}. 
The possibility of IR pumping of SiO was raised for \arpW\ by \citet{Rangwala15}.
Unfortunately, we did not detect lines from vibrationally excited SiO, such as  SiO($v=1, J=$8--7) at 344.916 GHz in our frequency coverages of both Arp 220 and NGC 4418, but a weak feature could be lost in blending (e.g., with adjacent \HthirteenCN). 
At least in \HCthreeN, vibrational excitation appears more significant in \arpW\ than in \arpE\ 
(Section \ref{s.lineIm.vibLines.individual}).

The radiative SiO excitation through the sub/millimeter continuum explains
the variation of SiO absorption among the three nuclei 
and the localized SiO excitation to J \about\ 15 in each nucleus.
That we detected SiO(16--15) only in absorption and toward NGC 4418 and \arpW\ implies that SiO excitation
to the J=15 (16) level, at 250 (283) K above the ground, is limited to the hottest or brightest nuclei.
Both nuclei have peak continuum brightness temperatures of \about500 K, while \arpE\ has $\lesssim 200$ K 
at 200--700 GHz \citesp{Paper1}. 
It is consistent with the SiO(16--15) absorption detected only toward \arpW\ and NGC 4418.
If this radiative excitation dominates the SiO excitation to J $\sim 15$, it would be the reason why
the SiO(16--15) absorption coincides with the continuum peaks (Section \ref{s.lineIm.absline.outflows}).

\paragraph{Isotopologues:}
There are absorption features of rarer isotopologues toward Arp 220W.
Among them are \HthirteenCOplus(4--3), \HthirteenCN(4--3), and \thirteenCS(15--14) in this work 
(Figs.~\ref{f.triNuc_spec.B7}, \ref{f.a220.absorption}), 
\CthirtyfourS(2--1) in \citet{Sakamoto17},
and \thirteenCO(4--3) absorption in \citet{Wheeler20}.
While the absorption line of \HthirteenCOplus(4--3) is in Fig.~\ref{f.a220.absorption}(c) at around 5530 \kms\ in the
SiO(8--7) spectrum, we identify this feature to a blueshifted absorption of \HthirteenCOplus(4--3) rather than 
a redshifted absorption of SiO(8--7). 
The minimum of this feature is then at about 50 \kms\ blueshifted from the systemic velocity of the nucleus.
The blueshift is consistent with that of the deepest absorption of other lines, including \HCOplus, \HthirteenCN, and SiO.
Both Arp 220E and NGC 4418 also show signs of weak \HthirteenCN(4--3) absorption in Fig.~\ref{f.outflowCheck}(a) and (d). 
The absorption of these rarer isotopologues reflects the large gas column density toward these nuclei.
It is because they are less prone to self-absorption than the primary isotopologues 
and also because the dust in the background needs to be opaque enough to emit bright from behind these molecules.

\subsection{Vibrationally Excited Lines}
\label{s.lineIm.vibLines}
Rotational lines from vibrationally excited molecules have been explored as promising probes of deeply buried galactic nuclei
since their first extragalactic detection for \HCthreeN\ \citep{Costagliola10} and HCN \citep{Sakamoto10}.
For short, we refer to these as HCN-vib, \HCthreeN-vib, and vib-lines.
Our new data added the following to the information on vib-lines and their usage.

\subsubsection{Individual Species}
\label{s.lineIm.vibLines.individual}
\begin{figure*}[t]
\centering
\includegraphics{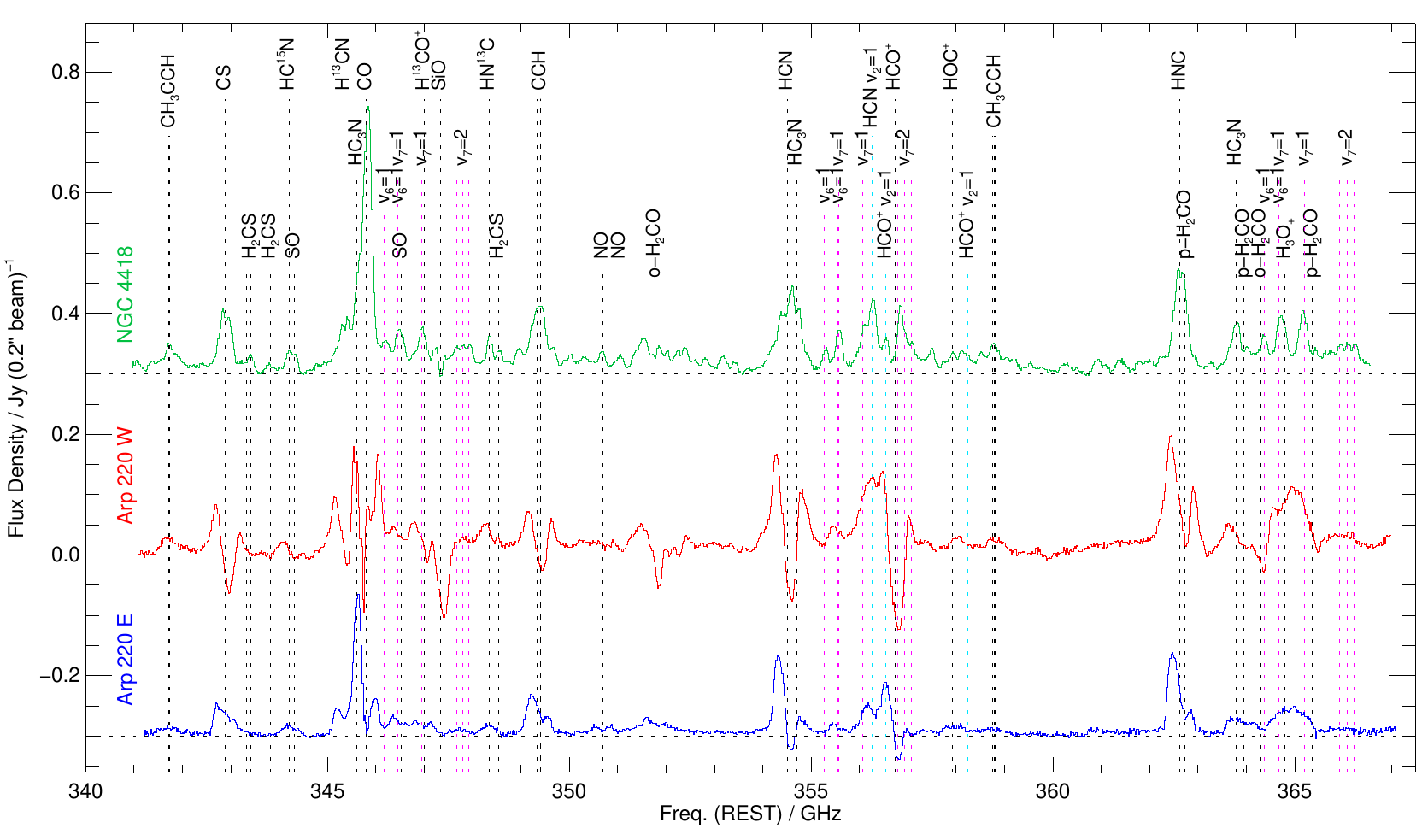}
\includegraphics{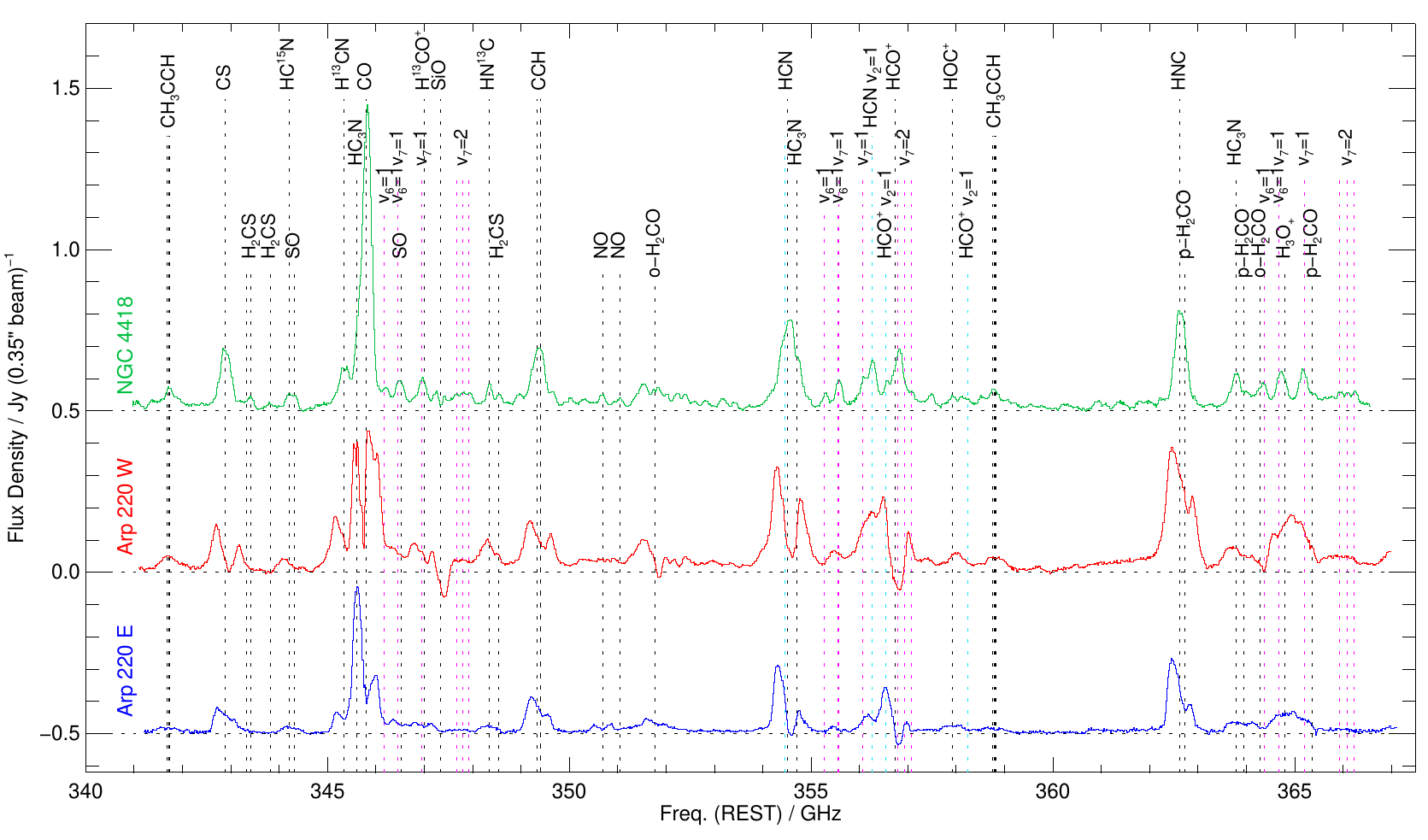} 
\caption{ \label{f.triNuc_spec.B7}
Spectra of the three nuclei in NGC 4418 and Arp 220
in the rest-frequency range of 339--362 GHz at 20 MHz resolution.
The spatial resolution is 0\farcs20 (top) and 0\farcs35 (bottom).
Continuum has been subtracted and constant offsets have been added for clarity.
Major lines and lines of interest are marked with the species 
except for lines from vibrationally excited \HCthreeN\ (magenta) for which only their vibrational levels are given.
Lines from vibrationally excited HCN and \HCOplus\ are plotted in cyan.
It is evident that at a higher resolution more lines go into absorption and do so deeper
and that emission lines from vibrationally excited molecules stand out more than with the emission lines 
from the vibrational ground state.
}
\end{figure*}

\paragraph{\HCthreeN: }
Our 0\farcs2 data in Fig.~\ref{f.triNuc_spec.B7} have
rotational lines of \HCthreeN\ in both vibrationally ground and excited states.
Their \rotJ=40--39 transitions are at $f_{\rm rest} \sim$ 364--365 GHz.
We note that the \HCthreeN\ \vseven=1 doublet, 
which is resolved in NGC 4418 but blended in the Arp 220 nuclei,
is brighter than the \HCthreeN\ $v$=0 line toward all three nuclei.
The former $v$=0 line has upper-state energy of 358 K while the latter \vseven=1 lines have $E_u = 680$ K.
They have the same statistical weights in their upper levels and virtually the same Einstein $A$ coefficients.
If their vibrational and rotational excitation temperatures are the same in the LTE condition in a uniform
gas cloud, the $v$=0 line should be stronger than or as strong as the \vseven=1 lines.
Therefore, the observed line ratio suggests the gas in these nuclei to be non-uniform or not in LTE or both. 
A plausible situation is a radial temperature gradient with a higher temperature inside. 
The ratio of the mean integrated intensity of the \vseven=1 doublet to that of the $v$=0 line is higher 
in Arp 220W than in Arp 220E and NGC 4418. 
Specifically, the ratio for \rotJ=40--39 is 2.42$\pm$0.04, 1.50$\pm$0.04, and 1.14$\pm$0.01, 
respectively for \arpW, \arpE, and NGC 4418 in the 0\farcs2 scale, 
where the integration of each line is over a velocity width of 500 \kms\ for the Arp 220 nuclei and 300 \kms\ for NGC 4418.
The ratios imply that \arpW\ has the steepest temperature gradient or infrared radiation, or both.

\begin{figure}[t]
\epsscale{0.4}
\plotone{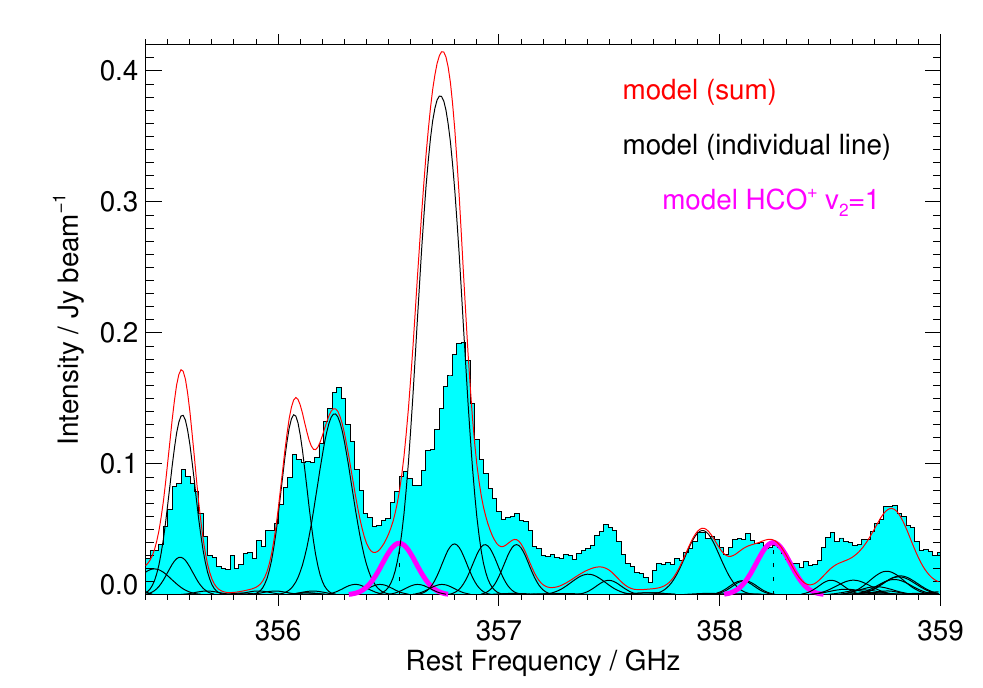} \\
\plotone{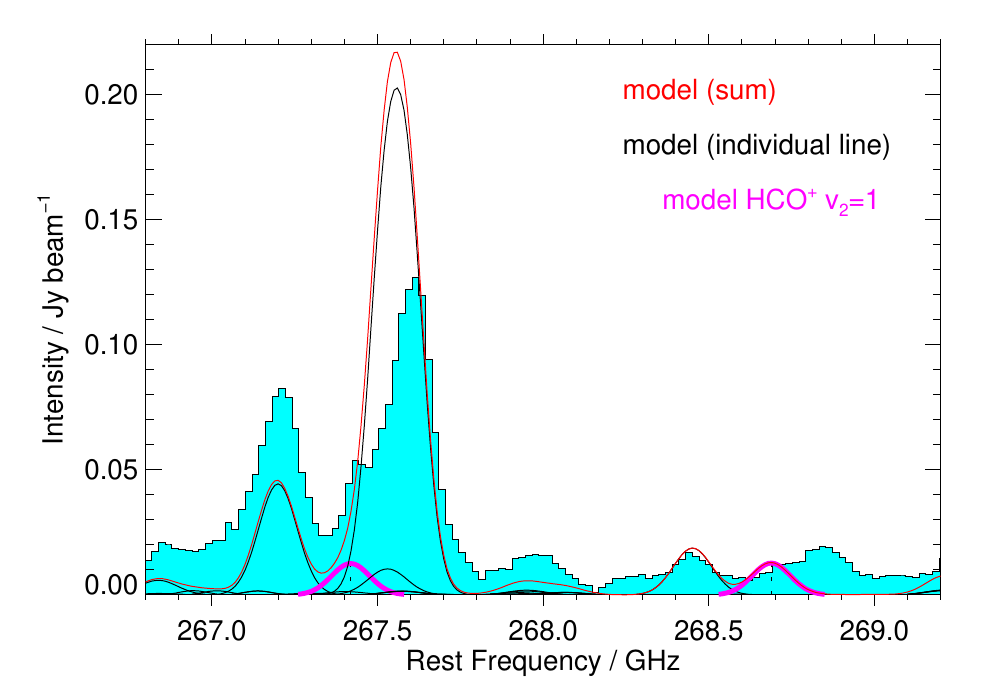}
\caption{ \label{f.n4418_HCO+vib}
NGC 4418 spectrum around \HCOplus\ J=4--3 and 3--2 transitions.
Our model fit in red is overlaid on the observed spectrum in light blue.
Individual lines in the model are shown in black, and magenta for the vibrationally excited \HCOplus.
}
\end{figure}

\paragraph{\HCOplus: }
Detection of \HCOplus-vib is possible but not yet certain toward the three nuclei.
The NGC 4418 spectra in Figs.~\ref{f.triNuc_spec.B7} and \ref{f.triNuc_spec.B6.350mas} have 
peaks at the frequencies of \HCOplus(\vtwo=1, \rotJ=4--3, $l$=1e) and \HCOplus(\vtwo=1, \rotJ=3--2, $l$=1e).
However, much of these features must be artifacts caused by the blueshifted absorption of the 
adjacent \HCOplus(4--3) and (3--2) lines. 
Indeed, our spectrum does not have isolated peaks at the frequencies of the other transitions 
of the \vtwo=1 doublets ($l$=1f). The intensities there are 2--5 times lower than at the $l$=1e transitions.
Our initial LTE modeling of the spectrum in Fig.~\ref{f.n4418_HCO+vib}  gives an idea of the possible contribution of 
the vibrationally-excited \HCOplus\ lines to the observed spectrum.
Better line de-blending and emission+absorption modeling are necessary for firmly constraining the line strengths.
In the nuclei of Arp 220, the line blending is much worse than in NGC 4418. 
Although the spectra show peaks at the \HCOplus(\vtwo=1, $l$=1e) frequencies, 
intensities are again much lower at the frequencies of \HCOplus(\vtwo=1, $l$=1f) transitions.

\paragraph{HNC: }
\citet{Costagliola13} detected HNC(\vtwo=1, \rotJ=3--2, $l$=1f) toward the nucleus of NGC 4418, but we 
only have the HNC(\vtwo=1, \rotJ=4--3) doublet in our spectral coverage.
We did not firmly detect the HNC(\vtwo=1, \rotJ=4--3, $l$=1f) line because it is heavily blended with another line.
Unfortunately, HNC(\vtwo=1, \rotJ--\rotJ\minus1, $l$=1f) lines always have
\HCthreeN(\vseven=1, 10\rotJ--10\rotJ\minus1, $l$=1f) nearby because 
their rotational constants have a ratio close to an integer, 10.\footnote{The rotational constant 
is 45.484 GHz for HNC(\vtwo=1, $l$=1f) \citep{Thorwirth00}
and 4.5635 GHz for \HCthreeN(\vseven=1, $l$=1f) \citep{Bizzocchi17}.}
With their separations of 39, 82, and 113 \kms\ for \rotJ=4, 3, and 2, respectively,
they tend to blend in extragalactic observations and most severely so for \rotJ=4 among those easily accessible transitions.
In our NGC 4418 spectrum in Figure~\ref{f.triNuc_spec.B7}, 
the \HCthreeN(\vseven=1, $J$=4--3) doublet lines, $l$=1e and 1f, have about the same peak intensities as they should
(because they have the same statistical weights and virtually the same Einstein $A$ coefficients.) 
The $l$=1f peak is only slightly higher. 
While this slight excess could be due to the HNC(\vtwo=1, \rotJ=4--3, $l$=1f) line, a firm conclusion waits for de-blending of
lines to exclude the possible contribution from any other low-level lines.
In the Arp 220 nuclei, the same vibrationally-excited HNC line (if any) is too heavily blended with 
the \HCthreeN(\vseven=1, $l$=1f) line for detection.
The other half of the HNC(\vtwo=1) doublet,  $l$=1e lines,  should be only 63 \kms\ offset from the vibrationally-ground HNC lines.
Therefore, both of the HNC(\vtwo=1) doublet lines are not immediately visible in our data.

\paragraph{Comparison with HCN:}
In the three galactic nuclei, rotational lines from vibrationally excited states are comparably strong in HCN and \HCthreeN\ 
and much weaker in \HCOplus\ and HNC. 
Indeed, the HCN-to-\HCOplus\ ratio in their $v_2$=1 lines is 3.5 in both J=4--3 and 3--2 in our spectral fit for NGC 4418.
In our Galaxy, hot molecular cores heated by young massive stars inside also show 
rotational lines from vibrationally excited HCN but not \HCOplus; it has been attributed to
abundance enhancement of HCN (but not \HCOplus) in the high-temperature environment \citep{Rolffs11}.
HCN enhancement in such an environment is also expected in the context of galactic nuclei \citep{Harada10}.
An elevated abundance of HCN has been suggested, from its 14 \micron\ absorption line, in deeply obscured 
nuclei of infrared galaxies \citep{Lahuis07, GA12}.

\subsubsection{Vib-lines as a Group}
\label{s.lineIm.vibLines.asgroup}
We saw in Section \ref{s.lineIm.size} that the line-emitting regions tend to be smaller 
for vibrationally-excited lines than for vibrationally-ground lines.
We also found in NGC 4418 that the line-emitting areas are smaller for transitions having 
higher upper-state energies even within each vibrational state.
The emitting regions of the vib-lines are found comparable in size and shape with continuum emission. 
This similarity is reasonable considering that the vib-lines are most likely due to radiative excitation by mid-IR continuum 
photons from warm dust.
These observations have the following implications.
One of them is that the vib-lines are suited for selectively studying the inner regions of the warm buried nuclei, 
as was proposed in our detection report of HCN-vib lines in NGC 4418 \citep{Sakamoto10}.
Another is that VIB lines tend to be more beam-diluted than lines in the vibrational ground state when observed with a large beam.
Another is that beam dilution is larger for vib-lines than for the vibrationally ground lines 
in an observing beam larger than both line-emitting regions.
Therefore, the ratio of vibrationally-excited-to-vibrationally-ground lines is biased (underestimated) 
in observations at a low angular resolution.
This bias lowers the vibrational excitation temperature $T_{\rm vib}$ estimated from the observed ratio 
unless correcting for the different beam-filling factors.
This difficulty also applies to the vibrational excitation analysis of multiple vibrationally-excited lines at different
vibrational levels, such as the one using $v_6$, $v_7$, and $v_4$ states in \HCthreeN.

While the beam-filling factors can be matched with a high enough spatial resolution to resolve
the emitting areas of the vibrationally-excited lines, there is another difficulty in the line-ratio analysis
with such observations. 
The vibrational-ground lines are often in absorption when the vibrationally-excited lines are in emission.
Their line ratios would be then negative numbers, and hence their logarithms do not give vibrational-excitation temperatures.
For a vibrational excitation temperature, one needs formal radiative-transfer modeling \citep[e.g.,][]{GS19} or 
ratio(s) between emission lines from different vibrationally excited states.
An example of the latter could use lines from the $v_6$ and $v_7$ states of \HCthreeN\ if both are optically thin.  
On the other hand, if the vib-lines are optically thick,
their observed brightness temperatures can set a lower limit to both the
rotational and vibrational excitation temperatures.

Another implication of the absorption of rotational lines within the vibrationally-ground state is 
that this absorption tends to mask the effect of IR radiative pumping to the rotational excitation 
within the vibrational ground state for the inner nucleus.
This masked area is where the IR pumping should have the maximum effect on the rotational level population.
The absorption of the  $v=0$ lines there means that this effect should appear reduced in low-resolution data integrating 
the entire nucleus. 
This effect, IR radiative pumping, has been a concern in the usage of HCN as a tracer of dense gas in
luminous galactic nuclei under the assumption of HCN excitation through collision with \HH\
\citep{Aalto95, Garcia-Carpio06}. 
The virtual masking of the innermost region reduces the error due to the omission of  IR radiative excitation.
Without correcting for the masking effect, the results of such excitation analysis will 
reflect the properties of the circumnuclear region rather than the innermost regions.

\section{Follow-up on Previous Reports}
\label{s.followup}
Our wide-band, high-resolution, and sensitive ALMA data allow us to verify some of the earlier observations.

\subsection{Arp 220E --- Outflow Emission in CO(3--2)?}
\label{s.followup.Wheeler20}
\begin{figure}[t]
\epsscale{0.27}
\plotone{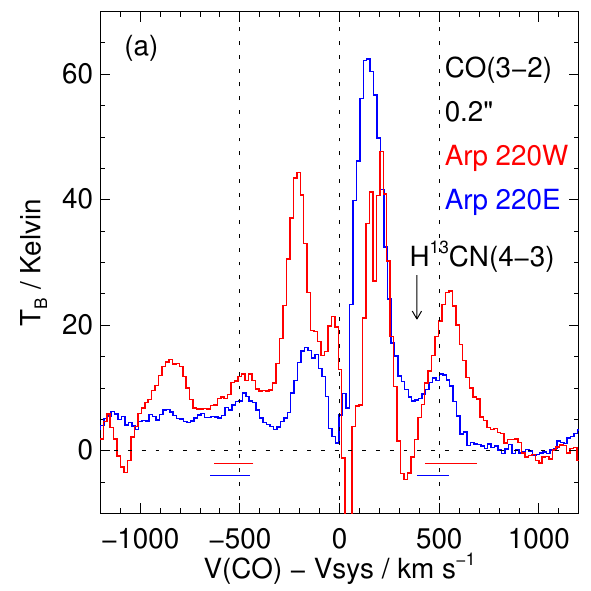}
\plotone{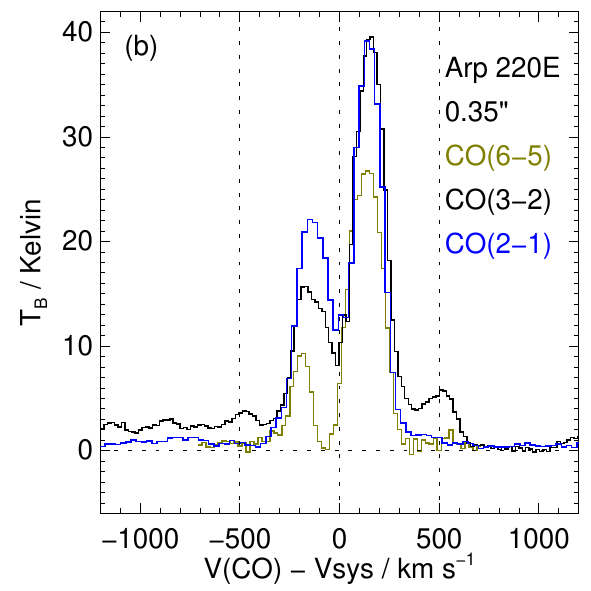}
\plotone{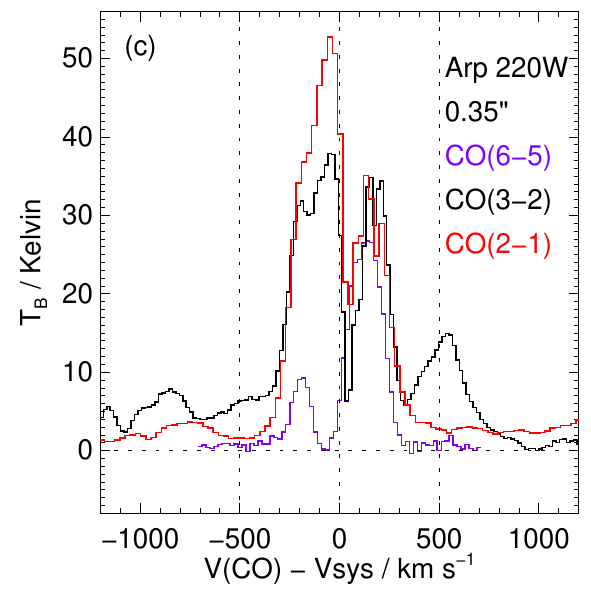}
\plotone{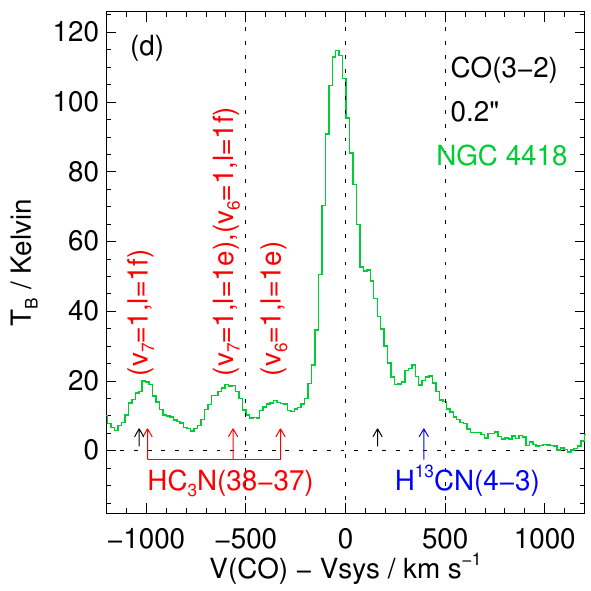}
\caption{ \label{f.outflowCheck}
Line-blending around CO(3--2) to affect the search for high-velocity gas.
(a) CO(3--2) spectra of the two Arp 220 nuclei; red for the W nucleus and blue for the E nucleus.
The offset velocities are from the systemic velocity of each nucleus.
The short horizontal bars around $\pm500$ \kms\ mark the putative high-velocity emission that \citet{Wheeler20} 
identified in their CO(3--2) data. Our spectra confirm the peaks in both nuclei.
A black arrow is where \HthirteenCN(4--3) would be centered if its profile were symmetric.
(b) Arp 220E spectra of CO in J=6--5, 3--2, and 2--1.
The $\pm500$ \kms\ features in CO(3--2) are absent in CO(2--1) and (6--5).
(c) Same as (b) but for Arp 220W.
(d) CO(3--2) spectrum of NGC 4418. 
The feature near \plus400 \kms\ is \HthirteenCN(4--3), whose double-peak may be due to a slight absorption.
In negative CO velocities are three peaks due to four transitions of vibrationally excited \HCthreeN. 
The main text explains how \HthirteenCN(4--3) and vibrationally-excited \HCthreeN(38--37) lines form 
the $\pm500$ \kms\ features in the CO(3--2) spectra of Arp 220. 
For completeness, two short black arrows indicate the locations of \HthirteenCOplus(4--3) around \minus1030 \kms\ and
\HCthreeN(38--37) at \plus160 \kms.
}
\end{figure}

\citet{Wheeler20} suggested a collimated outflow from Arp 220E 
on the basis of emission features at $V - V_{\rm sys} \approx \pm 500$ \kms\ in their CO(3--2) data.
The same features are also in our data (Fig.~\ref{f.outflowCheck}a).
However, their model has the opposite orientations of ours for the outflow and the eastern nuclear disk. 
While we both adopt a bipolar outflow normal to the nuclear disk, 
their model has the outflow redshifted (receding) to the northwest 
and the near side of the eastern nuclear disk also to the northwest. 
Both are the opposite of our model in Fig.~\ref{f.a220_illust}. 
Their disk and outflow configuration would produce continuum absorption to the southeast of the nucleus 
while we detected it to the northwest.

This discrepancy must be because much of the putative outflow emission is not CO(3--2) at high velocities 
but blended lines.
The first sign of this misidentification is that these features have peaks at around $\pm 500$ \kms\ of CO(3--2) rather than 
smoothly declining at a larger offset from the CO line center.
The second, stronger indication is the absence of the $\pm 500$ \kms\ features in CO(2--1) and (6--5), 
as shown in Fig.~\ref{f.outflowCheck}b.
Thirdly, the blended lines have identifications, \HthirteenCN(4--3) on the redshifted side 
and three J=38--37 transitions of vibrationally excited \HCthreeN\ on the blueshifted side.
We will explain how these lines form the $\pm500$ \kms\ features in the CO(3--2) spectrum.
Lastly, the $\pm500$ \kms\ features are also in the spectrum of the western nucleus, 
as expected in the case of blended lines (Fig.~\ref{f.outflowCheck}a and c).
These constitute strong evidence that the $\pm 500$ \kms\ features in the CO(3--2) spectrum toward Arp 220E
are due to line blending rather than high-velocity CO emission.

The \HthirteenCN(4--3) transition is on the redshifted side of CO(3--2) at \plus390 \kms\ in the CO velocity.
The known blueshifted absorption toward each nucleus makes the actual \HthirteenCN(4--3) line asymmetric 
and shifts its emission peak to a higher velocity in the CO(3--2) velocity.
As a reference, HCN(4--3) toward the E nucleus has its peak about 150 \kms\ redshifted from the systemic velocity. 
Therefore, it is reasonable that \HthirteenCN(4--3) has an intensity peak at $v \approx +$550 \kms\ in the CO(3--2) spectrum
toward the W nucleus.
The \HthirteenCN(4--3) identification toward the W nucleus is also based on its circumnuclear distribution
and convincing (Section \ref{s.lineIm.cautions}).
The line intensity on the E nucleus at the expected \HthirteenCN(4--3) peak frequency
is consistent with the \HthirteenCN(4--3) intensity on W and
the E--to--W ratio of other emissions (i.e., a factor 2--3 weaker toward E than toward W 
as in \CeighteenO(2--1) and 1.3--0.85 mm continuum), supporting our \HthirteenCN(4--3) identification.
We add that \HCthreeN(38--37), a transition within the vibrational ground state, 
must also be blended with CO(3--2) at \plus160 \kms\ though hardly discernible in our Arp 220 data.

On the blueshifted side of CO(3--2) around \minus500 \kms\ are three J=38--37 transitions of vibrationally excited \HCthreeN, 
namely \HCthreeN($v_7$=1, $l$=1e), \HCthreeN($v_6$=1, $l$=1f), and  \HCthreeN($v_6$=1, $l$=1e).
The $v_7$=1 line should be brighter because the upper states for $v_7$=1 and $v_6$=1 are 
646 K and 1042 K above the ground, respectively.
The three lines should be, respectively, at about \minus562, \minus554, and \minus322 \kms\ in the CO(3--2) spectrum. 
They are detected in NGC 4418 as two peaks with the first two transitions merged and being brighter than the third one (Fig.~\ref{f.outflowCheck}d).
In Arp 220 toward both nuclei, all three lines are blended and detected as a single peak
at $f_{\rm rest} \approx$ 355.4 GHz for J=39--38 and 264.4 GHz for J=29--28 
in Figs.~\ref{f.triNuc_spec.B7} and \ref{f.triNuc_spec.B6.350mas}.
From their information, the blended feature for J=38--37 is expected to be at $v \approx -500$ \kms\ in the CO(3--2) spectrum, 
having a \about3 K excess at 0\farcs2.
It is consistent with our observations and, therefore, explains the \minus500 \kms\ feature.

As we have seen, line-blending complicates the outflow search through broad line wings unless
the wing emission overwhelms adjacent lines.
Therefore, it is sensible to verify possible molecular outflow features using multiple lines, 
the spatial distribution of the emission, line de-blending, and careful baseline subtraction. 
The outflow from the western nucleus passed these tests, even though its $\pm500$ \kms\ features
in CO(3--2) in the central few tenths of arcsec must be mostly due to blended lines (Fig.~\ref{f.outflowCheck}c).
In contrast, the molecular outflow from Arp 220E has been seen only through absorption lines and not yet in emission.
In all likelihood, this outflow also has emissions in CO and other lines, 
and their imaging would constrain the outflow parameters more.

\subsection{HNC maser}
\label{s.followup.HNC}
\begin{figure}[t]
\epsscale{0.4}
\plotone{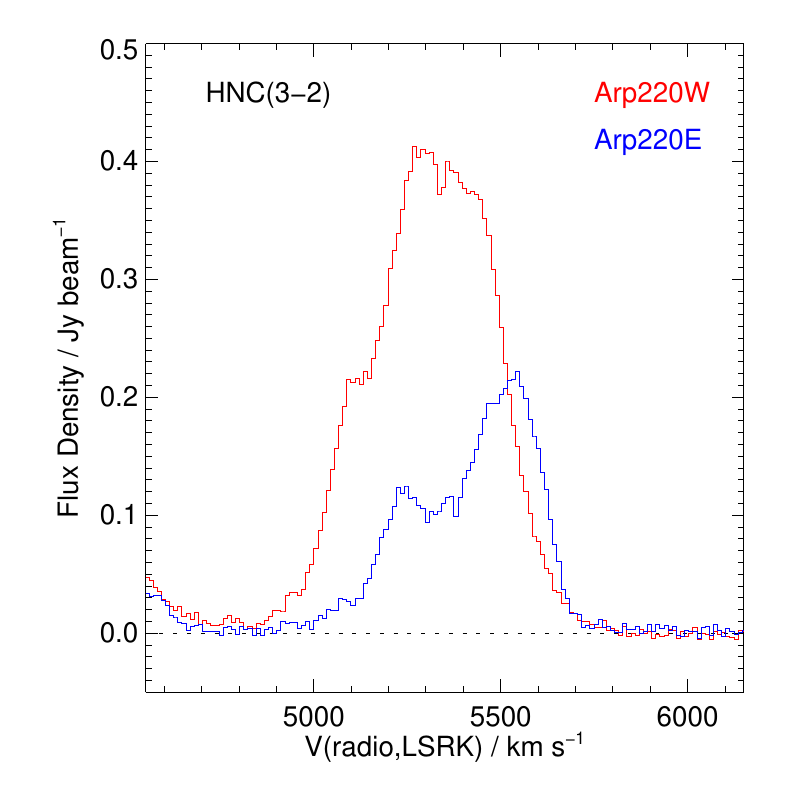} 
\caption{ \label{f.a220_HNC32}
HNC(3--2) line profiles toward the Arp 220 nuclei.
The continuum-subtracted data are from our companion spectral scan 
and have 0\farcs65 and 11 \kms\ resolutions.
The peak brightness temperature of the line is 16 K for the western nucleus.
}
\end{figure}

\citet{Aalto09} reported SMA observations of 
a bright (39 K at 0\farcs4 resolution), narrow (60 \kms) emission feature of HNC(3--2) from the western nucleus of Arp 220. 
This spike feature is on top of a broader spectral component, which may be from the rotating western nuclear disk. 
The spike is at $V$(radio, LSR) = 5358 \kms\ near the center of the broad component and 
has about three times higher peak intensity than the broad component.
It was argued that the narrow feature might be a weak, possibly transient maser emission. 
We observed HNC(4--3) at twice higher angular and spectral resolutions and 
more than ten times higher sensitivity in brightness temperature. 
Figure \ref{f.triNuc_spec.B7} (red line for Arp 220W) contains HNC(4--3) at \frest = 362.6 GHz
in the spectra sampled with 0\farcs20 and 0\farcs35 beams.
Neither spectra show the narrow emission spike reported in HNC(3--2).
Instead, they show a broader peak with peak brightness temperatures of about 47 and 30 K respectively 
at 0\farcs20 and 0\farcs35,
as well as the blueshifted absorption feature (or intensity decrement) that we attributed to an outflow. 
The bright emission spike is not seen either in the 0\farcs65 resolution HNC(3--2) spectrum 
from our companion spectral scan in Figure~\ref{f.a220_HNC32}
and the 0\farcs75 resolution spectrum of HNC(2--1) in \citet{Koenig17}.
The HNC(3--2) line profiles are similar to our HNC(4--3) profiles, showing signs of blueshifted decrement toward both nuclei.
They also have narrow dips near the systemic velocities of the individual nuclei; 
the one toward \arpW\ has \Vmin $\approx$ 5340 \kms. 
They are similar to the narrow CO absorption that we described in Section \ref{s.lineIm.absline.a220abssystem}.
Since the spike emission of HNC(3--2) is not confirmed in ALMA data, 
the putative maser feature must be either transient or an artifact. 
Our three galactic nuclei do not show spike-like emission features in any line in our dataset.

\subsection{\HthreeOplus\ detection}
\label{s.followup.H3O+}
\citet{vdTak08} reported detection of the \HthreeOplus($3^+_2$--$2^-_2$) line (\frest = 364.797 GHz) in emission 
toward Arp 220 using the James Clark Maxwell Telescope (JCMT).
They deduced from the line velocity that the line, tracing ionization of hot ($\gtrsim$100 K) gas, 
was from the western nucleus.
However, this association needs caution because the two nuclei of Arp 220 have 
line widths of \about500 \kms\ in FWHM and are
only 100 \kms\ different in their systemic velocities (see Fig.~\ref{f.vsys}). 
The lines from the two nuclei should overlap for the most part in the single-dish spectrum.
In addition, the reported line velocity of $V$(helio, radio) = 5473$\pm$8 \kms\
corresponds to $V$(LSR, radio) = 5488$\pm$8 \kms\ and 
is closer to the velocity of the eastern nucleus.

Our observations covered the frequency of the \HthreeOplus\ line, as shown in Figure \ref{f.triNuc_spec.B7}.
The Arp 220 data reveal a broad, \about1 GHz-wide, emission feature around the frequency of the \HthreeOplus\ transition. 
It is seen toward both nuclei in a similar shape, with its peak being a few tenths of GHz offset from the \HthreeOplus\ frequency.
We did not confirm the 113 \kms\ (0.14 GHz)-wide line at the \HthreeOplus\ frequency in the JCMT observations, 
whose bandwidth was 1 GHz.
Referencing our spectrum of NGC 4418, whose narrower lines suffer less blending, we see
bright lines of vibrationally excited \HCthreeN\ ($J_{\rm up}$=40; $v_6$=1 and $v_7$=1 lines) 
around the \HthreeOplus\ frequency. 
One must be careful about possible blending with these \HCthreeN-vib lines when searching for 
\HthreeOplus($3^+_2$--$2^-_2$) in galaxies.
In the Arp 220 nuclei, two lower rotational groups of the vibrationally excited \HCthreeN\ lines
are detected at lower intensities than the $J_{\rm up}$=40 group.
It is plausible that the \HthreeOplus($3^+_2$--$2^-_2$) line (or any other nearby lines)
contributes to the broad feature at \frest\ \about\ 365 GHz. 
Its confirmation and line fluxes need precise line de-blending.

\begin{figure}[t]
\epsscale{0.4}
\plotone{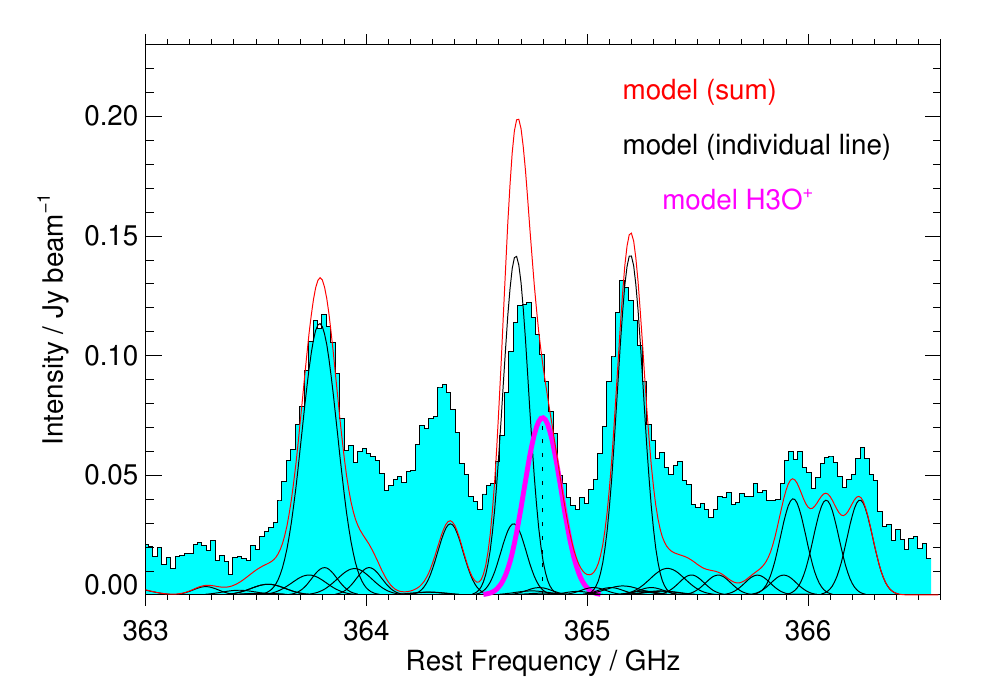} 
\caption{ \label{f.n4418_H3O+}
NGC 4418 spectrum around the \HthreeOplus($3^+_2$--$2^-_2$) line (\frest = 364.797 GHz).
Our model fit in red is overlaid on the observed spectrum in light blue.
Individual lines in the model are shown in black and, for the \HthreeOplus, magenta.
}
\end{figure}

Our NGC 4418 data also suggest excess emission around the \HthreeOplus\ frequency.
Comparing the channel maps of the \HCthreeN(J=40--39, $v_7$=1) doublet in Fig.~\ref{f.chmap.n4418.B7}(a) and (b), 
there is excess emission around 2000 \kms\ in the \HCthreeN(J=40--39, $v_7$=1, $l$=1e) maps.
The \HthreeOplus($3^+_2$--$2^-_2$) transition is at 2001 \kms\ there.
Accordingly, our spectral modeling, a portion of which is in Figure~\ref{f.n4418_H3O+}, attributes the 
line feature at \about364.7 GHz to a combination of \HCthreeN(40--39, $v_7$=1, $l$=1e) 
and \HthreeOplus($3^+_2$--$2^-_2$), although the current model poorly fits the 364.7 GHz peak intensity. 
\HthreeOplus\ has no other transitions in our data to verify this identification.
The flux of the presumed \HthreeOplus\ line must be much less than that of \HCthreeN(40--39, $v_7$=1, $l$=1e) 
judging from the comparison with the adjacent $v_7$=1 $l$=1f line and the $v_7=1$ doublets in other rotational groups.

\subsection{\CeighteenO-to-\thirteenCO\ ratios}
\label{s.followup.18O}
\begin{figure*}[t]
\centering
\includegraphics{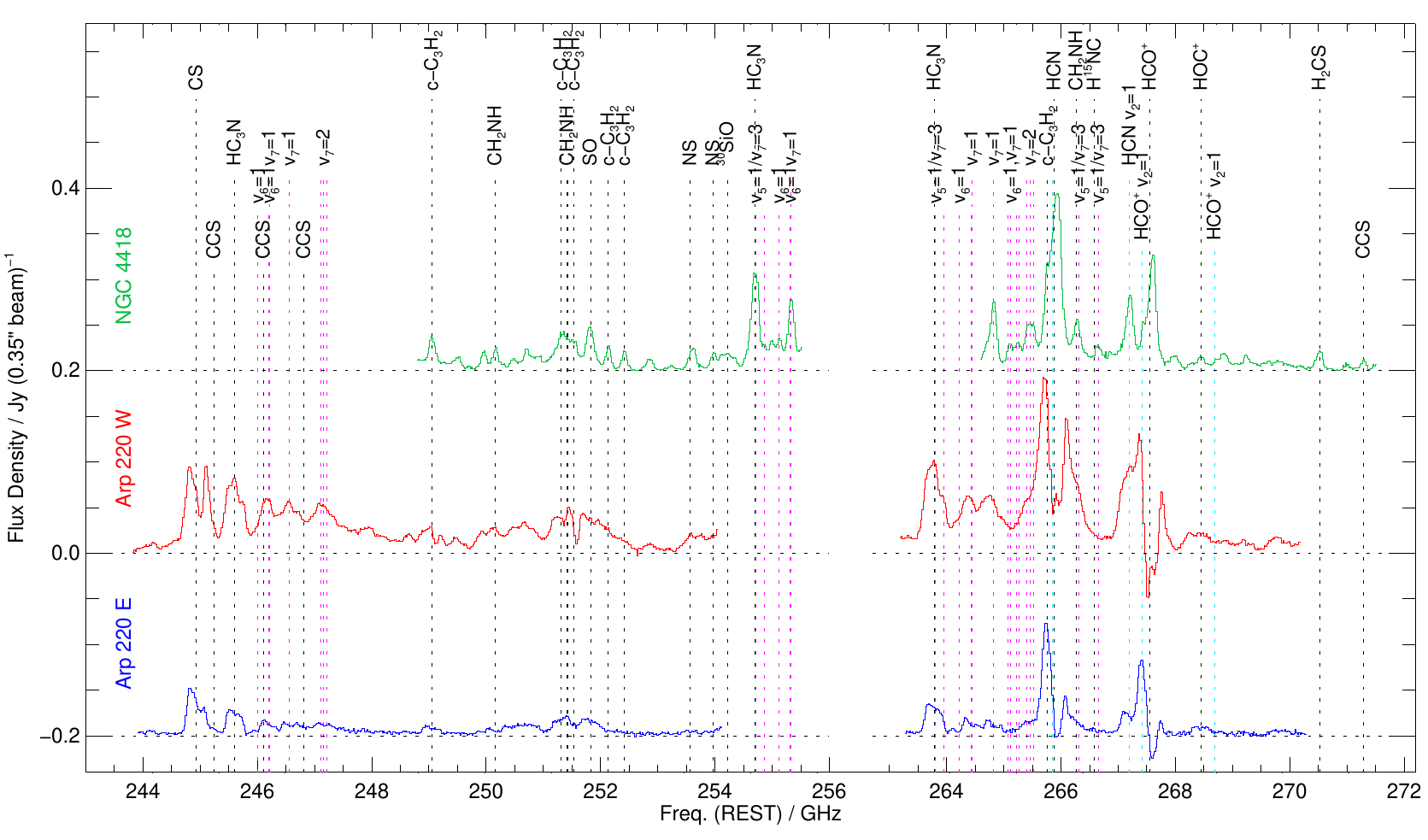}
\includegraphics{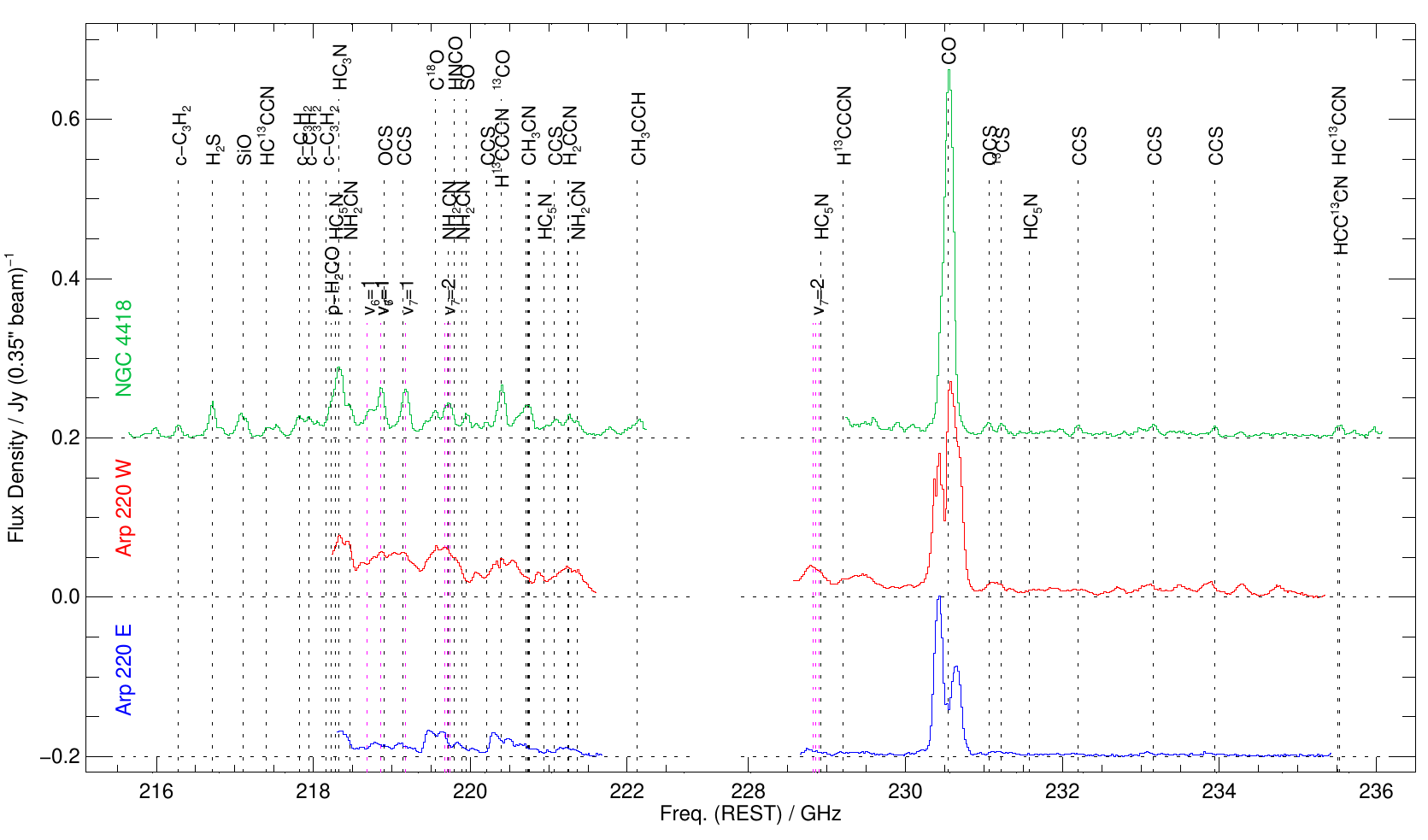}
\caption{ \label{f.triNuc_spec.B6.350mas}
Spectra of the three nuclei in NGC 4418 and Arp 220
in the rest-frequency range of 244--272 GHz (top) and 215--236 GHz (bottom)
sampled in a 0\farcs35 beam at 20 MHz resolution.
Continuum has been subtracted and constant offsets are added for clarity.
Major lines and lines of interest are marked with the species 
except for lines from vibrationally excited \HCthreeN\ (magenta) for which only their vibrational levels are given.
Lines from vibrationally excited HCN and \HCOplus\ are plotted in cyan.
}
\end{figure*}

There have been reports of anomalously high \CeighteenO-to-\thirteenCO\ line ratios of about unity 
in Arp 220 \citep{Greve09, Matsushita09, Martin11, Brown19}. 
Also reported is an enhancement of the [$^{18}$O/$^{16}$O] abundance ratio in Arp 220 by about a factor of 5 compared to the Galactic ones \citep{GA12}. 
The former is from millimeter-wave emission lines of the isotopologues,
while the latter is from far-IR absorption lines of  \water, OH, and their $^{18}$O isotopologues.
It has been argued that these unusual ratios in Arp 220 are due to the production of $^{18}$O in high-mass stars and their release to the ISM \citep{Matsushita09, GA12, Brown19}.
Our 1 mm spectra in Fig.~\ref{f.triNuc_spec.B6.350mas} confirm the observations of \citet{Martin11} that \CeighteenO(2--1) is
slightly stronger than \thirteenCO(2--1) in Arp 220. We now see this in both nuclei. 
(The two nuclei were unresolved in the 8\arcsec\ resolution data of \citet{Martin11}, and their assignment of two peaks in
spectra to the two nuclei was not upheld since both nuclei have much larger line widths than the velocity difference between them.) 
\citet{Brown19} already observed the  \CeighteenO-to-\thirteenCO\ ratio in the $J$=1--0 transition to be similar in both nuclei and slightly larger than unity.
The 1--0 lines of these isotopologues are no less blended with adjacent lines \citep[e.g., vibrationally-excited \HCthreeN; ][]{Sakamoto17} than the 2--1 lines are. 
However, similar observations in the two transitions (despite their different line-blending) lend credence to the claim that the two nuclei have a similar \CeighteenO-to-\thirteenCO\ line ratio slightly larger than unity in both transitions.
The \CeighteenO-to-\thirteenCO\ ratio of integrated intensities for the entire galaxy is 1.3 in J=2--1, according to the line flux measurements with de-blending of contaminating lines in \citet{Martin11}. 
The two nuclei should therefore have about the same value.

Regarding NGC 4418, \citet{GA12} found its [\eighteenO/\sixteenO] abundance ratio to be \about 500 (or $\gtrsim250$),
comparable to Galactic values.
It is then interesting to see in Fig.~\ref{f.triNuc_spec.B6.350mas} that \CeighteenO(2--1) is less bright than \thirteenCO(2--1) in NGC 4418, unlike in the Arp 220 nuclei. 
The \CeighteenO(2--1)-to-\thirteenCO(2--1) 
ratio is $0.730 \pm 0.009$ for line intensities integrated over 300 \kms\ 
and $0.507 \pm 0.015$ for peak intensities. (The uncertainties do not include those due to any line blending.)
The \CeighteenO-to-\thirteenCO\ ratio is less than unity also in $J$=1--0 \citep{Costagliola15}.
It appears that the \CeighteenO-to-\thirteenCO\ line ratio reflects the \eighteenO-to-\sixteenO\ abundance ratio to some extent and that the beam-averaged line fluxes reflect abundances, despite the likely saturation toward the centers of these nuclei.
It remains to be answered why the \eighteenO-to-\sixteenO\ abundance ratio is different between NGC 4418 and the Arp 220 nuclei
despite the number of properties they share. 
Possible reasons include the difference in star formation history, stellar initial mass function, and relative luminosity contribution between AGN and starburst in the sense that NGC 4418 has a larger AGN contribution.

\section{Summary and Concluding Remarks}
\label{s.summary}
We have presented the line information in our ALMA imaging spectroscopy of three deeply buried nuclei
in the infrared-luminous galaxies NGC 4418 and Arp 220.
Our scan covers 67 GHz from within \frest=215--697 GHz at \about0\farcs2 resolution.
The target nuclei are characterized by their   
very high obscuration (\NH $\gtrsim 10^{25}$ \persquarecm), 
high luminosity ($\Lbol \gtrsim 10^{11}$ \Lsun),
and compactness ($\lesssim 100$ pc).

\begin{enumerate}
\item The three nuclei have forests of lines in their spectra.
We reached the line-confusion limit at our sensitivity of $\sigma=0.1$ K for 50 \kms\  
in our 0\farcs35 spectra from \frest=215--367 GHz (ALMA Bands 6 and 7).
Our spectra from \frest=670--697 GHz (Band 9) are about twice less sensitive but still 
show many lines in both emission and absorption.

\item Toward the center of each nucleus, most of the bright lines are in absorption (i.e., going below the continuum level)
or at least show decrement of emission around the systemic velocity.
They even include lines of rare isotopologues such as \HthirteenCN, \HthirteenCOplus, and \thirteenCS.
Lines of vibrationally-excited HCN and \HCthreeN\ do not show absorption at the current resolution
(although they go into absorption at higher resolution.)

\item Our data show that the size of the emitting region varies among lines.
It tends to be smaller 
for lines with higher upper-level energies (including lines at vibrationally excited states) 
and lines of rarer isotopologues. 
This trend suggests a radial gradient of excitation conditions, such as temperature, in each nucleus.

\item As our initial attempt for line identification and modeling, we fitted our 0\farcs35 spectrum of NGC 4418 in Bands 6 and 7,
assuming isothermal and uniform gas and LTE. 
More than 200 lines from 55 species have been identified, although some of them are still tentative.

\item Lines show systematic velocity gradients approximately along the major axis of each continuum nucleus.
This gas motion is consistent with the known rotation of nuclear gas disks in Arp 220.
We confirmed that the velocity gradient of molecular gas in the center of NGC 4418 is 
opposite to that of the galactic disk rotation.

\item Many line profiles toward the nucleus of NGC 4418 are asymmetric in various degrees 
to indicate redshifted absorption. 
This redshifted absorption, consistent with previous far-IR and submillimeter observations,
indicates either an inward gas motion along our sightline or a collimated outflow slanted to the nuclear disk. 
We found small-scale spatial lopsidedness of the absorption to favor the latter and suggest further verification.
We also confirmed in the central 60 pc that lines of dense gas tracers are broader than the CO lines.

\item The molecular outflow from Arp 220W is imaged in several emission lines, most clearly 
in HCN(4--3) and CO(2--1). 
It is almost perpendicular to the major axis of the nuclear gas disk, redshifted to the north and blueshifted to the south.
The end-to-end extent of the bipolar outflow is about 1\arcsec\ (400 pc) in our images.
Many lines show broad, blueshifted, and spatially-lopsided absorption toward \arpW. 
They are consistent with the collimated outflow.

\item The molecular outflow from Arp 220E is found visible through broad line absorption offset from the continuum peak.
The offset of blueshifted absorption to the northwest suggests that the eastern nuclear disk has its far side to the northwest 
and that the northwestern outflow is approaching us.
A recent report of CO(3--2) detection of this outflow has an outflow+disk configuration that is the opposite of ours.
The detection must be due to misidentification of \HthirteenCN\ and vibrationally excited \HCthreeN.

\item 
The nuclei of Arp 220 also have narrow absorption lines near their systemic velocities, most clearly visible in CO and also in HCN.

\item 
The two nuclear disks of Arp 220 remain in counter-rotation in the three-dimensional space 
after reevaluating their major-axis position angles, inclinations, and near-far sides using our new ALMA data.
It is the western nuclear disk that is in counter-rotation with the kpc-scale gas around the binary nucleus. 
The counter-rotations of \arpW\ and the nucleus of NGC 4418 relative to their surroundings 
may be partly responsible for their extreme central gas concentrations of $\NH \sim 10^{26}$ \persquarecm\ 
and luminous nuclear activities.

\item 
We verified earlier observations on HNC and \HthreeOplus\ in Arp 220. 
We did not confirm a bright HNC spike in a previous report.
We found that the putative \HthreeOplus($3^+_2$--$2^-_2$) line overlaps with bright lines of vibrationally-excited \HCthreeN.
However, a minor contribution of the \HthreeOplus\ line is still possible in the three nuclei of Arp 220 and NGC 4418.

\item The \CeighteenO(2--1) to \thirteenCO(2--1) flux ratio is slightly larger than unity in Arp 220 toward both nuclei
while it is below unity toward the nucleus of NGC 4418.

\end{enumerate}

Overall, our spectral scan has shown many similarities among the three galactic nuclei in NGC 4418 and Arp 220.
Each of them has a rotating nuclear disk of molecular gas \about100 pc in size; 
the nuclear disks are also seen in dust emission in our continuum analysis in \citest{Paper1}. 
Line brightness temperatures indicate high gas temperatures of more than 100 K in the three nuclear disks.
Each nucleus must be warmer toward the center, judging from the size-energy correlation among the lines, sharply peaked dust continuum intensity toward the center, and many absorption lines against the continuum peak.
Vibrationally excited HCN and \HCthreeN\ are also detected among the smallest areas at the center of each nucleus.
It is consistent with the greenhouse effect (i.e., continuum photon-trapping) suggested for these nuclei
\citep{GS19}. 
Also common to all three nuclei is a radial motion of molecular gas around them, 
including the molecular outflows to the polar directions from both nuclear disks of Arp 220.
NGC 4418 may also have a small-scale, collimated outflow oblique to its nuclear disk
in addition to its known kpc-scale outflow perpendicular to the galactic disk.

Our data also highlighted some differences in the three nuclei. 
The eastern nucleus of Arp 220 is less bright in the 1 mm continuum than Arp 220W and  shows less vibrational excitation in \HCthreeN\
than the western nucleus. 
These differences suggest that the E nucleus has less luminosity concentration or less photon trapping, or both. 
The two are consistent with our continuum brightness temperatures and
dust opacity estimates from continuum spectral slopes \citesp{Paper1}.
The nuclear outflow from Arp 220W is much more prominent than those from the other two nuclei in our data.

Our spectral scan program also demonstrated the benefits of 
and need for wide-band observations in extragalactic studies in the ALMA era. 
The wide bandwidth helps to de-blend and identify overlapping lines, avoid misinterpretation in narrow-band data,
and allow more reliable calibration and continuum-subtraction to facilitate accurate data analysis and interpretation.
The virtue of wide-band {\em and} high angular resolution is evident in our findings of prevalent line absorption and 
systematic variations in the line-emitting areas. 
Both call for caution in the simple one-zone analysis of low-resolution spectral scans.
Our paired study of NGC 4418 and Arp 220 also showed the effectiveness of
the comparative study. 
The narrow and hence better-identified lines in NGC 4418 helped interpret the broad lines in the nuclei of Arp 220.
Differences in their spectra also highlight the differential properties of these otherwise similar nuclei.

\vspace{5mm}
\acknowledgements
We are grateful to the ALMA Observatory and its staff members for making the observations used here.
This paper makes use of the following ALMA data: 
ADS/JAO.ALMA\#2012.1.00377.S, 
ADS/JAO.ALMA\#2012.1.00317.S,
and
ADS/JAO.ALMA\#2012.1.00453.S.
ALMA is a partnership of ESO (representing its member states), NSF (USA), and NINS (Japan), 
together with NRC (Canada), MOST and ASIAA (Taiwan), and KASI (Republic of Korea), 
in cooperation with the Republic of Chile. 
The Joint ALMA Observatory is operated by ESO, AUI/NRAO, and NAOJ.
This research has made use of NASA's Astrophysics Data System Bibliographic Services.
This research has also made use of the NASA/IPAC Extragalactic Database (NED), 
which is operated by the Jet Propulsion Laboratory, California Institute of Technology, 
under contract with the National Aeronautics and Space Administration.
KS is supported by grants MOST 108-2112-M-001-015 and 109-2112-M-001-020 from the Ministry of Science and Technology, Taiwan.
We thank the reviewer for many constructive comments that help clarify this paper.

\facility{ALMA}
\software{CASA \citep{CASA07}, 
mpfit \citep{More77,More93,Markwardt09}, 
uvmultifit \citep{uvmultifit14},
MADCUBA \citep{MADCUBA}}

\clearpage
\bigskip

\begin{deluxetable}{ll}
\tabletypesize{\scriptsize}
\tablewidth{0pt}
\tablecaption{Summary of Observations \label{t.obsSummary} }
\tablehead{ 
       \colhead{Parameter} &
       \colhead{Value} 
} 
\startdata
ALMA Band & 6, 7, 9  \\
Frequency Coverage & 67 GHz in \frest=215--697 GHz  \\
Frequency Resolution & 0.98 MHz \\
Primary Beam Size & 24\arcsec--8\arcsec   \\
Angular resolution & 0\farcs14--0\farcs28 in $(\theta_\mathrm{maj}\theta_\mathrm{min})^{1/2}$  \\
				& 0\farcs35 in $\max \theta_\mathrm{maj}$ \\
Max. Recoverable Scale & 11\arcsec--3\arcsec \\
Sensitivity $\sigma_\mathrm{50\,km/s}$  			& 0.6, 1.0, 8 mJy (native beam)$^{-1}$ \\
{in Band 6, 7, 9 }								& 0.3, 0.6, 0.6 K for native beam \\
{		}									& 0.1, 0.1, 0.2 K at 0\farcs35 resolution \\
\enddata
\tablecomments{
Angular resolution is given in the geometrical mean of the major- and minor-axis FWHM as well as in the maximum
of the major-axis FWHM of the synthesized beam.
Sensitivity is for typical values in Band 6, 7, and 9, respectively. 
The native beam is the one from imaging at {\sf robust=0.5}.
See \citest{Paper1} for more information about the ALMA observations
and data reduction.
}
\end{deluxetable}

\begin{deluxetable}{clr}
\tabletypesize{\scriptsize}
\tablewidth{0pt}
\tablecaption{Frequency Coverage \label{t.frestCoverage} }
\tablehead{ 
       \colhead{Band} &
	\multicolumn{2}{c}{Rest frequency / GHz} 
	\\
	\colhead{}  &
	\colhead{ranges} &
	\colhead{width} 
}
\colnumbers  
\startdata
\sidehead{NGC 4418} 
9	& 678.309--685.206, 690.081--697.039  	& 13.9 \\
7	& 340.939--366.574 						& 25.6 \\
6     & 248.777--255.533, 264.570--271.520 		& 	\\
{ }	& 215.633--222.265, 229.225--236.096 		& 27.2 \\
\sidehead{Arp 220} 
9	& 669.675--676.606, 696.460--689.423 		& 14.0 \\
7	& 341.132--367.057 						& 25.9 \\
6	& 243.856--254.088, 263.224--270.251 		& 	\\
{ }	& 218.265--221.658, 228.595--235.403 		& 27.5 \\
\enddata
\tablecomments{
(1) ALMA receiver band.
(2) Ranges of rest frequency we observed. For the conversion to rest frequency 
we used $V$(radio, LSRK)=2100 \kms\ and 5350 \kms\ for NGC 4418 and Arp 220, respectively.
(3) Total frequency coverage in each band.
Our total coverage is 66.7 and 67.4 GHz for NGC 4418 and Arp 220, respectively. 
}
\end{deluxetable}



\begin{deluxetable}{llLCLRCcl}
\tabletypesize{\scriptsize}
\tablewidth{0pt}
\tablecaption{Parameters of the Line Channel Maps  \label{t.chmaps} }
\tablehead{ 
       \colhead{Species} &
       \colhead{Transtion} &       
       \colhead{\frest} &
       \multicolumn{2}{c}{$\sigma_{\rm 50\, km/s}$}  &
       \colhead{$\max \left| I_\nu \right|$} &
       \colhead{$p$} &
       \colhead{Figure} &
       \colhead{{\bf Nearby Lines}}     
       \\
	\colhead{ }  &
	\colhead{ }  &
	\colhead{GHz} &
	\colhead{mJy \perbeam} &
	\colhead{K} &
	\colhead{K} &
	\colhead{ } &
	\colhead{} &
	\colhead{}	
}
\decimals
\colnumbers   
\startdata
\sidehead{NGC 4418}
\HNthirteenC 	& J=8--7 				& 696.532 		& 8.5  	& 0.55 & 6.8 		& 1.0 	& \ref{f.chmap.n4418.B9}a 	&  \\
SiO             	& J=16--15 				& 694.275 		& 7.3 	& 0.45 & -7.7 	& 1.0 	& \ref{f.chmap.n4418.B9}b 	& \HthirteenCOplus@2271 \\
\thirteenCS   	& J=14--13 				& 693.234 		& 15.6  	& 1.56 & 15.3 	& 1.0 	& \ref{f.chmap.n4418.B9}c 	& \HthirteenCOplus@1824\\
CO             	& J=6--5 				& 691.473 		& 14.5 	& 1.45 & 120.3 	& 1.7 	& \ref{f.chmap.n4418.B9}d 	& \HthirteenCN@2497 \\
\HthirteenCN	& J=8--7 				& 690.552 		& 15.1 	& 1.58 & 20.9 	& 1.0 	& \ref{f.chmap.n4418.B9}e 	& CO@1703 \\
CN              	& N=6--5\tnm{a} 			& 680.165		& 12.2	& 1.23 & -36.0	& 1.5 	& \ref{f.chmap.n4418.B9}f 	& \CthirtythreeS@2268 \\
\HCthreeN & $v_7$=1,  J=40--39, $l$=1f & 365.195 		& 1.31 	& 0.50 & 34.6 	& 1.5 	& \ref{f.chmap.n4418.B7}a  &  \\
\HCthreeN & $v_7$=1,  J=40--39, $l$=1e & 364.676 		& 1.18 	& 0.45 & 32.0 	& 1.5 	& \ref{f.chmap.n4418.B7}b  & H$_3$O$^+$@2001, \HCthreeN($v_6$=1f)@2108 \\
\HCthreeN & J=40--39                             & 363.785 		& 1.18 	& 0.56 & 28.9 	& 1.5 	& \ref{f.chmap.n4418.B7}c  &  \\
HNC          & J=4--3                                 & 362.630 		& 1.65 	& 0.86 & 54.4 	& 1.5 	& \ref{f.chmap.n4418.B7}d  &  \\
\HCOplus  & J=4--3                                  & 356.734 		& 1.21 	& 0.63 & 38.5 	& 1.5 	& \ref{f.chmap.n4418.B7}e  & HCN($v_2$=1,$l$=1f)@2499 \\
HCN          & $v_2$=1, J=4--3, $l$=1f       & 356.256 		& 1.80 	& 0.85 & 43.0 	& 1.5 	& \ref{f.chmap.n4418.B7}f  & \HCOplus@1700 \\
HCN          & J=4--3                                 & 354.505 		& 1.66 	& 0.82 & 55.1 	& 1.5 	& \ref{f.chmap.n4418.B7}g  & \HCthreeN(39--38)@1939, HCN($v_2$=1,$l$=1e)@2138  \\
\CtwoH     & N=4--3                                 & 349.365\tnm{b} & 1.06 	& 0.54 & 35.9 	& 1.5 	& \ref{f.chmap.n4418.B7}h &  \\
\HNthirteenC & J=4--3                             & 348.340 		& 1.15 	& 0.63 & 28.3 	& 1.5 	& \ref{f.chmap.n4418.B7}i  &  \\
SiO           & J=8--7                                  & 347.331 		& 1.11 	& 0.54 & 19.5 	& 1.5 	& \ref{f.chmap.n4418.B7}j & \HthirteenCOplus@2385 \\
\HthirteenCOplus & J=4--3                       & 346.998 		& 1.01 	& 0.48 & 29.9 	& 1.5 	& \ref{f.chmap.n4418.B7}k & SiO@1815 \\
CO            & J=3--2                                 & 345.795 		& 1.01 	& 0.47 & 144.0 	& 2.2 	& \ref{f.chmap.n4418.B7}l  & \HCthreeN(38--37)@2261, \HthirteenCN@2493 \\
\HthirteenCN & J=4--3                             & 345.340 		& 1.05 	& 0.51 & 33.0 	& 1.5 	& \ref{f.chmap.n4418.B7}m & CO@1707 \\
CS            & J=7--6                                  & 342.883 		& 1.35 	& 0.65 & 37.3 	& 1.5 	& \ref{f.chmap.n4418.B7}n &  \\
\HCOplus  & J=3--2                                  & 267.558 		& 0.99 	& 0.28 & 26.6 	& 1.7 	& \ref{f.chmap.n4418.B6}a   & HCN($v_2$=1, $l$=1f)@2499 \\
HCN         & $v_2$=1, J=3--2, $l$=1f        & 267.199 		& 0.94 	& 0.26 & 21.3 	& 1.7 	& \ref{f.chmap.n4418.B6}b   & \HCOplus@1701 \\
HCN         & J=3--2                                  & 265.886 		& 0.81 	& 0.22 & 40.2 	& 1.7 	& \ref{f.chmap.n4418.B6}c   & HCN($v_2$=1, $l$=1e)@2137 \\                
\HCthreeN & $v_7$=1, J=28--27, $l$=1e  & 255.325 		& 0.58 	& 0.24 & 28.9 	& 1.7 	& \ref{f.chmap.n4418.B6}d    & \HCthreeN($v_6$=1f)@2109 \\
\HCthreeN & J=28--27                             & 254.700 		& 0.47 	& 0.19 & 32.4 	& 1.7 	& \ref{f.chmap.n4418.B6}e    &  \\
CO           & J=2--1                                  & 230.538 		& 0.63 	& 0.24 & 123.9 	& 2.2 	& \ref{f.chmap.n4418.B6}f    &  \\
\thirteenCO  & J=2--1                              & 220.399 		& 0.54 	& 0.22 & 22.7 	& 1.7 	& \ref{f.chmap.n4418.B6}g &  \\
\CeighteenO & J=2--1                              & 219.560 		& 0.70 	& 0.31 & 13.6 	& 1.7 	& \ref{f.chmap.n4418.B6}h &  \\
\HCthreeN & $v_7$=1, J=24--23, $l$=1f  & 219.174 		& 0.81 	& 0.35 & 25.2 	& 1.7 	& \ref{f.chmap.n4418.B6}i &  \\
\HCthreeN & J=24--23                           	& 218.325 		& 0.80 	& 0.32 & 31.7 	& 1.7 	& \ref{f.chmap.n4418.B6}j &  \\
SiO         & J=5--4                                  	& 217.105 		& 0.55 	& 0.21 & 16.8 	& 1.7 	& \ref{f.chmap.n4418.B6}k &  \\
\HtwoS   & J=$2_{2,0}$--$2_{1,1}$ 		& 216.710 		& 0.69 	& 0.23 & 16.8 	& 1.7 	& \ref{f.chmap.n4418.B6}l &  \\                     
\sidehead{Arp 220}
SiO & J=16--15 						& 694.275 	& 11.3 	& 0.97 	& -24.4	& 1.4 	& \ref{f.chmap.a220.B9}a &  \HthirteenCOplus@5519, \thirteenCS@5792 \\
CO & J=6--5 							& 691.473 	& 9.8 	& 0.83 	& 66.5	& 1.6 	& \ref{f.chmap.a220.B9}b &  \HthirteenCN@5742 \\
\HtwoCO & J=$9_{1,8}$--$8_{1,7}$ 		& 674.810 	& 9.5 	& 0.78 	& -13.0	& 1.0 	& \ref{f.chmap.a220.B9}c &  \\
HNC          & J=4--3                                 & 362.630 	& 1.78 	& 0.70 	& 72.3 	& 1.5 	& \ref{f.chmap.a220.B7}a            &  \\
\HCOplus  & J=4--3                                  & 356.734 	& 1.28 	& 0.52 	& -54.5 	& 1.5 	& \ref{f.chmap.a220.B7}b            & HCN($v_2$=1, $l$=1f)@5745 \\
HCN          & $v_2$=1, J=4--3, $l$=1f       & 356.256 	& 1.37 	& 0.56 	& 41.6	& 1.5 & \ref{f.chmap.a220.B7}c & \HCOplus@4954 \\
HCN          & J=4--3                                 & 354.505 	& 1.36 	& 0.61 	& 71.3 	& 1.5 	& \ref{f.chmap.a220.B7}d             & \HCthreeN(39--38)@5191, HCN($v_2$=1, $l$=1e)@5387 \\
\CtwoH     & N=4--3                                 & 349.365\tnm{b} & 1.29 & 0.53 & 36.2 & 1.5 	& \ref{f.chmap.a220.B7}e &  \\
SiO           & J=8--7                                  & 347.331 	& 1.04 	& 0.43 	& -45.9 	& 1.5 	& \ref{f.chmap.a220.B7}f             &  \HthirteenCOplus@5632 \\
CO            & J=3--2                                 & 345.795 	& 1.15 	& 0.48 	& 106.0 	& 1.8 	& \ref{f.chmap.a220.B7}g           &  \HCthreeN(38--37)@5509, \HthirteenCN@5738 \\
CS            & J=7--6                                  & 342.883 	& 1.26 	& 0.56 	& 39.0 	& 1.5 	& \ref{f.chmap.a220.B7}h             &  \\
\HCOplus  & J=3--2                                  & 267.558 	& 0.94 	& 0.41 	& 37.7 	& 1.5 	& \ref{f.chmap.a220.B6}a             & HCN($v_2$=1, $l$=1f)@5744 \\
HCN         & $v_2=$1, J=3--2, $l$=1f        & 267.199 	& 0.89 	& 0.38 	& 31.4	& 1.5 & \ref{f.chmap.a220.B6}b  & \HCOplus@4955 \\
HCN         & J=3--2                                  & 265.886 	& 0.89 	& 0.36 	& 55.4 	& 1.7 	& \ref{f.chmap.a220.B6}c           & HCN($v_2$=1, $l$=1e)@5387 \\             
CS            & J=5--4                                  & 244.936 	& 0.57 	& 0.27 	& 39.5 	& 1.7 	& \ref{f.chmap.a220.B6}d          &  \\
CO           & J=2--1                                  & 230.538 	& 0.68 	& 0.31 	& 77.2 	& 1.8 	& \ref{f.chmap.a220.B6}e             &  \\
\enddata
\tablecomments{
(3) Line rest frequency from {\sf Splatalogue}\footnote{\sf https://splatalogue.online//}. 
(4)--(5) Noise rms in 50 \kms\ channel maps. 
(6) Peak Rayleigh-Jeans brightness temperature of the line in our channel maps. 
Brighter blending lines are excluded from the search. 
A negative value is listed when the deepest absorption has a larger absolute intensity than the brightest emission. 
(7) The power-law index used for contouring. 
(8) Figure and panel ID of the channel maps.
(9) {\bf Major species} whose lines {\bf may be blended with} or close to the line in Columns (1) and (2). 
Not all of them are firmly detected, and we do not list all possible blending either.
Numbers after @ are the velocities at which the blending lines at the reference velocity of the galaxy (2100 \kms\ for NGC 4418 and 5350 \kms\ for Arp 220) 
would appear in the data of the blended lines.
For example,  a transition of H$_3$O$^{+}$ at 2100 \kms\ should be at 2001 \kms\ of \HCthreeN($v_7$=1,  J=40--39, $l$=1e).
}
\tablenotetext{a}{Weighted mean frequency of twelve hyper-fine transitions in CN(N=6--5, J=11/2--9/2) and CN(N=6--5, J=13/2--11/2. 
The wight is $g_u A$ where $g_u$ is the statistical weights of the upper state and $A$ is the Einstein's A coefficient.}
\tablenotetext{b}{The four strongest transitions in Splatalogue are in the range of \frest=349.337--349.400 GHz
and are blended in our data. We use their strength-weighted mean frequency.  }
\end{deluxetable}


\begin{figure}[t]
\figurenum{\ref{f.chmap.n4418.B7} (continued)}
\fig{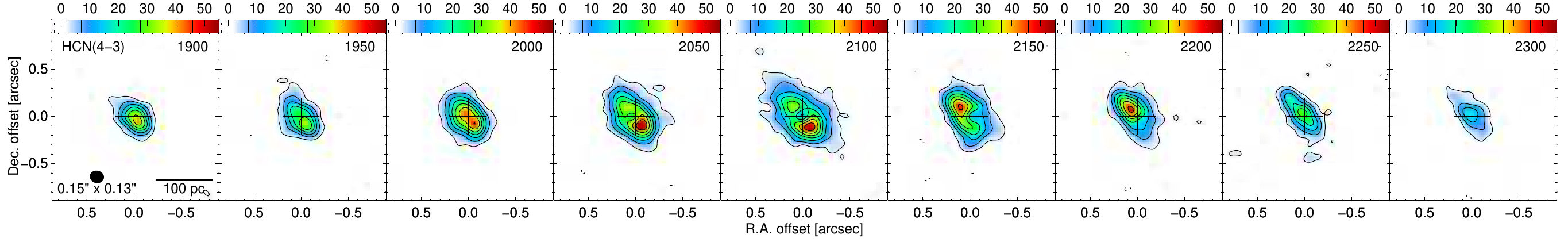}{1.0\textwidth}{(g) HCN(4--3)}
\fig{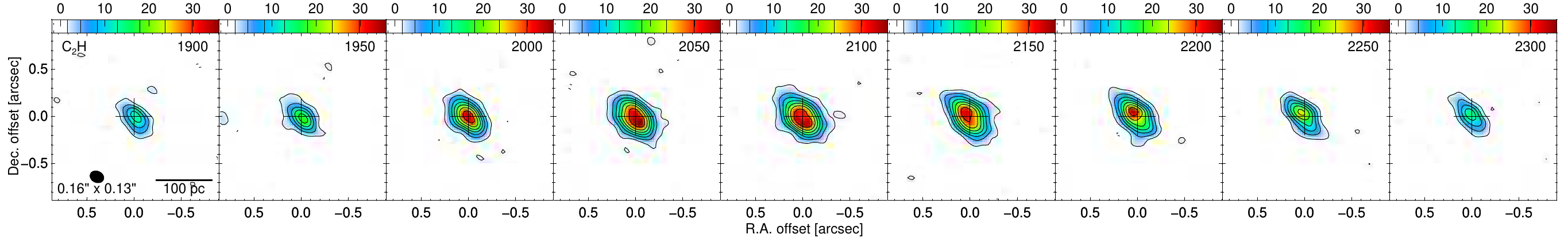}{1.0\textwidth}{(h) \CtwoH(N=4--3) }
\fig{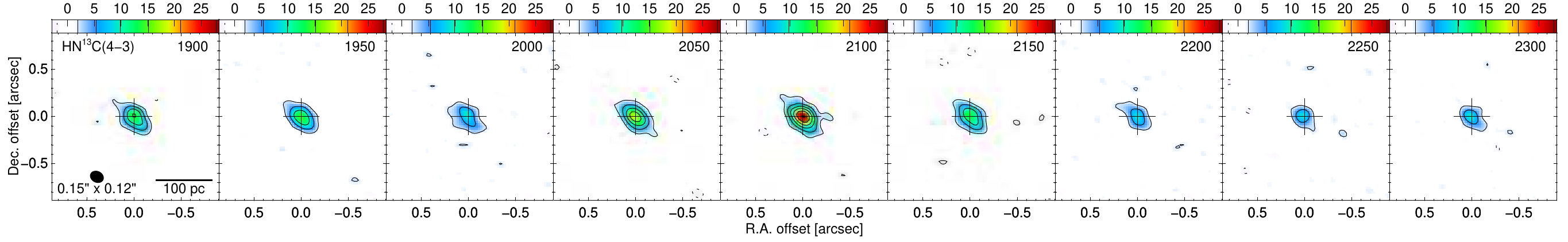}{1.0\textwidth}{(i) \HNthirteenC(4--3)}
\fig{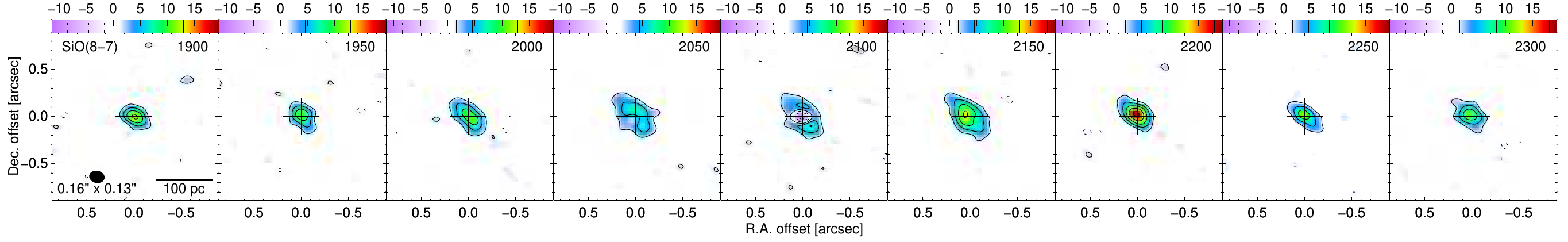}{1.0\textwidth}{(j) SiO(8--7)}
\fig{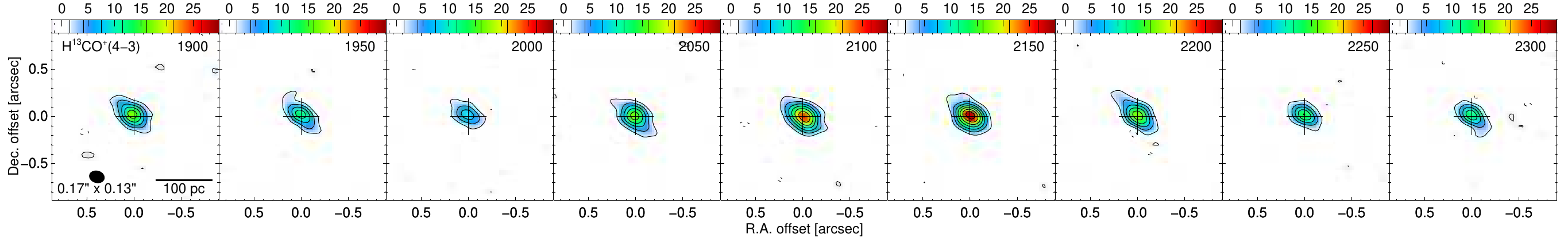}{1.0\textwidth}{(k) \HthirteenCOplus(4--3)}
\fig{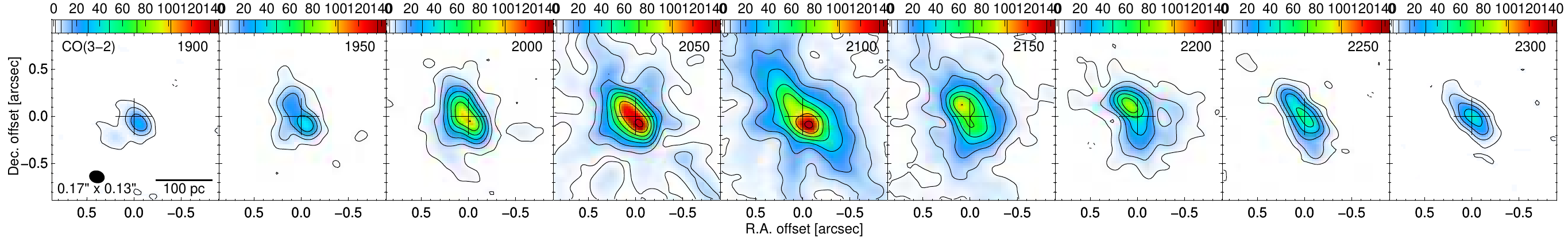}{1.0\textwidth}{(l) CO(3--2)}
\caption{ }
\end{figure}
\begin{figure}[t]
\figurenum{\ref{f.chmap.n4418.B7} (continued)}
\fig{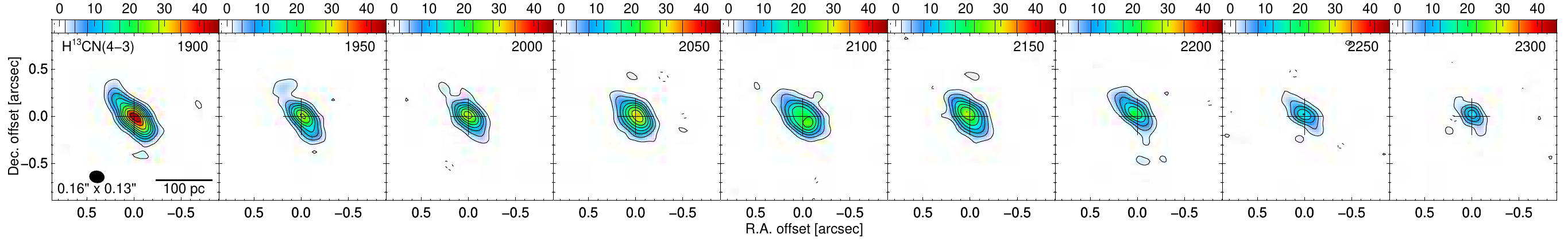}{1.0\textwidth}{(m) \HthirteenCN(4--3)}
\fig{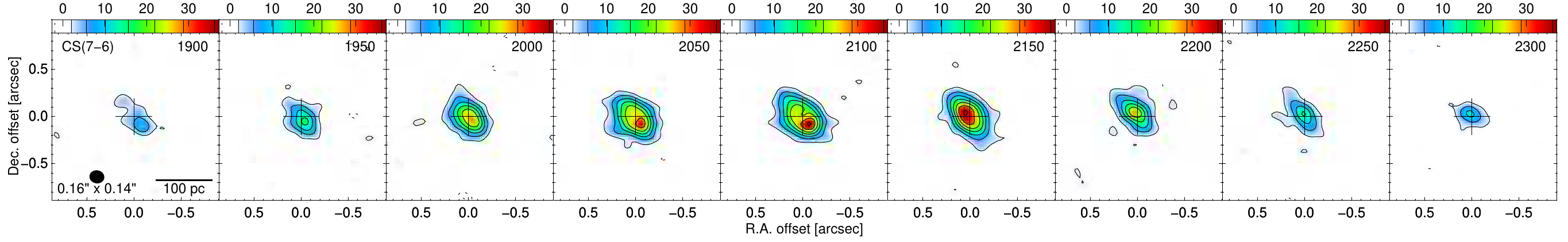}{1.0\textwidth}{(n) CS(7--6)}
\caption{ }
\end{figure}
\begin{figure}[t]
\figurenum{\ref{f.chmap.n4418.B6} (continued)}
\fig{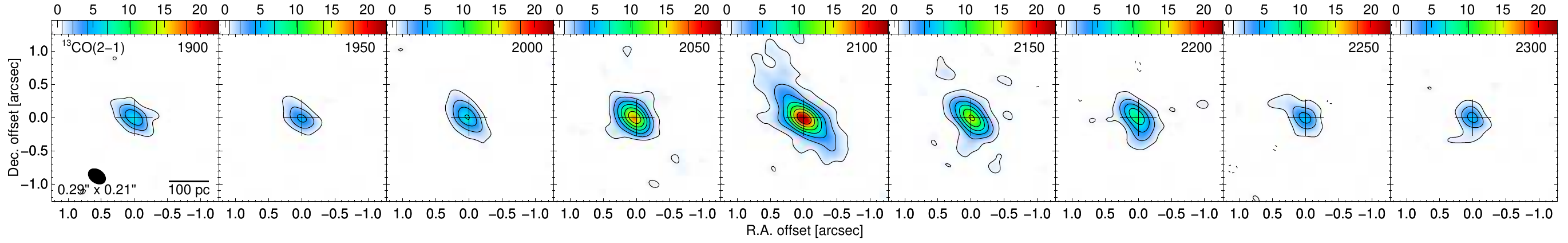}{1.0\textwidth}{(g) \thirteenCO(2--1)}
\fig{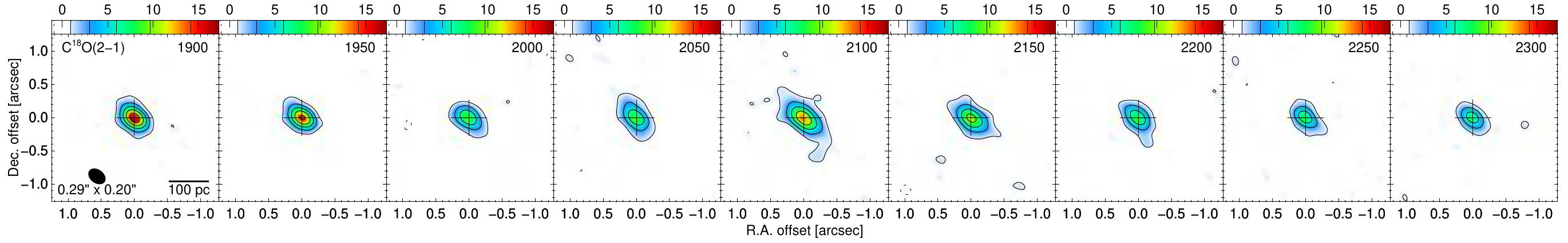}{1.0\textwidth}{(h) \CeighteenO(2--1) [blended with \NHtwoCN]}
\fig{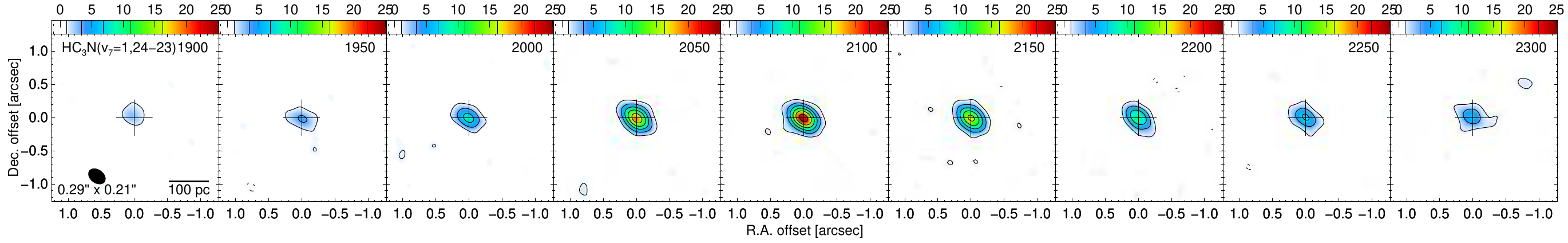}{1.0\textwidth}{(i) \HCthreeN($v_7=1$, 24--23, $l=1f$)}
\fig{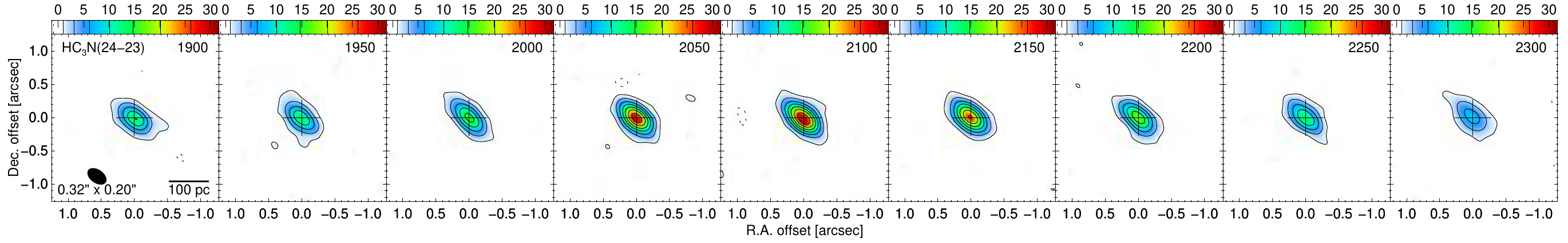}{1.0\textwidth}{(j) \HCthreeN(24--23)}
\fig{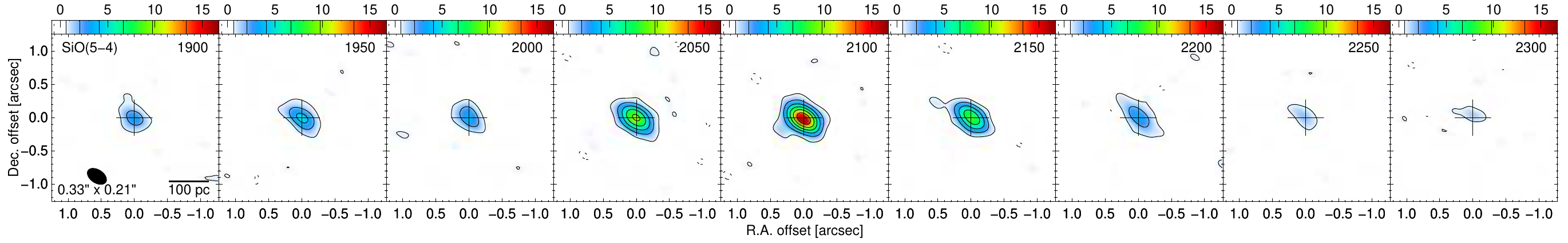}{1.0\textwidth}{(k) SiO(5--4)}
\fig{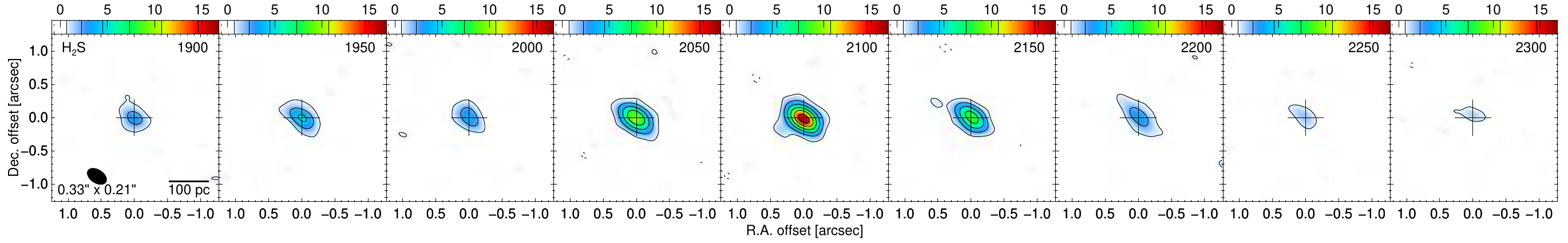}{1.0\textwidth}{(l) \HtwoS($2_{2,0}$--$2_{1,1}$) }
\caption{ }
\end{figure}

\begin{figure}[t]
\figurenum{\ref{f.chmap.a220.B9} (continued)}
\fig{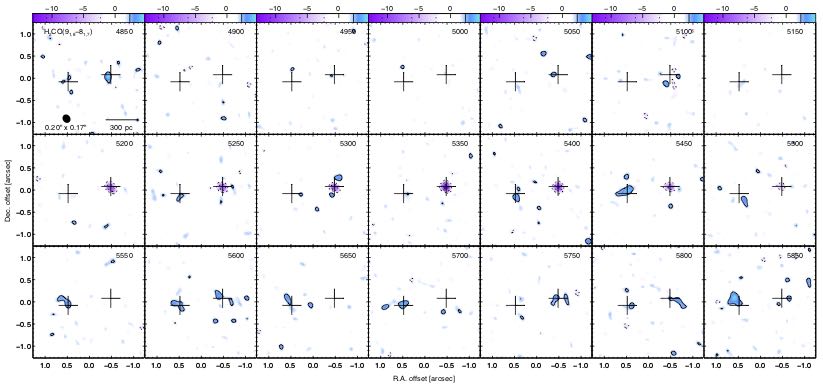}{1.0\textwidth}{(c) \HtwoCO($9_{1,8}$--$8_{1,7}$) }
\caption{ }
\end{figure}
\begin{figure}[t]
\figurenum{\ref{f.chmap.a220.B7} (continued)}
\fig{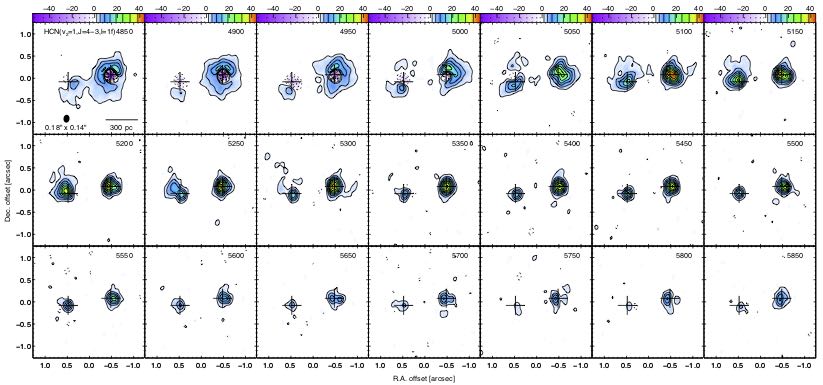}{1.0\textwidth}{(c) HCN($v_2=1$, 4--3, $l=1f$) [blended with \HCOplus(4--3)]}
\fig{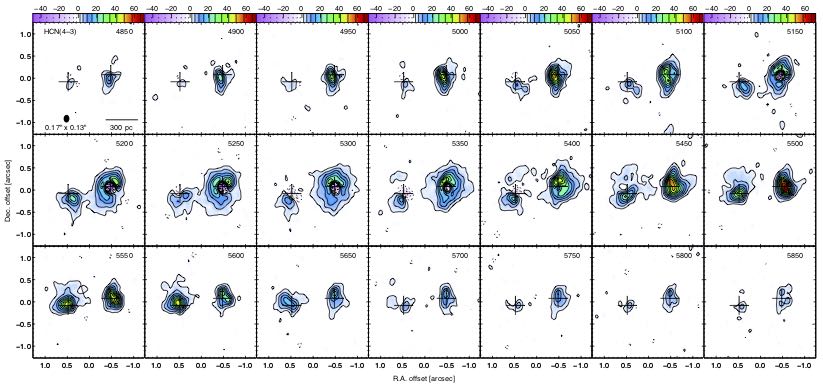}{1.0\textwidth}{(d) HCN(4--3) }
\caption{ }
\end{figure}
\begin{figure}[t]
\figurenum{\ref{f.chmap.a220.B7} (continued)}
\fig{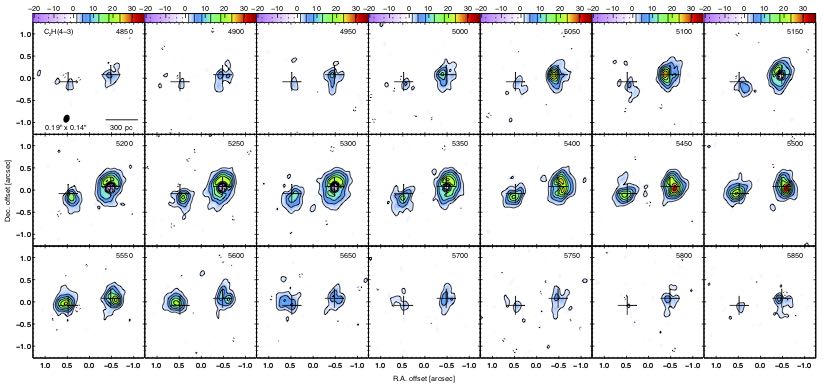}{1.0\textwidth}{(e) \CtwoH(4--3)}
\fig{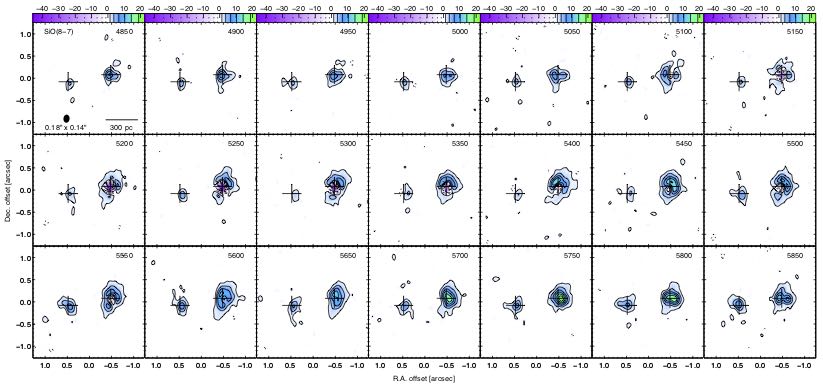}{1.0\textwidth}{(f) SiO(8--7) [blended with \HthirteenCOplus(4--3)]}
\caption{ }
\end{figure}
\begin{figure}[t]
\figurenum{\ref{f.chmap.a220.B7} (continued)}
\fig{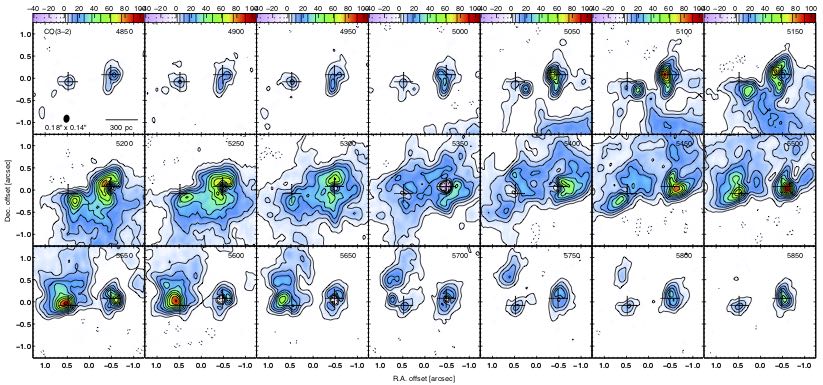}{1.0\textwidth}{(g) CO(3--2) [blended with \HthirteenCN(4--3)]}
\fig{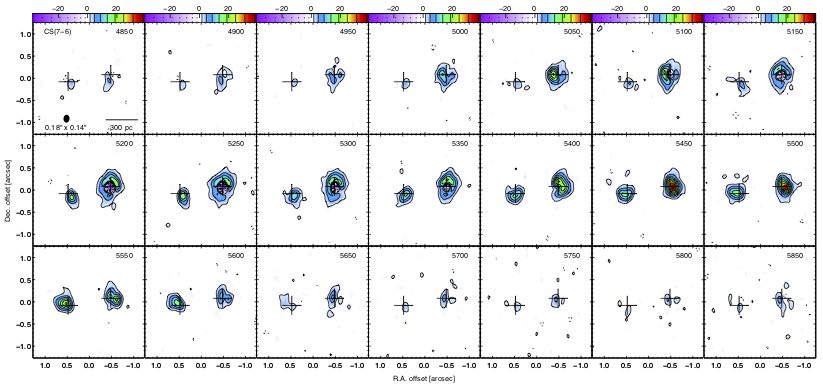}{1.0\textwidth}{(h) CS(7--6)}
\caption{ }
\end{figure}
\begin{figure}[t]
\figurenum{\ref{f.chmap.a220.B6} (continued)}
\fig{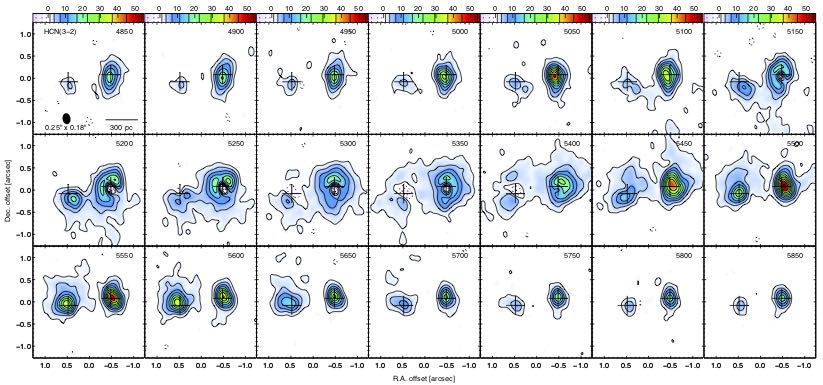}{1.0\textwidth}{(c) HCN(3--2)}
\fig{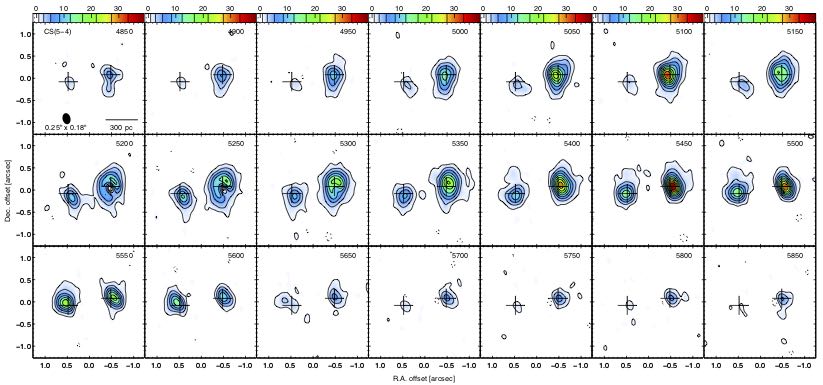}{1.0\textwidth}{(d) CS(5--4)}
\caption{ }
\end{figure}
\begin{figure}[t]
\figurenum{\ref{f.chmap.a220.B6} (continued)}
\fig{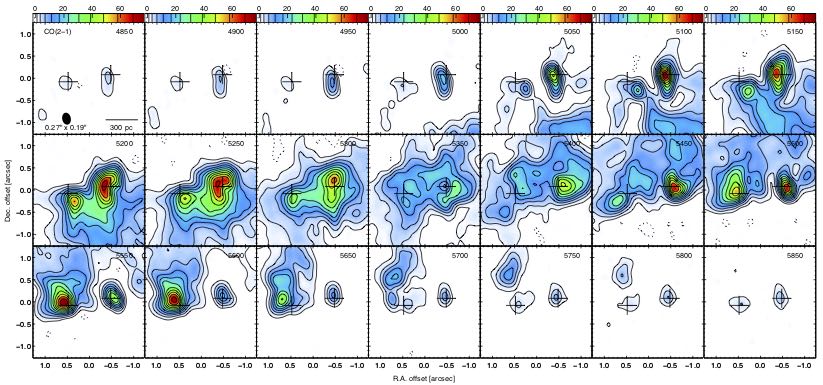}{1.0\textwidth}{(e) CO(2--1)}
\caption{ }
\end{figure}

\clearpage
\begin{figure}[t]
\figurenum{\ref{f.sqash.n4418} (continued)}
\includegraphics[width=0.233\textwidth]{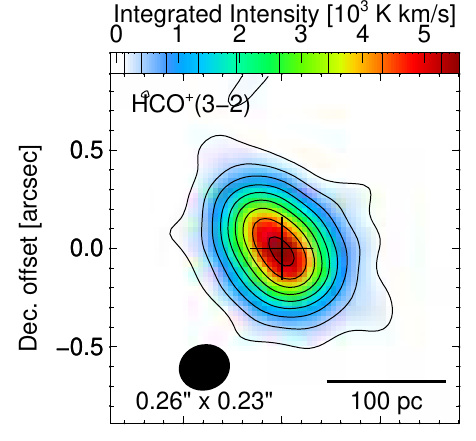} 
\includegraphics[width=0.185\textwidth]{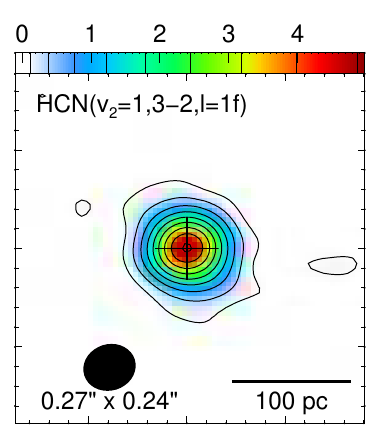} 
\includegraphics[width=0.185\textwidth]{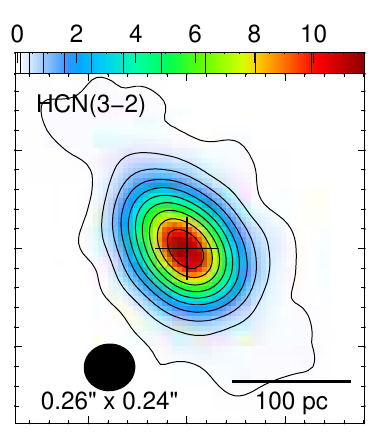} 
\includegraphics[width=0.185\textwidth]{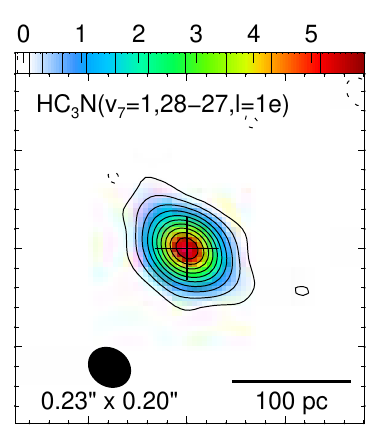} 
\includegraphics[width=0.185\textwidth]{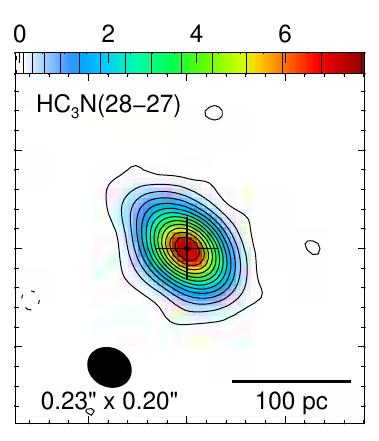} 
\\
\includegraphics[width=0.233\textwidth]{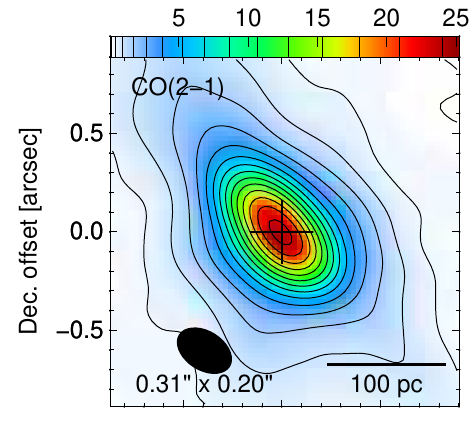} 
\includegraphics[width=0.185\textwidth]{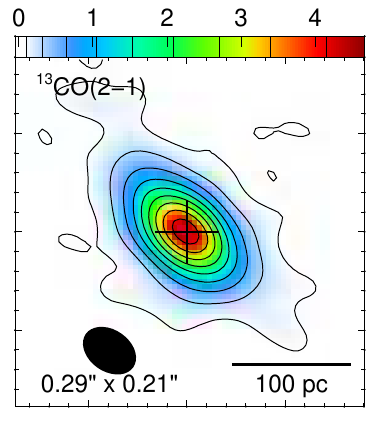} 
\includegraphics[width=0.185\textwidth]{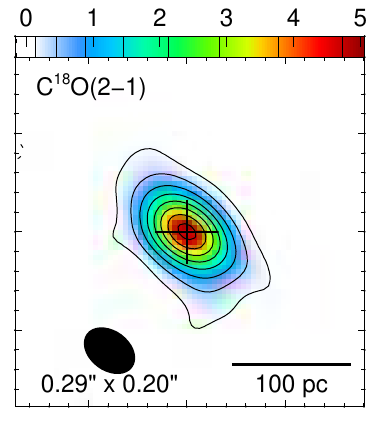} 
\includegraphics[width=0.185\textwidth]{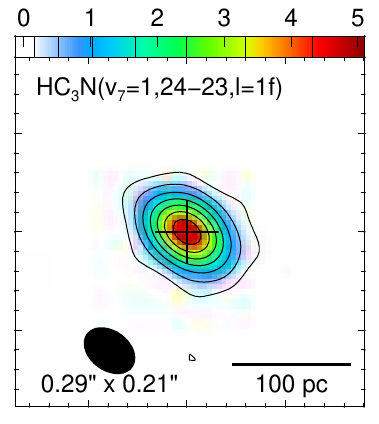} 
\includegraphics[width=0.185\textwidth]{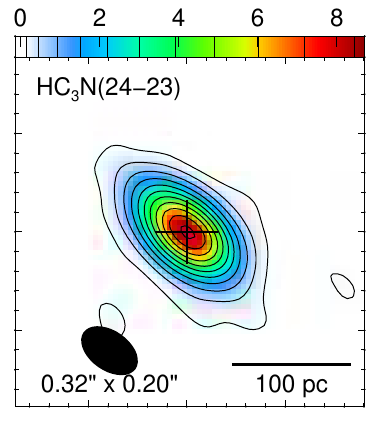} 
\\
\includegraphics[width=0.233\textwidth]{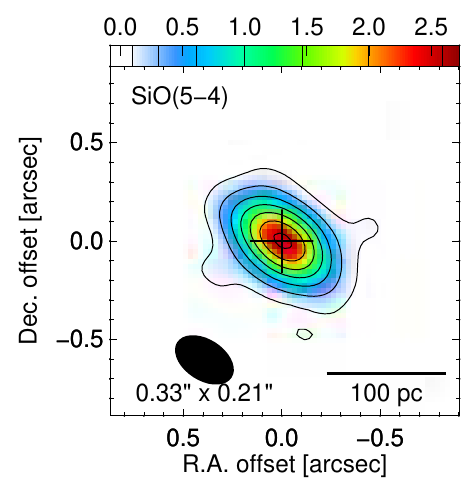} 
\includegraphics[width=0.185\textwidth]{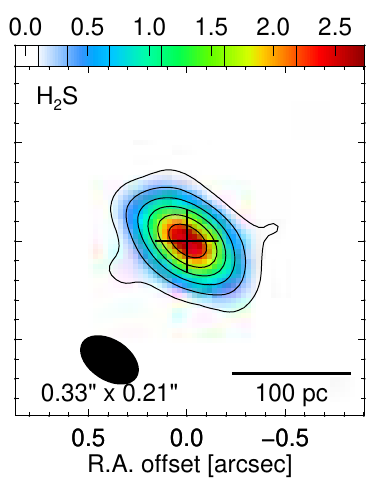} 

\caption{ }
\end{figure}

 		
\clearpage

\clearpage


\clearpage
\appendix

\section{Initial Spectral Fitting of NGC 4418}
\label{ap.spid}
\begin{figure}[t]
\epsscale{1.1}
\plotone{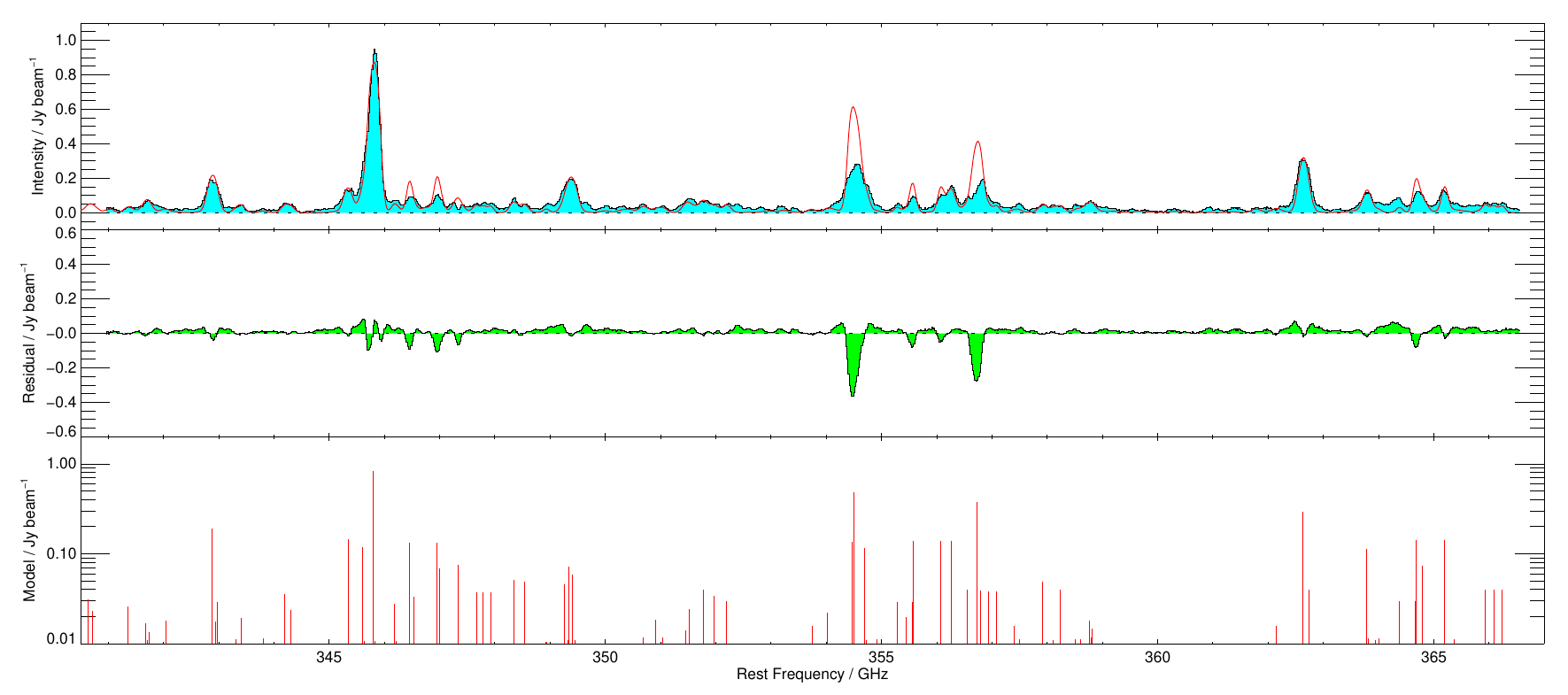}
\plotone{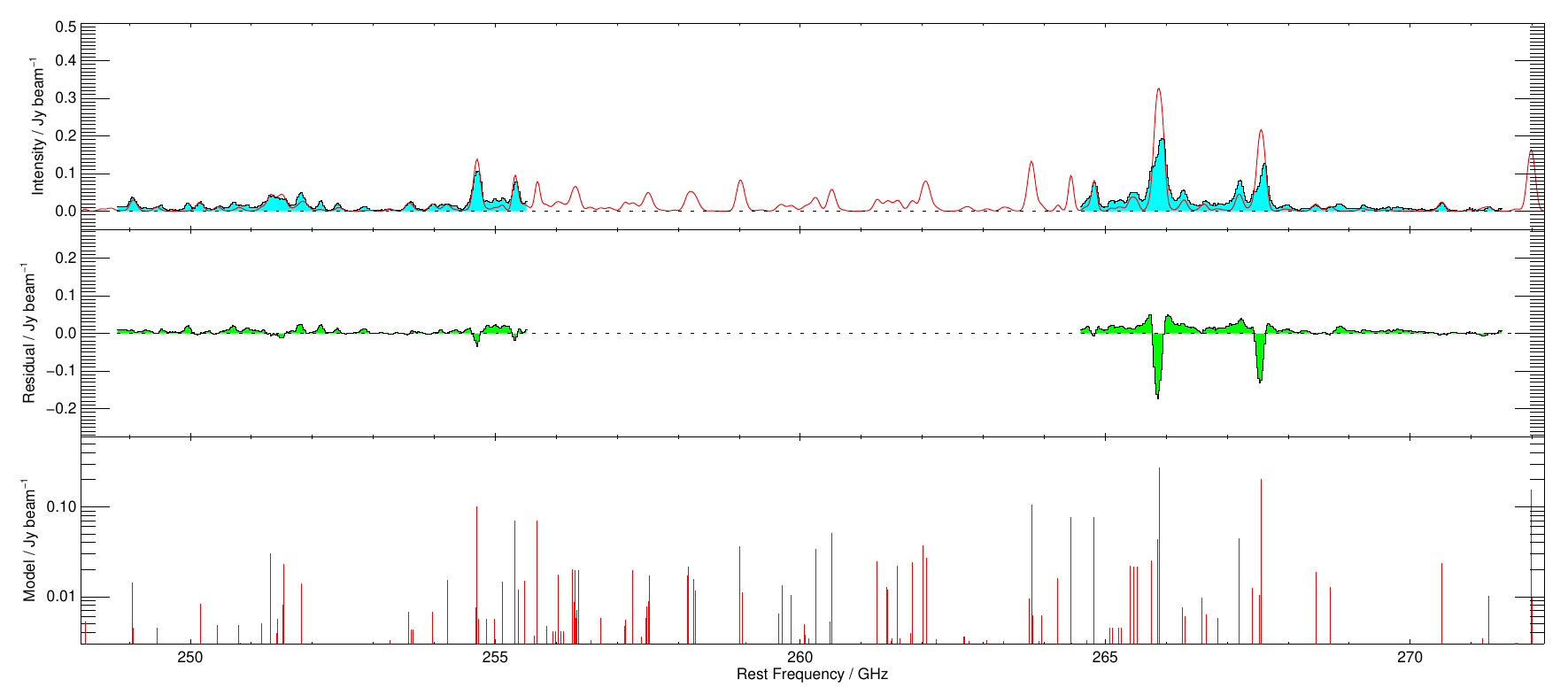}
\plotone{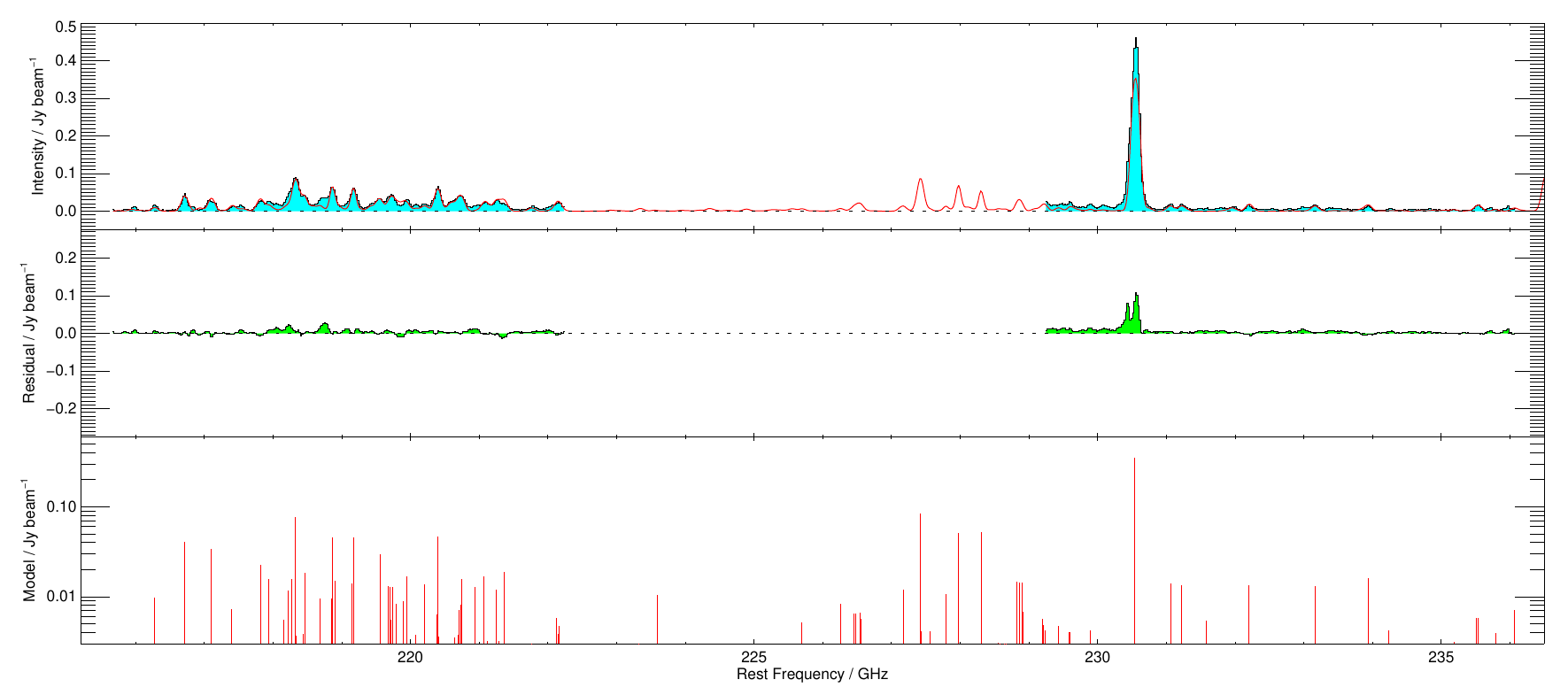}
\caption{ \label{f.specfit}
Spectral fitting of NGC 4418.
In each panel, the top sub-panel shows in light-blue the continuum-subtracted spectrum at 0\farcs35 and 20 MHz resolutions;
it also shows the model spectrum fitted to the data in red.
The middle sub-panel shows the fitting residual (i.e., observed $-$ model spectrum).
The bottom sub-panel shows peak intensities of the individual lines in the model.
}
\end{figure}

Figure~\ref{f.specfit} shows our initial fitting to the 0\farcs35-beam spectrum toward the nucleus of NGC 4418.
In each of the three panels, the top sub-panel shows the observed (light blue) and model (red) spectra.
The middle sub-panel shows their difference (i.e., observed $-$ model). 
The bottom sub-panel shows individual line transitions at their rest frequencies without smoothing them with the line widths. 
The modeling procedure is in Section \ref{s.Vsys_lineID.lineID}, 
and the model line-emitting region was a Gaussian of 0\farcs3 FWHM without continuum emission.
Table \ref{t.species} lists 55 species in our model. Four of them have their vibrationally excited states included.
Table \ref{t.specfit} lists 255 lines in the frequency ranges of our observations.
It shows those lines having peak model intensities $\geq$ 10 mJy \perbeam\ in Band 7 and $\geq$ 3 mJy \perbeam\ in Band 6;
the thresholds are about 3 $\sigma$ in our data. Our model spectral synthesis had no cutoffs.
The table also gives the velocity-integrated intensity in our model for each line.
Note that an entry here does not necessarily mean detection of the particular transition in our data.
Since we fitted our entire spectrum rather than individual lines, the entry may be due to 
other brighter transitions of the same species and excitation modeling.

\begin{deluxetable*}{ll}{!bht}
\tablewidth{0pt}
\tablecaption{Species in the NGC 4418 spectral model \label{t.species}}
\tablecolumns{2}
\tablehead{
	\colhead{$N_{\rm atom}$} &
	\colhead{Species} }
\startdata
2 & CO, \thirteenCO, \CeighteenO, CO$^{+}$, CS, $^{13}$CS, $^{13}$CN, 
NO, NS, N$^{34}$S, 
SiO, $^{29}$SiO, $^{30}$SiO,  
SO
\\
3 & 
H$_2$S, 
HCN, H$^{13}$CN, HC$^{15}$N, 
HNC, HN$^{13}$C, H$^{15}$NC,
\HCOplus, H$^{13}$CO$^{+}$,  HC$^{18}$O$^{+}$, \\ 
  & HCO, HOC$^{+}$, 
HCS$^{+}$,   
CCH, $^{13}$CCH, C$^{13}$CH, CCS,  
OCS
\\
4 & H$_3$O$^+$, H$_2$CS, HNCO, p-H$_2$CO, o-H$_2$CO, C$_3$H, HOCO$^{+}$
\\
5 & CH$_2$NH, \HCthreeN, c-C$_3$H$_2$, H$_2$CCN, NH$_2$CN, H$^{13}$CCCN, HC$^{13}$CCN, HCC$^{13}$CN 
\\
6 & CH$_3$OH, CH$_3$CN, CH$_3$$^{13}$CN, $^{13}$CH$_3$CN
\\
7 & CH$_3$CCH, C$_2$H$_3$CN
\\
9 & C$_2$H$_5$CN, C$_2$H$_5$OH 
\\
\enddata
\tablecomments{
Major molecules can be absent if they have no transition in our spectral coverage. 
Vibrationally excited states are in our model for HCN, \HCOplus, \HCthreeN, and CH$_3$CN.
}
\end{deluxetable*}
 
\begin{deluxetable*}{crrl}
\tablewidth{0pt}
\tablecaption{Initial Spectral Model of the nucleus of NGC 4418 \label{t.specfit}}
\tablecolumns{4}
\tablehead{
	\colhead{$f_{\rm rest}$} & 
	\colhead{$I_\nu$ } &
	\colhead{$S$} & 
	\colhead{Species} 
}
\decimals
\colnumbers
\startdata
341.35083 &     26.2 &     4.3 &  HCS+                  \\
341.68257 &     16.8 &     2.7 &  CH3CCH                \\
341.71509 &     10.9 &     1.8 &  CH3CCH                \\
341.73461 &     12.8 &     2.1 &  CH3CCH                \\
341.74111 &     13.5 &     2.2 &  CH3CCH                \\
342.03807 &     18.2 &     2.9 &  NH2CN                 \\
342.88300 &    192.3 &    37.2 &  CS                    \\
342.94437 &     17.7 &     2.8 &  H2CS                  \\
342.97911 &     29.1 &     4.7 &  Si-29-O               \\
343.31964 &     11.3 &     1.8 &  H2CS                  \\
343.40812 &     19.1 &     3.1 &  H2CS                  \\
343.41233 &     19.1 &     3.1 &  H2CS                  \\
343.81076 &     11.3 &     1.8 &  H2CS                  \\
344.20032 &     35.6 &     5.8 &  HCN-15                \\
344.31061 &     23.6 &     3.8 &  SO                    \\
345.33976 &    144.1 &    26.0 &  HC-13-N               \\
345.60901 &    118.8 &    19.4 &  HC3N,v=0              \\
345.63213 &     10.7 &     1.3 &  HC3N,v5=1/v7=3        \\
345.79599 &    831.9 &   192.0 &  CO                    \\
345.82329 &     10.6 &     1.7 &  NS                    \\
345.82467 &     10.7 &     1.3 &  HC3N,v5=1/v7=3        \\
346.17440 &     27.8 &     3.3 &  HC3N,v6=1             \\
346.22014 &     10.6 &     1.7 &  NS                    \\
346.44620 &     27.8 &     3.3 &  HC3N,v6=1             \\
346.45573 &    132.5 &    15.7 &  HC3N,v7=1             \\
346.52848 &     33.4 &     5.4 &  SO                    \\
346.94912 &    132.9 &    15.7 &  HC3N,v7=1             \\
346.99834 &     69.3 &    11.6 &  HC-13-O+              \\
347.33063 &     76.1 &    12.8 &  SiO                   \\
\vdots    &  \vdots  &  \vdots & \vdots                 \\
\enddata
\tablecomments{
(1) Rest frequency in GHz.
(2) Peak line intensity (in mJy \perbeam) of the model for our 0\farcs35 spectrum.
(3) Velocity integrated line intensity (in Jy \perbeam\ \kms) in the model.
(4) Species (with vibrational states when necessary). 
The full contents of this table are provided online.
}
\end{deluxetable*}


\begin{thebibliography}{}
\bibitem[Aalto et al.(1995)]{Aalto95} 
	Aalto, S., Booth, R.~S., Black, J.~H., et al.\ 
	1995, \aap, 300, 369
\bibitem[Aalto et al.(2009)]{Aalto09} 
	Aalto, S., Wilner, D., Spaans, M., et al.\ 
	2009, \aap, 493, 481
\bibitem[Aalto et al.(2015)]{Aalto15} 
	Aalto, S., Mart{\'{\i}}n, S., Costagliola, F., et al.\ 
	2015, \aap, 584, A42	
\bibitem[Armus et al.(2009)]{Armus09} 
	Armus, L., Mazzarella, J.~M., Evans, A.~S., et al.\ 
	2009, \pasp, 121, 559	
\bibitem[Baan et al.(1982)]{Baan82} 
	Baan, W.~A., Wood, P.~A.~D., \& Haschick, A.~D.\ 
	1982, \apjl, 260, L49 
\bibitem[Barcos-Mu{\~n}oz et al.(2015)]{Barcos-Munoz15} 
	Barcos-Mu{\~n}oz, L., Leroy, A.~K., Evans, A.~S., et al.\ 
	2015, \apj, 799, 10		
\bibitem[Barcos-Mu{\~n}oz et al.(2018)]{Barcos-Munoz18} 
	Barcos-Mu{\~n}oz, L., Aalto, S., Thompson, T.~A., et al.\ 
	2018, \apjl, 853, L28	
\bibitem[Bizzocchi et al.(2017)]{Bizzocchi17} 
	Bizzocchi, L., Tamassia, F., Laas, J., et al.\ 
	2017, \apjs, 233, 11
\bibitem[Boettcher et al.(2020)]{Boettcher20} 
	Boettcher, E., et al.\
	2020, \aap, in press
\bibitem[Brown \& Wilson(2019)]{Brown19} 
	Brown, T., \& Wilson, C.~D.\ 
	2019, \apj, 879, 17
\bibitem[Condon et al.(1990)]{Condon90}  
	Condon, J.~J., Helou, G., Sanders, D.~B., et al.\ 
	1990, \apjs, 73, 359	
\bibitem[Condon et al.(1991)]{Condon91}  
	Condon, J.~J., Huang, Z.-P., Yin, Q.~F., et al.\ 
	1991, \apj, 378, 65	
\bibitem[Costagliola \& Aalto(2010)]{Costagliola10}  
	Costagliola, F., \& Aalto, S.\ 
	2010, \aap, 515, A71 			
\bibitem[Costagliola et al.(2013)]{Costagliola13}  
	Costagliola, F., Aalto, S., Sakamoto, K., et al.\ 
	2013, \aap, 556, A66	
\bibitem[Costagliola et al.(2015)]{Costagliola15} 
	Costagliola, F., Sakamoto, K., Muller, S., et al.\ 
	2015, \aap, 582, A91	
\bibitem[Downes \& Solomon(1998)]{Downes98} 
	Downes, D. \& Solomon, P.~M.\ 
	1998, \apj, 507, 615 		
\bibitem[Downes \& Eckart(2007)]{Downes07} 
	Downes, D. \& Eckart, A.\ 	
	2007, \aap, 468, L57 	
\bibitem[Dwek \& Arendt(2020)]{Dwek20} 
	Dwek, E. \& Arendt, R.~G.\ 
	2020, \apj, 901, 36				
\bibitem[Evans et al.(2003)]{Evans03} 
	Evans, A.~S., Becklin, E.~E., Scoville, N.~Z., et al.\ 
	2003, \aj, 125, 2341
\bibitem[Fadda \& Rodighiero(2014)]{Fadda14} 
	Fadda, D. \& Rodighiero, G.\ 
	2014, \mnras, 444, L95 
\bibitem[Fluetsch et al.(2019)]{Fluetsch19}  
	Fluetsch, A., Maiolino, R., Carniani, S., et al.\ 
	2019, \mnras, 483, 4586	
\bibitem[Garc{\'{\i}}a-Burillo et al.(2014)]{GarciaBurillo14} 
	Garc{\'{\i}}a-Burillo, S., Combes, F., Usero, A., et al.\ 2014, \aap, 567, A125	
\bibitem[Graci{\'a}-Carpio et al.(2006)]{Garcia-Carpio06}{2006ApJ...640L.135G} 
	Graci{\'a}-Carpio, J., Garc{\'\i}a-Burillo, S., Planesas, P., et al.\ 
	2006, \apjl, 640, L135 
\bibitem[Gonz{\'a}lez-Alfonso et al.(2004)]{GA04} 
	Gonz{\'a}lez-Alfonso, E., Smith, H.~A., Fischer, J., et al.\ 
	2004, \apj, 613, 247 	
\bibitem[Gonz{\'a}lez-Alfonso et al.(2012)]{GA12} 
	Gonz{\'a}lez-Alfonso, E., Fischer, J., Graci{\'a}-Carpio, J., et al.\ 
	2012, \aap, 541, A4		
\bibitem[Gonz\'{a}lez-Alfonso \& Sakamoto (2019)]{GS19}	
	Gonz\'{a}lez-Alfonso, E. \& Sakamoto, 
	2019, \apj, 882, 153
\bibitem[Greve et al.(2009)]{Greve09} 
	Greve, T.~R., Papadopoulos, P.~P., Gao, Y., et al.\ 
	2009, \apj, 692, 1432	
\bibitem[Harada et al.(2010)]{Harada10}  
	Harada, N., Herbst, E., \& Wakelam, V.\ 
	2010, \apj, 721, 1570	
\bibitem[Imanishi et al.(2004)]{Imanishi04} 
	Imanishi, M., Nakanishi, K., Kuno, N., et al.\ 
	2004, \aj, 128, 2037 
\bibitem[Jarrett et al.(2000)]{Jarrett00}  
	Jarrett, T.~H., Chester, T., Cutri, R., et al.\ 
	2000, \aj, 119, 2498	
\bibitem[Kawara et al.(1990)]{Kawara90} 
	Kawara, K., Taniguchi, Y., Nakai, N., et al.\ 
	1990, \apjl, 365, L1. 	
\bibitem[K{\"o}nig et al.(2017)]{Koenig17} 
	K{\"o}nig, S., Mart{\'\i}n, S., Muller, S., et al.\ 
	2017, \aap, 602, A42 	
\bibitem[Lahuis et al.(2007)]{Lahuis07} 
	Lahuis, F., Spoon, H.~W.~W., Tielens, A.~G.~G.~M., et al.\ 
	2007, \apj, 659, 296
\bibitem[Lutz et al.(2020)]{Lutz20} 
	Lutz, D., Sturm, E., Janssen, A., et al.\ 
	2020, \aap, 633, A134	
\bibitem[Markwardt(2009)]{Markwardt09}
	Markwardt, C.~B.\ 
	2009, Astronomical Data Analysis Software and Systems XVIII, 411, 251 
\bibitem[Mart{\'{\i}}n et al.(2011)]{Martin11} 
	Mart{\'{\i}}n, S., Krips, M., Mart{\'{\i}}n-Pintado, J., et al.\ 
	2011, \aap, 527, A36 		
\bibitem[Mart{\'{\i}}n et al.(2016)]{Martin16} 
	Mart{\'{\i}}n, S., Aalto, S., Sakamoto, K., et al.\ 
	2016, \aap, 590, A25	
\bibitem[Mart{\'\i}n et al.(2019)]{MADCUBA} 
	Mart{\'\i}n, S., Mart{\'\i}n-Pintado, J., Blanco-S{\'a}nchez, C., et al.\ 
	2019, \aap, 631, A159
\bibitem[Martin-Pintado et al.(1992)]{MP92} 
	Martin-Pintado, J., Bachiller, R., \& Fuente, A.\ 
	1992, \aap, 254, 315				
\bibitem[Mart{\'{\i}}-Vidal et al.(2014)]{uvmultifit14} 
	Mart{\'{\i}}-Vidal, I., Vlemmings, W.~H.~T., Muller, S., \& Casey, S.\ 
	2014, \aap, 563, A136	
\bibitem[Matsushita et al.(2009)]{Matsushita09}  
	Matsushita, S., Iono, D., Petitpas, G.~R., et al.\ 
	2009, \apj, 693, 56				
\bibitem[McMullin et al.(2007)]{CASA07} 
	McMullin, J.~P., Waters, B., Schiebel, D., Young, W., \& Golap, K.\ 
	2007, Astronomical Data Analysis Software and Systems XVI, 376, 127
\bibitem[Mor\'{e}(1977)]{More77}
	Mor\'{e}, J.
	1977, 
	``The Levenberg-Marquardt Algorithm: Implementation and Theory,'' in Numerical Analysis, 
	vol. 630, ed. G. A. Watson (Springer-Verlag: Berlin), 105
\bibitem[Mor\'{e} \& Wright(1993)]{More93} 	
	Mor\'{e}, J. \& Wright, S. 
	1993, 
  	Optimization Software Guide, Frontiers in Applied Mathematics, vol. 14, 
  	(Philadelphia, PA: SIAM)
\bibitem[Norris(1988)]{Norris88} 
	Norris, R.~P.\ 
	1988, \mnras, 230, 345 	
\bibitem[Ohyama et al.(2019)]{Ohyama19} 
	Ohyama, Y., Sakamoto, K., Aalto, S., et al.\ 
	2019, \apj, 871, 191		
\bibitem[Pearson(1999)]{Pearson99} 
	Pearson, T.~J.\ 
	1999, Synthesis Imaging in Radio Astronomy II, 180, 335 
\bibitem[Rangwala et al.(2011)]{Rangwala11} 
	Rangwala, N., Maloney, P.~R., Glenn, J., et al.\ 
	2011, \apj, 743, 94 	
\bibitem[Rangwala et al.(2015)]{Rangwala15} 
	Rangwala, N., Maloney, P.~R., Wilson, C.~D., et al.\ 
	2015, \apj, 806, 17	
\bibitem[Rieke et al.(1985)]{Rieke85} 
	Rieke, G.~H., Cutri, R.~M., Black, J.~H., et al.\ 
	1985, \apj, 290, 116		
\bibitem[Roche et al.(1986)]{Roche86} 
	Roche, P.~F., Aitken, D.~K., Smith, C.~H., \& James, S.~D.\ 
	1986, \mnras, 218, 19P		
\bibitem[Rolffs et al.(2011)]{Rolffs11} 
	Rolffs, R., Schilke, P., Wyrowski, F., et al.\ 
	2011, \aap, 527, A68	
\bibitem[Rosenberg et al.(2015)]{Rosenberg15} 
	Rosenberg, M.~J.~F., van der Werf, P.~P., Aalto, S., et al.\ 
	2015, \apj, 801, 72 	
\bibitem[Rowan-Robinson \& Crawford(1989)]{RRC89} 
	Rowan-Robinson, M., \& Crawford, J.\ 
	1989, \mnras, 238, 523 
\bibitem[Sakamoto et al.(1999)]{Sakamoto99} 
	Sakamoto, K., Scoville, N.~Z., Yun, M.~S., et al.\ 
	1999, \apj, 514, 68	
\bibitem[Sakamoto et al.(2008)]{Sakamoto08} 
	Sakamoto, K., Wang, J., Wiedner, M.~C., et al.\ 
	2008, \apj, 684, 957-977
\bibitem[Sakamoto et al.(2009)]{Sakamoto09} 
	Sakamoto, K., Aalto, S., Wilner, D.~J., et al.\ 
	2009, \apjl, 700, L104		
\bibitem[Sakamoto et al.(2010)]{Sakamoto10}   
	Sakamoto, K., Aalto, S., Evans, A.~S., Wiedner, M.~C., \& Wilner, D.~J.\ 
	2010, \apjl, 725, L228			
\bibitem[Sakamoto et al.(2013)]{Sakamoto13} 
	Sakamoto, K., Aalto, S., Costagliola, F., et al.\ 
	2013, \apj, 764, 42 
\bibitem[Sakamoto et al.(2017)]{Sakamoto17} 
	Sakamoto, K., Aalto, S., Barcos-Mu{\~n}oz, L., et al.\ 
	2017, \apj, 849, 14 	
\bibitem[Sakamoto et al.(2021)Paper I]{Paper1} 
	Sakamoto, K., Gonzalez-Alfonso, E., Martin, S., et al.\ 
	2021, ApJ in press (paper I), arXiv:2109.06695 
\bibitem[Sanders \& Mirabel(1996)]{SM96} 
	Sanders, D. B., and Mirabel, I. F.
	1996, \araa, 34, 749		
\bibitem[Scoville et al.(1997)]{Scoville97} 
	Scoville, N. Z., Yun, M. S., and Bryant, P. M.
	1997, \apj, 484, 702
\bibitem[Scoville et al.(1998)]{Scoville98} 
	Scoville, N.~Z., Evans, A.~S., Dinshaw, N., et al.\ 
	1998, \apjl, 492, L107		
\bibitem[Scoville et al.(2015)]{Scoville15} 
	Scoville, N., Sheth, K., Walter, F., et al.\ 
	2015, \apj, 800, 70	
\bibitem[Scoville et al.(2017)]{Scoville17} 
	Scoville, N., Murchikova, L., Walter, F., et al.\ 
	2017, \apj, 836, 66 
\bibitem[Smith et al.(1998)]{Smith98} 
	Smith, H.~E., Lonsdale, C.~J., Lonsdale, C.~J., \& Diamond, P.~J.\ 
	1998, \apjl, 493, L17		
\bibitem[Soifer et al.(1984)]{Soifer84} 
	Soifer, B.~T., Helou, G., Lonsdale, C.~J., et al.\ 
	1984, \apjl, 283, L1 	
\bibitem[Soifer et al.(1999)]{Soifer99} 
	Soifer, B.~T., Neugebauer, G., Matthews, K., et al.\ 
	1999, \apj, 513, 207	
\bibitem[Thorwirth et al.(2000)]{Thorwirth00} 
	Thorwirth, S., M{\"u}ller, H.~S.~P., Lewen, F., et al.\ 
	2000, \aap, 363, L37		
\bibitem[Tunnard et al.(2015)]{Tunnard15} 
	Tunnard, R., Greve, T.~R., Garcia-Burillo, S., et al.\ 
	2015, \apj, 800, 25	
\bibitem[Van der Tak et al.(2008)]{vdTak08} 
	Van der Tak, F.~F.~S., Aalto, S., \& Meijerink, R.\ 
	2008, \aap, 477, L5			
\bibitem[Varenius et al.(2014)]{Varenius14} 
	Varenius, E., Conway, J.~E., Mart{\'\i}-Vidal, I., et al.\ 
	2014, \aap, 566, A15	 	
\bibitem[Varenius et al.(2016)]{Varenius16} 
	Varenius, E., Conway, J.~E., Mart{\'\i}-Vidal, I., et al.\ 
	2016, \aap, 593, A86	
\bibitem[Varenius et al.(2017)]{Varenius17} 
	Varenius, E., Costagliola, F., Kl{\"o}ckner, H.-R., et al.\ 
	2017, \aap, 607, A43	
\bibitem[Varenius et al.(2019)]{Varenius19} 
	Varenius, E., Conway, J.~E., Batejat, F., et al.\
	2019, \aap, 623, A173	
\bibitem[Veilleux et al.(2009)]{Veilleux09} 
	Veilleux, S., Rupke, D.~S.~N., Kim, D.-C., et al.\ 
	2009, \apjs, 182, 628 
\bibitem[Veilleux et al.(2013)]{Veilleux13} 
	Veilleux, S., Mel{\'e}ndez, M., Sturm, E., et al.\ 
	2013, \apj, 776, 27 
\bibitem[Wheeler et al.(2020)]{Wheeler20} 
	Wheeler, J., Glenn, J., Rangwala, N., et al.\ 
	2020, \apj, 896, 43	
\bibitem[Wilson et al.(2014)]{Wilson14}  
	Wilson, C.~D., Rangwala, N., Glenn, J., et al.\ 
	2014, \apjl, 789, L36 	
\bibitem[Wilson et al.(2009)]{ToRA6} 
	Wilson, T.~L., Rohlfs, K., \& H{\"u}ttemeister, S.\ 
	2009, Tools of Radio Astronomy, by Thomas L. Wilson; Kristen Rohlfs and Susanne H{\"u}ttemeister. 
	ISBN 978-3-540-85121-9. Published by Springer-Verlag, Berlin, Germany, 2009.
\bibitem[Wynn-Williams \& Becklin(1993)]{WWB93} 
	Wynn-Williams, C.~G. \& Becklin, E.~E.\ 
	1993, \apj, 412, 535. doi:10.1086/172941	
\bibitem[Zschaechner et al.(2016)]{Zschaechner16} 
	Zschaechner, L.~K., Ott, J., Walter, F., et al.\ 
	2016, \apj, 833, 41
\end{thebibliography}
\end{document}